\documentclass{aa}  

\usepackage{graphicx}
\usepackage{xcolor}
\usepackage{txfonts}
\newcommand{\totalnumberwithcontamination}{439 }
\newcommand{\totalnumber}{301 }
\newcommand{\surfacemodulationnumber}{147 }
\newcommand{\gdornumber}{24 }
\newcommand{\dsctnumber}{35 }
\newcommand{\EBnumber}{5 }
\newcommand{\gdornumberwithcleargmodepattern}{11 }
\newcommand{\bprp}{G_\mathrm{BP}-G_\mathrm{RP}}
\newcommand{\clusterage}{$102\pm15$\,Myr}
\newcommand{\clusterextinction}{$0.53\pm0.04$\,mag}
\newcommand{\echelle}{{\'e}chelle}
\newcommand{\Echelle}{{\'E}chelle}

\newcommand{\numax}{\mbox{$\nu_{\rm max}$}}
\newcommand{\Dnu}{\mbox{$\Delta\nu$}}
\newcommand{\muhz}{\mbox{$\mu$Hz}}

\newcommand{\Teff}{\mbox{$T_{\rm eff}$}}
\newcommand{\Msun}{\mbox{$\text{M}_{\odot}$}}
\newcommand{\Lsun}{\mbox{$\text{L}_{\odot}$}}

\usepackage[]{hyperref}
\usepackage{siunitx}

\usepackage{CJKutf8}
\newcommand{\CNnames}[1]{{\begin{CJK}{UTF8}{gbsn}~(#1)~\end{CJK}}}

\newcommand{\ca}[1]{\textcolor{black}{#1}}

\begin{document}

   \title{Asteroseismology of the young open cluster NGC 2516\\ I: Photometric and spectroscopic observations}

%\subtitle{Internal rotation for g-mode pulsators}

   \author{Gang Li \CNnames{李刚}\inst{1}
          \and
          Conny Aerts\inst{1,2,3}
          \and
          Timothy R. Bedding \inst{4}
          \and
          Dario J. Fritzewski \inst{1}
          \and
          Simon J. Murphy \inst{5}
          \and
          Timothy Van Reeth \inst{1}
          \and
          Benjamin T. Montet \inst{6,7}
          \and
          Mingjie Jian \CNnames{简明杰}\inst{8} 
                \and
        Joey S. G. Mombarg \inst{9}
        \and
          Seth Gossage\inst{10}
          \and
          K. R. Sreenivas\inst{4}
          }

   \institute{Institute of Astronomy (IvS), Department of Physics and Astronomy, KU Leuven, Celestijnenlaan 200D, 3001 Leuven, Belgium\\
     \email{gang.li@kuleuven.be, conny.aerts@kuleuven.be}
           \and
      Department of Astrophysics, IMAPP, Radboud University Nijmegen, PO Box 9010, 6500 GL Nijmegen, The Netherlands
      \and
      Max Planck Institut für Astronomie, Königstuhl 17, 69117 Heidelberg, Germany
         \and
             Sydney Institute for Astronomy (SIfA), School of Physics, University of Sydney, NSW 2006, Australia. 
        \and
            Centre for Astrophysics, University of Southern Queensland, Toowoomba, QLD 4350, Australia
        \and
             School of Physics, University of New South Wales, Sydney, NSW 2052, Australia
             \and
             UNSW Data Science Hub, University of New South Wales, Sydney, NSW 2052, Australia
             \and
            Department of Astronomy, Stockholm University, AlbaNova
            University Center, 106 91 Stockholm, Sweden
        \and
             IRAP, Université de Toulouse, CNRS, UPS, CNES, 14 Avenue Édouard Belin, 31400 Toulouse, France
                  \and
      Center for Interdisciplinary Exploration and Research in Astrophysics (CIERA), Northwestern University, 1800 Sherman Ave,
      Evanston, IL 60201, USA
             }

   %\date{Received September 15, 1996; accepted March 16, 1997}

% \abstract{}{}{}{}{} 
% 5 {} token are mandatory
 
  \abstract
   {Asteroseismic modelling of isolated stars presents significant challenges due to the difficulty in accurately determining stellar parameters, particularly the stellar age. These challenges can be overcome by observing stars in open clusters, whose coeval members share an initial chemical composition. The light curves from the all-sky survey by the Transiting Exoplanet Survey Satellite (TESS) allow us to investigate and analyse stellar variations in clusters with an unprecedented level of detail for the first time. }
   {We aim to detect gravity-mode oscillations in the early-type main-sequence members of the
     young open cluster NGC\,2516 to deduce their internal rotation rates.}
   {We selected the \totalnumber of member stars with no more than mild contamination as our sample. We analysed the full-frame image (FFI) light curves, which provide
     nearly continuous observations in the first and third years of
     TESS  monitoring. We also collected high-resolution spectra using the Fiber-fed Extended Range Optical Spectrograph (FEROS) for the g-mode pulsators, with the aim to assess the Gaia effective temperatures and gravities, and to prepare for future seismic modelling.}
   {By fitting the theoretical isochrones to the colour-magnitude diagram (CMD) of a cluster, we
     determined an age of \clusterage~and
     inferred the extinction at 550\,nm ($A_0$) is \clusterextinction. We identified
     \surfacemodulationnumber stars with surface brightness modulations,
     \gdornumber with gravity (g-)mode pulsations ($\gamma$\,Doradus or
     Slowly Pulsating B stars), and \dsctnumber with pressure (p-)mode pulsations
     ($\delta$\,Sct stars).  %We identified a regular frequency spacing of 6.21\,$\mathrm{d^{-1}}$ in one $\delta$\,Sct star.
     When sorted by colour index, the amplitude spectra of the $\delta$\,Sct stars show a distinct ordering and reveal a discernible
     frequency-temperature relationship.  The near-core rotation
     rates, measured from period spacing patterns in two SPB and nine $\gamma$\,Dor
     stars, reach up to $3\,\mathrm{d^{-1}}$. This is at the high end
     of the values found from {\it Kepler\/} data of field stars of
     similar variability type. The $\gamma$\,Dor stars of NGC\,2516
     have internal rotation rates as high as 50\% of their critical value, whereas the SPB stars exhibit rotation rates close to their critical rate. Although the B-type stars are rotating rapidly, we did not find long-term brightness and colour variations in the
     mid-infrared, which suggests that there are no disk or shell formation events in our sample. We also discussed the results of our spectroscopic observations for the g-mode pulsators. }
   %{We provide the first application of ensemble g-mode asteroseismology for a young open cluster. Future seismic modelling with both spectroscopic and cluster constraints will provide calibrations of stellar models of unprecedented level and will greatly boost our understanding of stellar interiors of fast rotating young stars. }
   {}

   \maketitle
%
%-------------------------------------------------------------------

\section{Introduction}

Clusters serve as excellent laboratories for stellar astrophysics due
to the advantage that member stars within the same cluster typically
share similar distance, age, and initial metallicity. Nevertheless, recent
studies have complicated these notions: some works reveal extended
coronae in nearby star clusters, where their sizes greatly exceed
those of their cluster cores \citep{Meingast2019, Meingast2021, Bouma2021}. Furthermore, both globular and open clusters have shown
evidence of multiple populations, indicating complex star formation
histories \citep{Gratton2012, LiChengYuan2016Natur,
  WangChen2022}. Additionally, uncertainties arise in stellar
evolutionary tracks and isochrones due to the complex input physics
and poor calibration of stellar models \citep[e.g.][]{Martins2013, Johnston2019A&A}.

Rotation, a process of stellar physics that remains poorly calibrated,
plays a crucial role in stellar evolution, as recognised for many
decades \citep[e.g.][]{Shajn1929MNRAS, Maeder2009}. For instance,
rotation enhances element mixing and transports more fuel to
the core, resulting in increased luminosity and extended
lifetimes. Rapid rotation also causes a star to become oblate and
generates a surface temperature gradient, affecting the observed
colour and brightness \citep[e.g.][known as the gravity darkening
  effect]{von_Zeipel1924, Zahn2010, Espinosa_Lara2011,
  Bouchaud2020}. Consequently, the rotational effects strongly
influence the locations of stars on the colour-magnitude diagram
(CMD), leading to scatter in the observed CMD. Therefore, rotation is
considered the most plausible mechanism for the extended main-sequence turn-off (eMSTO) phenomenon \citep[e.g.][among many other
  studies]{Bertelli2003AJ, Glatt2009, Bastian2009MNRAS,
  Girardi2013MNRAS, Correnti2015, DAntona2015MNRAS, Brandt2015ApJ,
  Bastian2016MNRAS, Bastian2018MNRAS, ChLi2019ApJ, Gossage2019ApJ,
  Lim2019NatAs}.

There is a limited availability of observed rotation rates for hot eMSTO
stars, but numerous surface rotation measurements are available for cool
main-sequence stars \citep[spectral class F8V or later,][]{Kraft1967}. The rotation rates
of these cool stars can be easily and accurately measured through
their surface modulations. \textcolor{black}{These are quasi-periodic brightness variations caused by magnetism-induced surface \ca{inhomogeneities} in rotating stars} \citep[e.g.][]{Irwin2009, McQuillan2014ApJS}. 
These observations have paved the way for gyrochronology, a method that determines the ages of cool
main-sequence stars based on their masses and rotation rates, while
other physical properties remain relatively unchanged during their
long main-sequence stage \citep{Skumanich1972ApJ, Barnes2003,
  Barnes2007}. However, gyrochronology is not applicable to hot
main-sequence stars since they lack thick convective envelopes,
resulting in no or weak magnetic braking. Consequently, hot
main-sequence stars tend to rotate rapidly, with
surface rotation frequencies of the order of $1\,\mathrm{d^{-1}}$ or $v\sin i \sim
100\,\mathrm{km/s}$ \citep{Royer2007,Li2020MNRAS_611}. For these
stars, surface modulation is often absent or weak, but their internal
rotation rates can be obtained through 
asteroseismology.

Asteroseismology, the field dedicated to the interpretation of stellar
oscillations, has proven to be a potent tool for peering into the
interior physics of stars \citep[e.g.,][]{Aerts2021RvMP}. The fundamental concept revolves around standing
waves that penetrate different layers of stars. These waves undergo
modifications due to the local environment such as sound speed and
density. Consequently, the eigenfrequencies carry information about
the internal structure of stars \citep[for more detailed information,
  we refer to textbooks such as][]{Unno1989book, Aerts2010book,
  Basu2017asda.book}. Notably, the temperature range at which
gyrochronology loses its effectiveness aligns with the cool boundary
of the classical instability strip (IS) for pulsating stars \citep[roughly
  from 7000 to 10000\,K, where $\delta$\,Scuti ($\delta$\,Sct) stars and $\gamma$\,Doradus ($\gamma$\,Dor)
  stars
  reside][]{Dupret2005A&A,Bouabid2013,Xiong2016,Murphy2019}. Slowly Pulsating B-type (SPB) stars and $\beta$\,Cephei ($\beta$\,Cep) stars appear at
higher temperature ranges (hotter than $\sim$10000\,K). As a result,
various types of pulsating stars are being studied asteroseismically, which are hotter than Sun-like stars with solar-like oscillations.

At the temperature range of eMSTO in young open clusters, stars more massive than the Sun are
the main objects for asteroseismology. Some of these stars pulsate in
pressure (p) modes, including $\beta$\,Cep stars
\citep{Sterken1993, Aerts2003} and $\delta$\,Sct stars
\citep{Goupil2005, Handler2009, Bedding2020}. These modes are
particularly sensitive to the outer envelopes of stars. Conversely,
some stars pulsate in gravity (g) modes, like $\gamma$\,Dor stars
\citep{Balona1994, Kaye1999, VanReeth2015ApJS} and SPB stars
\citep{Waelkens1991, DeCat2002A&A, Pedersen2021}, allowing us to probe
deeper into the stellar structure, reaching down to the boundary of
the convective core. The \textit{Gaia} space mission confirmed that many
pulsating stars exist between the SPB and $\delta$\,Sct instability
strips \citep{DeRidder2023A&A}.

The excitation mechanisms behind
the pulsations in this part of the Hertzsprung-Russell diagram (HRD) are 
\ca{numerous and quite diverse. First of all, the $\kappa$ mechanism is operational in stars of spectral types O, B, A, and the hottest F stars along the main sequence
\citep{Pamyatnykh1999AcA}. Flux blocking at the bottom of the thin convective envelope excites g modes in the $\gamma$\,Dor stars \citep{Guzik2000,Dupret2005,Xiong2016}.
The role of turbulent pressure in the A- and F-type stars has also been investigated but is less established as a common excitation mechanism \citep{Grassitelli2015,houdek2000,antocietal2014}. This is also the case for the so-called edge-bump mechanism operational in chemically peculiar stars, notably in stars with strong helium depletion \citep{stellingwerf1979,murphyetal2020d}. While mode excitation is fairly well established, its interplay with damping mechanisms is less understood. Aside from radiative damping, additional physical processes affect the waves. The wave excitation, propagation, and damping are notably affected by a multitude of transport processes induced by rotation and magnetism \citep[e.g.,][for reviews of these processes]{Mathis2013LNP,Aerts2019ARA&A}.} A main goal of asteroseismology is to better understand \ca{all these processes, starting from a well-characterised large sample of pulsators }that can be accurately modelled. From an observational perspective, the ability to detect non-radial pulsators all along the main
sequence significantly broadens the temperature coverage for such
asteroseismic investigations \citep{Kurtz2023}.

Research on internal stellar rotation is the most attractive topic
among many of those driven by asteroseismology, with large progress in
stellar physics based on rotationally induced processes
\citep{Aerts2019ARA&A}. Rotation lifts the degeneracy among mode
frequencies, splitting modes into multiplets
\citep{Ledoux1951}. Rotational splittings have been observed in many
kinds of variable stars, such as the Sun \citep{Deubner1979}, white
dwarfs \citep{Winget1991ApJ}, subdwarfs \citep{Reed2000BaltA},
$\beta\,$Cep stars \citep{Aerts2003Sci}, slowly-rotating $\gamma$\,Dor
and SPB stars \citep{Papics2014A&A,Kurtz2014,Saio2015,murphyetal2016a,Li2019_splitting_gdor}, and red giant stars
\citep[e.g.][]{Mosser2012,Deheuvels2014,Gehan2018}. The splittings observed from different p or g modes carry information on rotation rates in different stellar layers, providing the possibility to re-construct core-to-surface differential rotation profiles \citep{Corbard1999,Deheuvels2014,Deheuvels2015,Triana2015,DiMauro2016,Triana2017,Deheuvels2020}.

Many of the variable stars situated in hot eMSTOs rotate rapidly, notably
with a rotation frequency close to the frequencies of g modes. Such
modes are often in the sub-inertial regime of the frequency range
\citep{Aerts2019ARA&A} and are gravito-inertial modes. The traditional
approximation of rotation (TAR) offers an appropriate formalism to
treat the rotational effects properly when computing such mode
frequencies \citep{Lee1987, Lee1997, Townsend2003, Mathis2009,
 VanReeth2016_TAR, Saio2018}. Under the approximation of the TAR, the
mode period spacing $\Delta P \equiv P_{n+1, l, m}-P_{n, l, m}$,
defined as the period difference of two modes with consecutive radial
order $n$ but the same angular degree $l$ and azimuthal order $m$, is
no longer constant as predicted by asymptotic theory
\citep{Shibahashi1979}. Instead, the period spacings exhibit
decreasing or increasing trends as a function of period, depending on
the mode identification
\citep{Bouabid2013,VanReeth2015,Ouazzani2017}. Prograde ($m=1$) dipole ($l=1$) modes are the most common in real observations and have been
observed in hundreds of $\gamma$\,Dor or tens of SPB stars \citep{Li2020MNRAS_611, Pedersen2021}. They show decreasing period spacings with increasing period and allow us to measure the stars'
near-core rotation rates and calibrate internal mixing processes
\citep{VanReeth2016_TAR,Li2020MNRAS_611, Szewczuk2018,
 Pedersen2021,Mombarg2021}. Recently discovered coupling between an
inertial mode in the convective core and a gravito-inertial mode in
the envelope has opened a new window for exploring the rotation
and internal physics of convective cores, where g modes do not propagate
\citep{Ouazzani2020,Saio2021,Tokuno2022,Aerts2023A&A}.

Our primary aim is to compile a comprehensive list of various types of pulsating stars
within NGC\,2516, which is a bright, young, and solar-metallicity open cluster. In this study, we analysed space photometry of the members of NGC\,2516 from the
Transiting Exoplanet Survey Satellite (TESS) mission \citep{Ricker2015} and conducted high-resolution spectroscopic follow-up
observations of the g-mode pulsators in the open cluster NGC\,2516. 
The paper is organized as follows: Section~\ref{sec:cluster_introduction_sample_selection} introduces the basic information of NGC\,2516 and the sample selection criteria. Section~\ref{sec:isochrone_fitting} discusses the fitting of theoretical isochrones to the observed CMD. The TESS photometry and the research on variable stars are presented in Sect.~\ref{sec:photometry}. Spectroscopic observations and results are described in Sect.~\ref{sec:spectra}. Finally, conclusions are provided in Sect.~\ref{sec:conclusions}.

%--------------------------------------------------------------------
\section{The cluster and sample selection}\label{sec:cluster_introduction_sample_selection}
\subsection{Cluster information}
NGC\,2516 \citep[$\alpha = 119^\circ.5270,\,\delta
= -60^\circ.8000$][]{Tarricq2021}, known also as the Southern
Beehive, is a bright, young, and solar-metallicity open cluster
located in the southern hemisphere, with distance of $409^{+17}_{-16}\,\mathrm{pc}$ \citep{Cantat-Gaudin2018}. Its ecliptic latitude
($-75^\circ50^\mathrm{m}$) leads this cluster to fall within the
southern continuous viewing zone (CVZ) of TESS. Therefore, TESS provides 1-year near-continuous light curves for NGC\,2516. It is worth noting that there is another open cluster located in the northern CVZ, which is UBC1, as studied by \cite{Fritzewski2024}.

Previous investigations of open clusters have typically been limited by the lack of sufficiently long light curves. These studies predominantly focused on p-mode oscillations, which exhibit relatively large frequency separations and thus require lower frequency resolution. Such analyses have been conducted for solar-like oscillators in the Kepler field \citep{Stello2010, Basu2011, Hekker2011}, as well as those in the K2 fields \citep{Ripepi2015, Lund2016, Sandquist2020, Murphy2022Pleiades}, and with TESS data \citep{Murphy2021, Bedding2023ApJ, Palakkatharappil2023, Pamos-Ortega++2023}. The only research for $\gamma$~Dor asteroseismology in young open clusters using TESS that we are aware of is UBC\,1 by \cite{Fritzewski2024}.

In the case of NGC\,2516, its previous study also mainly focused on short-period oscillation signals, such as \cite{Antonello1986A&A} and \cite{Zerbi1998PASP}. Interestingly, both \cite{Antonello1986A&A} and \cite{Zerbi1998PASP} claimed that they found some long-period variable stars at the red edge of the instability strip, which might be the gravity-mode oscillations of $\gamma$\,Dor stars just defined in 1994 \citep[see][]{Balona1994}. The availability of 1-year light curves presents an unparalleled opportunity to study gravity-mode oscillations. These oscillations are characterised by period spacings, necessitating
Fourier analysis with high-frequency resolution.

The first measurements of distance, reddening, and extinction of this cluster were reported by \cite{Cox1955ApJ} using ground-based photometry. Abundant following photometry observations were conducted, such as \cite{Feinstein1973}, which reported the age of NGC\,2516 of 60\,Myr. After that, the age of NGC\,2516 has been consistently estimated as 150\,Myr via different methods (isochrone fitting: \citealt{Meynet1993} and \citealt{Sung2002AJ}; gyrochronology and lithium depletion: \citealt{Bouma2021} and \citealt{Fritzewski2020A&A}). It shows a very similar age as the Pleiades (whose age is estimated around 130\,Myr with a large spread, see the introduction by
\citealt{Murphy2022Pleiades}). However, the gyrochronology of NGC\,2516 shows a slower upper envelope of the rotation period distributions in Sun-like stars, implying that NGC\,2516 is slightly older than the Pleiades.

Apart from their similar ages, \cite{Eggen1964IAUS} discovered that NGC\,2516, $\alpha$\,Persei, and the Pleiades share common space motions and exhibit similar colour-magnitude diagrams (CMDs), suggesting a common origin. \cite{Abt1969AJ_rotational_velocities} conducted spectroscopic measurements of the equatorial projected velocities ($v \sin i$) of the 30 brightest member stars in NGC\,2516 and found that the mean $v \sin i$ value is also similar to that of the Pleiades, especially after removing the numerous Ap stars (chemical peculiar stars of spectrum type A). %This similarity in rotational velocities may further substantiate the common origin hypothesis proposed by \cite{Eggen1964IAUS}.

Metallicity is an important property in cluster age determinations since it impacts stellar evolution. NGC\,2516 has near-solar metallicity: \cite{Terndrup2002} reported a metallicity difference between NGC\,2516 and the Pleiades of $\Delta \mathrm{[Fe/H]}=0.04\pm0.07$ and conclude that the
metallicity of NGC\,2516 is [Fe/H]=$0.01\pm0.07$. \cite{Bailey2018} took multi-epoch high-dispersion optical spectra of 126 Sun-like member stars in NGC\,2516 and reported a metallicity of [Fe/H]=$-0.08\pm0.01$. Therefore, in our following isochrone fitting, we will use isochrones with solar metallicity based on the protosolar abundances of \cite{Asplund2009}.

Ap stars have been discovered in NGC\,2516 \citep[e.g.][]{Maitzen1981A&A}, and the surface magnetic field strength has been measured in one of these stars \citep{Bagnulo2003}. Ap stars, considering their ages within the cluster, provide constraints on the evolution of fossil fields, which are typically stable over decades \citep{Abt1979ApJ, Thompson1987ApJS}. Additionally, the discovery of white dwarfs in NGC\,2516 presents an opportunity to test the high-mass end of the initial mass function \citep{Reimers1982, Koester1996}.

\subsection{Selection of target stars}
\cite{Meingast2021} reported the membership identification of
NGC\,2516 and revealed that it has an extended corona spanning $\sim
500\,\mathrm{pc}$. \cite{Bouma2021} confirmed the existence of the
corona of NGC\,2516 and found that the corona is coeval with its core
by isochronal, rotational, and lithium dating. We therefore used the
result reported by \citep{Meingast2021} as our membership
input list.

Firstly, we adopted the TESS magnitude criterion of $T<13\,\mathrm{mag}$ for our selection. This threshold ensures the photometric quality of the TESS data because the targets are bright enough. Our scientific objective is to investigate early-type stars; therefore, we need to focus on stars with effective temperatures above $\sim$6000\,K. In fact, the criterion of $T<13\,\mathrm{mag}$ is quite conservative, as subsequent research revealed that this threshold corresponds to a temperature of approximately 5230\,K for main-sequence stars. We still retained those lower-temperature main-sequence stars (between 6000\,K and 5230\,K) because we want to observe the transition of the rotation rate at the Kraft break. Although three red giants in this cluster are brighter than $T=13$\,mag, their temperatures are lower than those of the early-type cluster members. The presence of these red giants offers additional constraints on the cluster's age during the isochrone fitting process.

Secondly, we calculated the absolute magnitude of all the member stars in the sample by \cite{Meingast2021}, applying a cut-off at a \textit{Gaia} G band absolute magnitude brighter than 5\,mag to include supplementary stars in the sample. In this step, we found additional four stars. Finally, we obtained a sample comprising \totalnumberwithcontamination stars. 

In the target selection procedure, we have not implemented any criteria to exclude binary stars, which can deviate from the single-star isochrone and potentially introduce systematic errors in the subsequent isochrone fitting described in Sect.~\ref{sec:isochrone_fitting}. Previous reports have identified spectroscopic binaries \citep[e.g.][]{Abt1972, Gieseking1982, Gonzalez2000AJ}, and in this work, we have identified five eclipsing binaries. Nevertheless, the absence of a distinct binary main sequence in the Colour-Magnitude Diagram (CMD) suggests their limited impact on the isochrone fitting that follows. 

\section{Isochrone fitting}\label{sec:isochrone_fitting}

\begin{figure*}
    \centering
        \includegraphics[width=\linewidth]{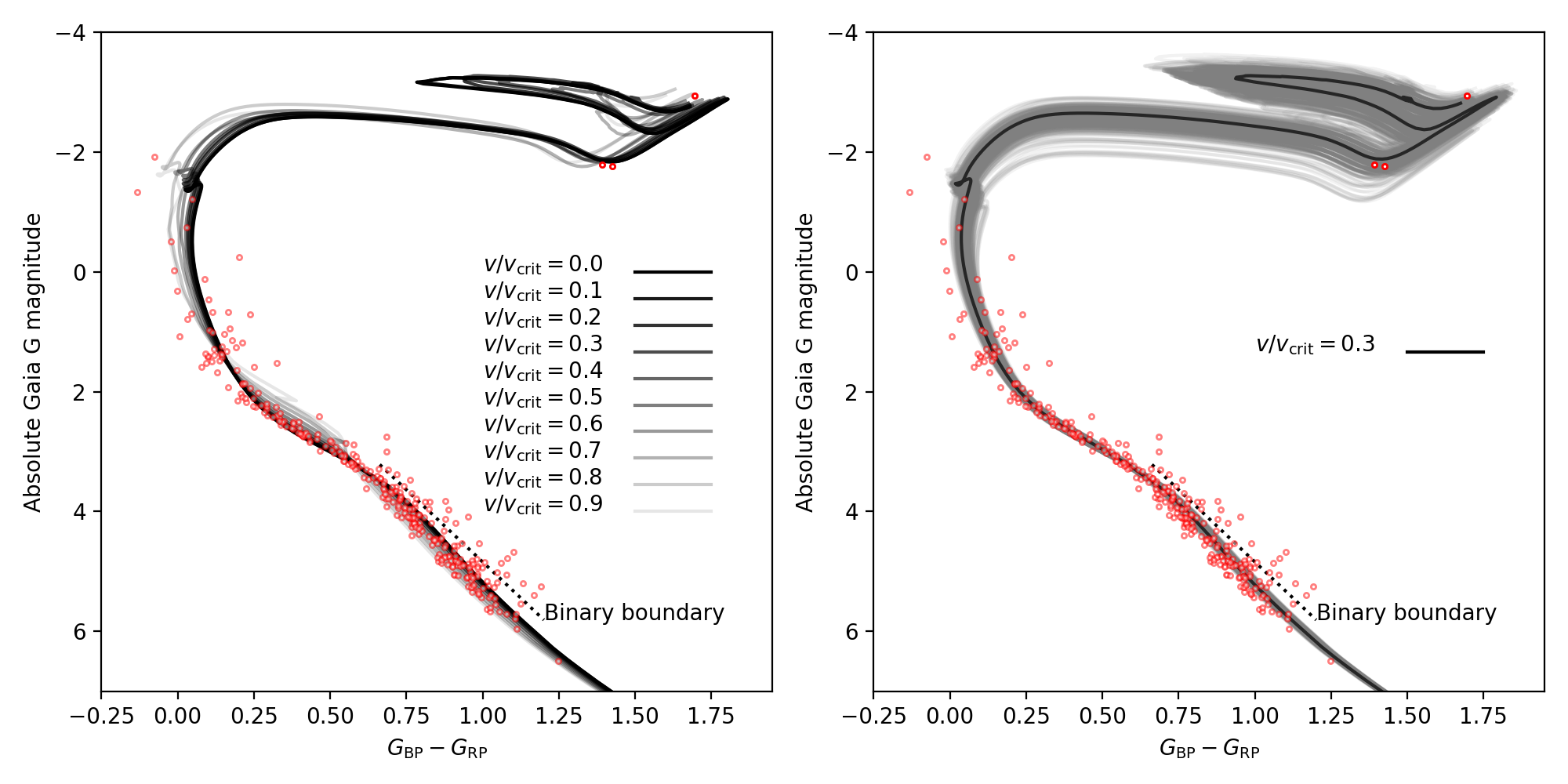}
    \caption{Observed Gaia DR3 CMD of
      NGC\,2516. Left panel: fits for isochrones with varying rotation
      rates. Isochrones with $v/v_\mathrm{crit} \leq 0.4$
      effectively reproduce the observed data. Right panel: the solid
      black isochrone represents the best-fitting result and
      $v/v_\mathrm{crit} = 0.3$. The grey background tracks
      collectively represent the uncertainty, which is determined
      using a set of 500 isochrones randomly selected during a
      Monte-Carlo approach as explained in the text. The dotted line marks the boundary of the binary sequence. }\label{fig:CMD_isochrone}
\end{figure*}

\begin{figure}
    \centering
    \includegraphics[width=\linewidth]{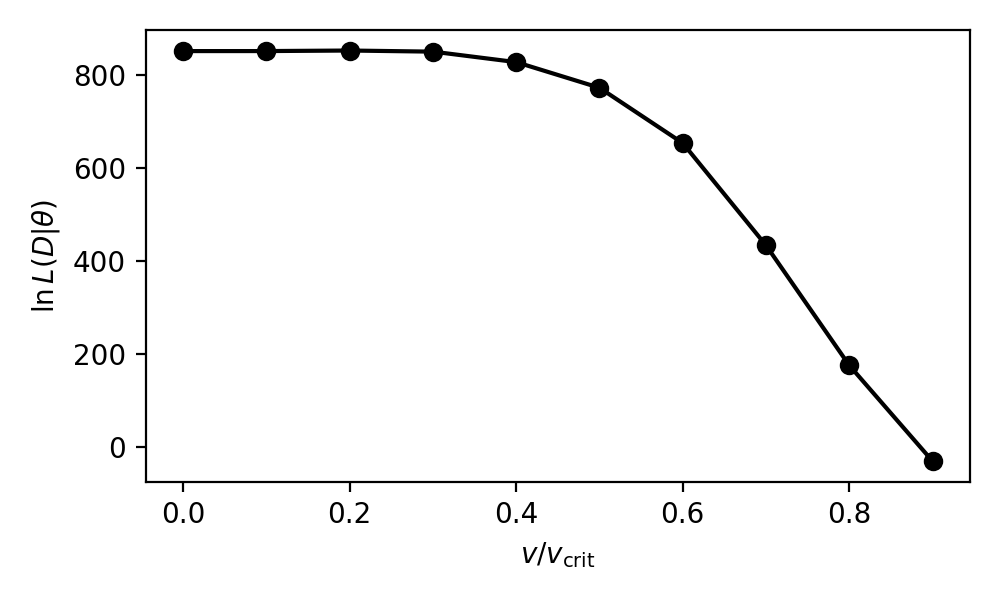}
    \caption{Relation between the log of the likelihood function (eq.~\ref{equ:isochrone_likelihood_whole} of the best-fitting isochrone and rotation rates $v/v_\mathrm{crit}$. Isochrones with $v/v_\mathrm{crit}> 0.5$ show significantly small likelihood values, implying worse fitting to the observational data.}
    \label{fig:likelihood_vs_Omega}
\end{figure}

\begin{figure}
    \centering
    \includegraphics[width=\linewidth]{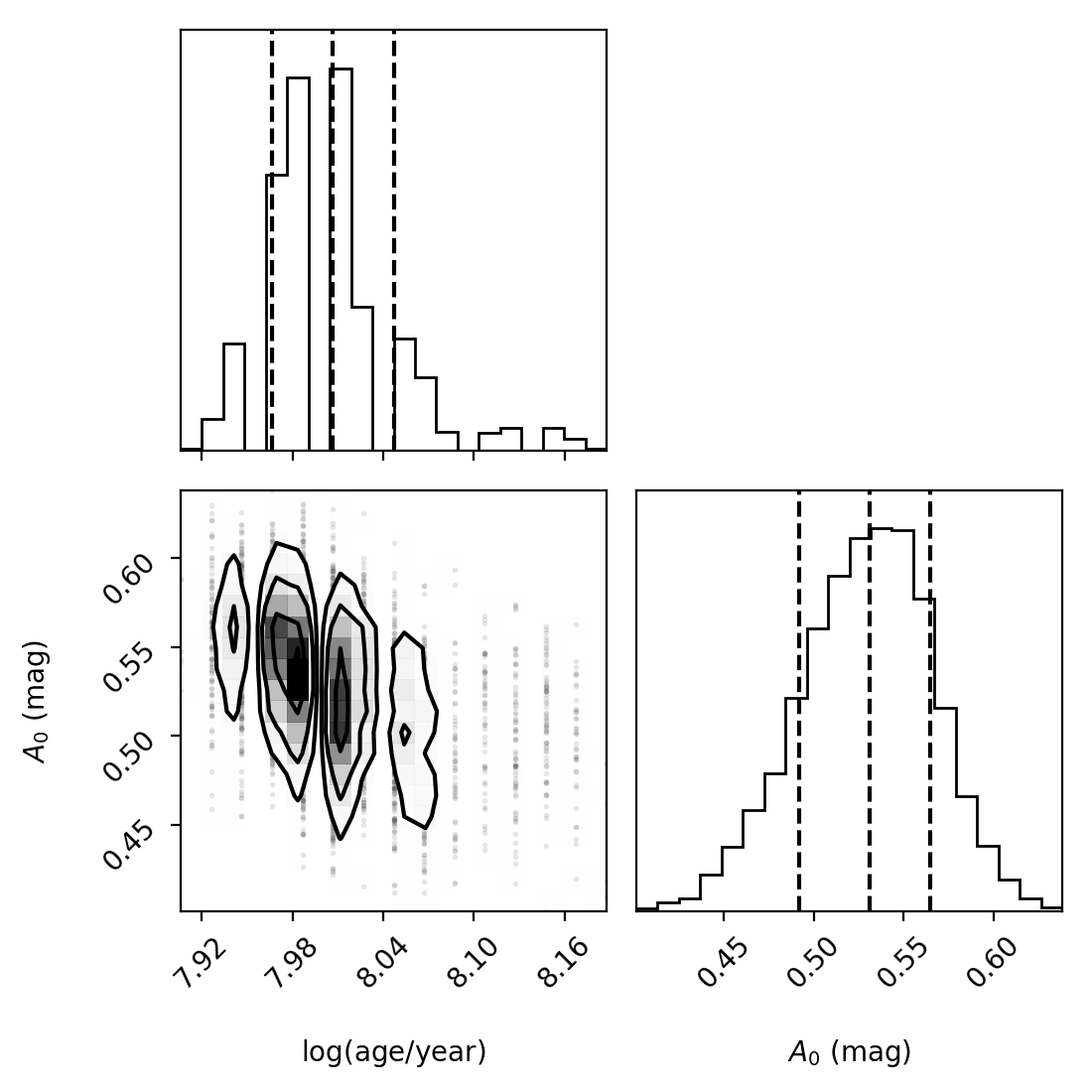}
    \caption{Posterior distributions of the isochrone fitting. }
    \label{fig:posterior_isochrone_fitting}
\end{figure}

\begin{figure*}
    \centering
    \includegraphics[width=\linewidth]{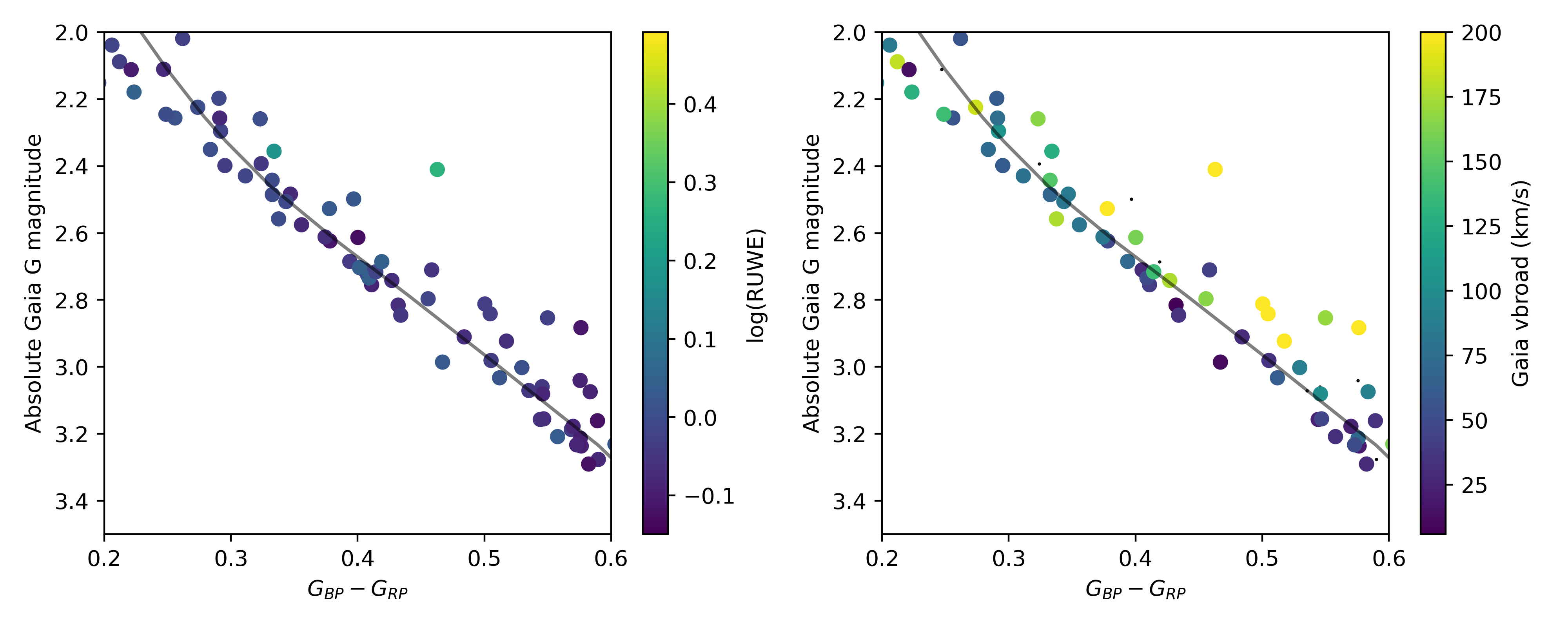}
    \caption{Zoom-in view of the CMD of NGC\,2516 showing how binarity and rotation lead a star deviating from the main sequence. The ranges of the
      x- and y-axes are: $0.2\,\mathrm{mag} < G_{BP}-G_{RP} < 0.6\,\mathrm{mag}$ and $2.0\,\mathrm{mag} < M_G <
      3.5\,\mathrm{mag}$. The stars are color-coded by RUWE (showing binarity) and line broadening (representing rotation rates),
      respectively. The small black dots are stars without broadening
      values in the \textit{Gaia} DR3 database.  }
    \label{fig:CMD_zoom_in}
\end{figure*}

Recent studies have examined how rotation affects stellar evolution and consequently the isochrones \citep[e.g.][]{Paxton2019ApJS, Gossage2019ApJ}. Therefore, we aim to identify the best-fitting isochrone for NGC\,2516, and from it, derive the age and extinction of the cluster to more effectively consider the effects of rotation. we used MIST isochrones \citep[MESA
Isochrones and Stellar Tracks,][]{Choi2016ApJ_MIST,
  Dotter2016ApJS_MIST} to fit our data. These
are based on the Modules for Experiments in Stellar Astrophysics
(MESA) code \citep[version v7503,][]{Paxton2011ApJS, Paxton2013ApJS,
  Paxton2015ApJS, Paxton2018ApJS}.  The original MIST isochrones\footnote{\url{https://waps.cfa.harvard.edu/MIST/index.html}}
only include two rotation rates, specifically \textcolor{black}{
$v/v_\mathrm{crit}=0.0$ and $0.4$, where
$v_\mathrm{crit}$} represents the critical rotation at which the
centrifugal force equals gravity at the equator of the isobar
\citep{Paxton2019ApJS}. However, our objective is to comprehensively
account for the impact of rotation. Therefore, we have explored
isochrones encompassing a whole range of rotation rates, ranging from
0.0 to 0.9 \textcolor{black}{$v / v_\mathrm{crit}$}, with a step of
0.1 \citep{Gossage2019ApJ}. The physical processes
related to rotation applied in the isochrone computations are done for
the shellular approximation \citep{Kippenhahn1970}. The chemical and
angular momentum transport induced by rotation is computed from the
diffusion equations from \citep{Endal1978}. Gravity darkening was
included according to \citep{Espinosa_Lara2011} as implemented in
\citet{Paxton2019ApJS}. Given that the temperature and luminosity
decrease due to the gravity darkening effect depend on the inclination of the
rotation axis with respect to the line-of-sight, \cite{Gossage2019ApJ} calculated the isochrones at a range of inclination angles and randomly sampled from them to create synthetic stellar populations. The isochrones used in this work represent an effect of gravity darkening averaged over inclination angles. Therefore, the scatter resulting from the gravity-darkening effect on the CMD should be distributed randomly around the best-fitting isochrone, rather than being concentrated on one side.

Due to the presence of main-sequence turn-offs and post-main-sequence
stars, the shape of isochrones is rather intricate, that is, one
colour index corresponds to multiple absolute
magnitudes. Consequently, traditional least-squares fitting is not
applicable here. Therefore, we consider an alternative fitting
approach. We hypothesise that each observed data point follows a
two-dimensional Gaussian distribution (without correlation) around the best-fitting point:
\begin{equation}
    P(C, G) = \frac{1}{2\pi \sigma_C \sigma_G}\exp \left( -\frac{1}{2}\left[\left(\frac{C-C_\mathrm{model}}{\sigma_C}\right)^2+\left(\frac{G-G_\mathrm{model}}{\sigma_G}\right)^2\right]   \right),\label{equ:isochrone_likelihood_single}
\end{equation}
where $C$ is the \textit{Gaia} colour index $\bprp$, and G is the
absolute \textit{Gaia} G band magnitude with extinction. $C_\mathrm{model}$ and $G_\mathrm{model}$ are
the colour index and absolute magnitude of the closest point in the
isochrone after extinction correction, while $\sigma_C$ and $\sigma_G$
are the standard deviations of the colour index and of the absolute G
band magnitude. We interpolate the isochrones, inserting ten values
between each pair of adjacent points, to calculate the theoretically
predicted points closest to a given observed point. The observed
$\sigma_C$ and $\sigma_G$ values obtained from \textit{Gaia} are
notably smaller than the scatter observed in the CMD. This discrepancy
could potentially be attributed to uncertainties in input physics,
including poor calibration of internal mixing, variations in
inclinations, differential extinction, various initial rotation rates, and unknown physical causes of
the eMSTO. Therefore, we have opted to use significantly larger values
for $\sigma_C$ and $\sigma_G$ instead of the \textit{Gaia}'s observation uncertainties to represent the residual's standard
deviations. By calculating a fourth-degree polynomial fit between the absolute magnitude and the colour index, these
values of the uncertainties are set to $\sigma_C = 0.056\,\mathrm{mag}$ and $\sigma_G = 0.332\,\mathrm{mag}$.

The final likelihood function is the product of a series of probabilities.
\begin{equation}
L(D|\theta) = \prod_{i} P(C_i, G_i),\label{equ:isochrone_likelihood_whole}
\end{equation}
where $D$ denotes the observed data, $C_i$ and $G_i$ are the colour
index and absolute G magnitude of the $i^\mathrm{th}$ star, $\theta
=\{ \mathrm{log(age)}, A_0\}$ is the vector of input parameters which
contains the age and extinction at 550\,nm.
The input colour index and absolute magnitude derived from the MIST
isochrones represent their intrinsic values, so we needed to calculate
the values after extinction correction for the $G$, $G_\mathrm{BP}$, and
$G_\mathrm{RP}$ bands. Specifically, we applied the extinction law as
described by \cite{Danielski2018}, which is calculated as
\begin{equation}
\begin{split}
    A_m/A_0 & = a_1+a_2X+a_3X^2+a_4X^3+a_5A_0+a_6 A_0^2+a_7A_0^3 \\
            & + a_8A_0X+a_9 A_0X^2+a_{10}XA_0^2,\label{equ:extinction_law}
\end{split}
\end{equation}
where $A_m$ stands for the extinction in the G, BP, and RP bands,
$A_0$ is the extinction at 550\,nm, and $X$ is the intrinsic colour
index $\bprp$. The coefficients $a_i$ for the \textit{Gaia} DR3 passband were
given by
\citet{Riello2021}\footnote{\url{https://www.cosmos.esa.int/web/gaia/edr3-extinction-law}}. Considering
the positions of our stars on the HRD, we use the coefficients adapted
for the upper regions of the HRD, which encompass giants and the upper
segment of the main sequence, up to approximately $G\sim5$\,mag.

We find that the data points exhibit a large spread in the low-temperature area, where $\bprp$ is approximately greater than 0.65\,mag (in Fig.\ref{fig:CMD_isochrone}). This spread is attributed to binarity,
\ca{although we do not observe a clean binary sequence. This is understood by realising that binarity increases the apparent luminosity for all system configurations, while a shift to lower effective temperatures occurs for particular binaries. Since we do not have a characterisation of the multiplicity at this stage}, we have excluded the data points \textcolor{black}{that are located above }the dotted line in Fig.\ref{fig:CMD_isochrone}, which was manually determined to begin at ($\bprp = 0.6626, G = 3.217$) and end at ($\bprp = 1.2011, G = 5.810$).

A pre-search of the best-fitting model was done by calculating a
coarse grid, using $7 < \mathrm{log(age/yr)} < 9$ with an evolution step
of 0.02 and $0\,\mathrm{mag} < A_0 < 1\,\mathrm{mag}$ with step of 0.025\,mag. After finding
the initial best-fitting values of log(age) and $A_0$, we 
estimate their uncertainties using a Monte-Carlo approach. In each
iteration, we added a Gaussian random noise perturbation to the
original data using $\sigma_C$ and $\sigma_G$, and searched for the
best-fitting model in a fine grid by calculating the likelihood values. The parameter spaces are: ages within the first eight and last eight steps of the
initial age range, and extinction within a range of 0.1\,mag
before and after the initial extinction, with a step size of 0.01\,mag. The age and $A_0$ corresponding to the largest likelihood
in each iteration were recorded, and their uncertainties were
calculated after 1000 iterations.

Figure~\ref{fig:CMD_isochrone} illustrates the best-fitting
isochrones. In the left panel, we find that the isochrones with
\textcolor{black}{$v/v_\mathrm{crit}\leq0.4$} successfully reproduce the
observed CMD, while those with \textcolor{black}{$v/v_\mathrm{crit}\geq0.5$}
exhibit discrepancies from the observations. Additionally, as shown in Fig.~\ref{fig:likelihood_vs_Omega}, isochrones within \textcolor{black}{$v/v_\mathrm{crit}\leq0.4$} yield similar
maximum likelihood values, but the likelihood begins to decrease
rapidly when \textcolor{black}{$v/v_\mathrm{crit}\geq0.5$}.
Obviously, not all stars in the cluster will have been born with the same value of \textcolor{black}{$v/v_\mathrm{crit}$}. 
All the isochrones with \textcolor{black}{$v/v_\mathrm{crit}\leq0.4$}
produce satisfactory results in the CMD.

The right panel of Fig.~\ref{fig:CMD_isochrone} presents one of the
best-fitting outcomes, featuring \textcolor{black}{$v/v_\mathrm{crit}=0.3$} for
all stars simultaneously. Under this premise, the grey background
represents the uncertainty range obtained from randomly selected 500 isochrones from the iterations with \textcolor{black}{$v/v_\mathrm{crit}\leq0.4$}. The presence of the three
post-main-sequence stars imposes additional constraints on isochrone
age, yet considerable spread persists in the post-main-sequence
phases. Finally, under equal weighting of the isochrone fitting with
\textcolor{black}{$v/v_\mathrm{crit}\leq0.4$}, we derived
$\mathrm{log(age/year) = 8.01\pm0.06}$ (equivalent to $102 \pm
15$\,Myr) and an extinction value of $A_0 = 0.53\pm
0.04\,\mathrm{mag}$. The posterior distributions are shown in Fig.~\ref{fig:posterior_isochrone_fitting}. Our determined age for NGC\,2516 is somewhat younger than that of the Pleiades, opposite to the prior study using gyrochronology \citep{Fritzewski2020A&A}. The age discrepancy might arise from different approaches. Applying the same isochrone-fitting approach to both NGC\,2516 and the Pleiades could still yield a similar age for both open clusters.

Regarding the extinction, $A_0 = 0.53\,\mathrm{mag}$ lead to a reddening of 0.25\,mag given a intrinsic \textit{Gaia} colour index $\left(\bprp\right)_0 = 0.2\,\mathrm{mag}$. If we assume $A_0
\approx A_V$ and $A_V = 3.1 E(B-V)$, where $A_V$ represents extinction
at the V band, we obtained a reddening value of $E(B-V) \approx
0.17\,\mathrm{mag}$. Alternatively, using
equation~\ref{equ:extinction_law} and the extinction coefficients from
\cite{Casagrande2018}, which state that $E(\mathrm{BP}-\mathrm{RP}) =
(3.374-2.035)E(B-V)$, we obtain a similar reddening value of $E(B-V)
\approx 0.18\,\mathrm{mag}$. It is worth noting that this reddening
value is slightly higher than the one reported in a previous study
($0.112 \pm 0.024\,\mathrm{mag}$) by \cite{Sung2002AJ}. Additionally,
the \textit{Gaia} total galactic extinction map provides an average
extinction value of $A_0 = 0.4358\,\mathrm{mag}$ within a radius of
0.25'. Even after transforming this value to $E(B-V)$ (which is 0.14\,mag), it remains
slightly higher than the literature value reported by
\cite{Sung2002AJ}. 

We also investigate the impact of different extinction values on age determination. We find that the values suggested by \cite{Sung2002AJ} and those from the \textit{Gaia} total galactic extinction map both lead to the same age estimate of $\log (\mathrm{age/yr}) = 7.99$. This estimate is just one step before our best-fitting age and still falls within the uncertainty range. Therefore, it appears that extinction does not significantly change the age determination.

%https://www.astro.umd.edu/~richard/ASTRO620/A620_Dust.pdf

Figure~\ref{fig:CMD_zoom_in} displays a zoom-in of the CMD, which roughly corresponds to the region of the classical instability strip. In the left panel, we can observe that \textit{Gaia}'s RUWE binary indicator values \citep[see definition in ][]{GaiaColl2023} show little variation in this region, making it difficult to distinguish binary systems based solely on RUWE values. Only one star
exhibits a significantly higher RUWE value, and interestingly, this star happens to fall to the right of the best-fit isochrone. In the right panel, we colour-coded the points with the \textit{Gaia} \texttt{vbroad}
parameter \citep[see definition in ][]{Fremat2023A&A}, revealing that many stars have high \texttt{vbroad} values (greater than 200\,km\,s$^{-1}$). These stars tend to be located on the right of the best-fitting isochrone, indicating the influence of gravity darkening caused by fast rotation and the flattening it induces \citep{Perez_Hernandez_1999}.

\section{Light curve reduction and variability identification}\label{sec:photometry}

\subsection{Light curve reduction}

The unique advantage of NGC\,2516 is its location near
the edge of the southern continuous viewing zone of the TESS
satellite. Therefore, TESS offers nearly continuous observations
spanning 11 sectors, for a total of approximately 297 days. Our analysis
used the data from Cycle 1 and Cycle 3, encompassing sectors 1, 4, and
7 to 11, as well as sectors 27, 31, and 34 to 37. A one-year gap is
seen during Cycle 2 when the TESS satellite observed the northern
celestial hemisphere. In Cycle 1, observations were taken at a cadence
of 30 minutes, whereas in Cycle 3, the cadence was reduced to 10
minutes.

We used the asteroseismic reduction pipeline developed by
\cite{Garcia2022} to download and process the TESS photometric
data\footnote{\url{https://github.com/IvS-Asteroseismology/tessutils}}. The TESS Full Frame Images were obtained from the Mikulski
Archive for Space Telescope (MAST) using the \texttt{TESScut} API
\citep{Brasseur2019}. For each star, a cutout of size 25 pixels
$\times$ 25 pixels was applied. This size allows for the examination
of neighbouring stars, and assessment of contamination, and provides an
adequate number of background pixels for subtraction.

Flux extraction using custom apertures was performed using the
\texttt{Python} package \texttt{Lightkurve}
\citep{Lightkurve2018}. While the apertures in the pipeline by
\cite{Garcia2022} were optimised to mitigate severe contamination, it
is challenging to completely avoid contamination in the crowded
NGC\,2516 field. For stars significantly affected by contamination, we
employed aperture radii smaller than those automatically determined by
the code. We calculated contamination levels by fitting Gaussian
profiles of the target star and nearby stars. A star was excluded from
the analysis if the contaminated flux constituted more than 10\% of
the total flux. After excluding the contaminated stars, our sample
contains \totalnumber of stars for further analysis. Subsequently,
detrending was performed on a sector-by-sector basis, involving
background subtraction and principal component analysis.

After obtaining the light curves, we calculated their Fourier
transform to identify their light variability. The Lomb-Scargle
algorithm as implemented in \textsc{astropy} \citep{2013A&A...558A..33A} was used for our stars to calculate their amplitude spectra
\citep{Lomb1976, Scargle1982}, which is suitable for unevenly spaced
data. %For those three red giant stars, we provide their power density
%spectra, but there is no fundamental difference compared to the
%amplitude spectra.

\subsection{Variable star classification}

We identified three red giant stars in the cluster, while all the
other stars are main-sequence stars. Therefore, we primarily focus on
the types of variables that occur among main-sequence stars, namely
p-mode pulsators (mainly $\delta$\,Sct stars), g-mode pulsators
($\gamma$\,Dor and SPB stars), stars with surface modulation, and
eclipsing binaries. The g-mode and p-mode pulsators typically exhibit
numerous narrow frequency peaks within their respective frequency
regimes. The frequencies of p-mode pulsations are generally above
$\sim$10\,$\mathrm{d^{-1}}$, while g-mode pulsations have frequencies below
$\sim$5\,$\mathrm{d^{-1}}$. Eclipsing binaries often exhibit a fundamental
frequency along with a series of harmonics in their power
spectra. These distinct features aid us in identifying various types
of variable stars. Rotational modulation occurs along the main
sequence, notably in cool stars subject to magnetic braking and in
hotter counterparts revealing chemical or temperature spots. Specifically, we search for a hump in the low-frequency range
caused by the rotational frequency and differing from additional
narrow frequency peaks due to pulsations, and we also require the
presence of harmonics of the rotational frequency.%, the so-called
%hump-and-spike stars following \citet{Henriksen2023MNRAS}.

\begin{figure*}
    \centering
    \includegraphics[width=0.8\linewidth]{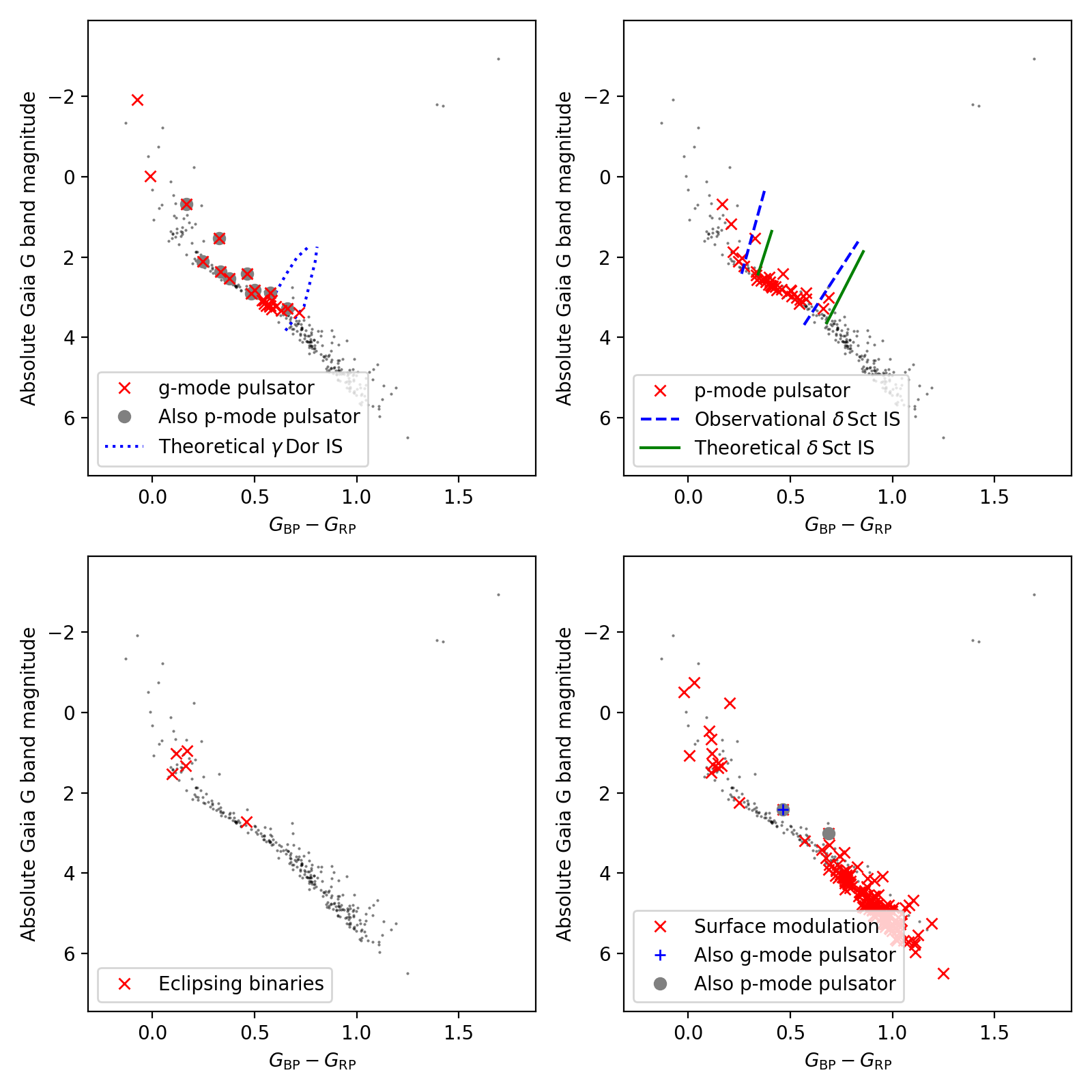}
    \caption{Locations of variable stars on the CMD of NGC\,2516.  }
    \label{fig:variable_on_CMD}
\end{figure*}

\begin{figure}
    \centering
    \includegraphics[width=\linewidth]{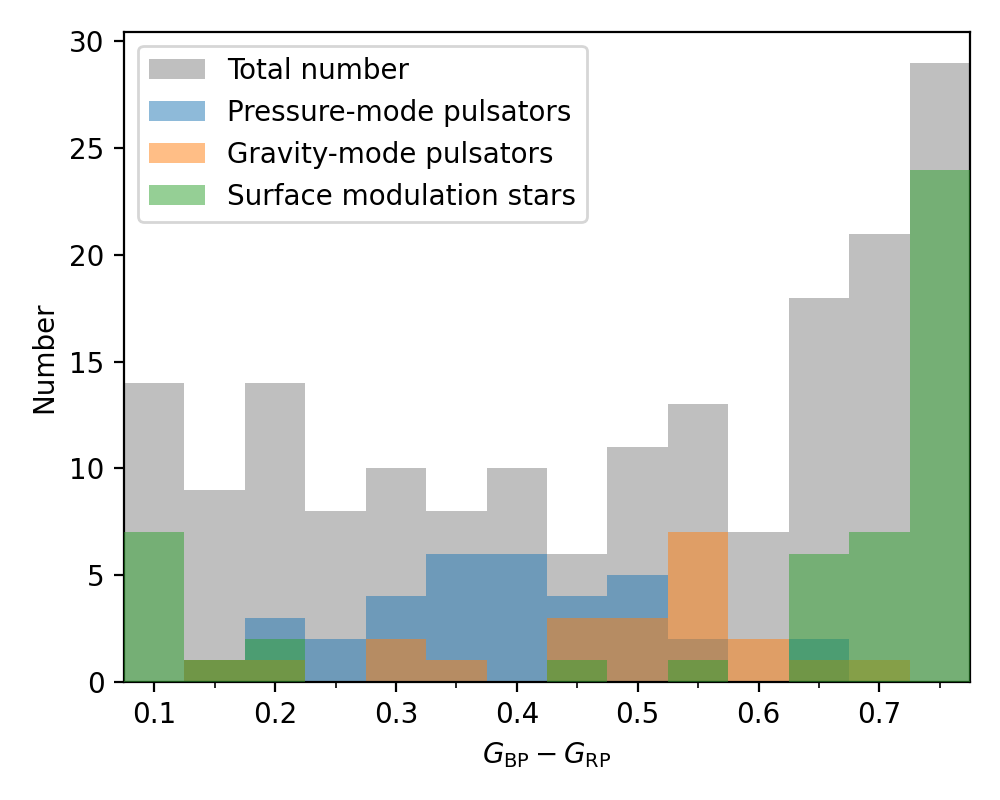}
    \caption{Histogram of the numbers of the p-mode pulsator, g-mode pulsator, surface modulations, and the total star in our sample. The bin size of the colour index $\bprp$ is 0.05. Pulsating stars (p- and g-mode pulsators) have high fractions around $\bprp$ of 0.5\,mag while surface modulation stars appear with small or large $\bprp$.  }
    \label{fig:hist_number}
\end{figure}

We find \gdornumber g-mode pulsators, \dsctnumber p-mode pulsators,
\EBnumber eclipsing binaries, and \surfacemodulationnumber surface
modulation stars. Figure~\ref{fig:variable_on_CMD} illustrates the
positions of these variables in the Gaia DR3 CMD of NGC\,2516. For
isolated stars, determining the positions of variable stars on the HRD
and evolutionary stages remains challenging due to various factors,
including extinction, inaccurate temperature and luminosity
measurements, or large uncertainty on mass. Consequently, defining the
observational instability strip (IS) of the variables in the cluster
is non-trivial. The CMD of a cluster offers a better opportunity to
investigate the instability strips of variable stars, as their
locations and parameters are well-defined in the CMD.

In the case of g-mode pulsators (shown in the upper left panel of
Fig.~\ref{fig:variable_on_CMD}), we observe a group of stars within a
range of $\bprp$ values from approximately 0.4\,mag to 0.6\,mag, which is
identified as the IS of $\gamma$\,Dor stars. We plotted the theoretical IS by \cite{Dupret2005A&A} after transforming the effective temperature and luminosity to the observed \textit{Gaia} colour index $\bprp$ and absolute G band magnitude with extinction. We find that the theoretical IS is slightly redder than the dense area of g-mode pulsators, and many g-mode pulsators appear above the blue edge of the theoretical IS. In Fig.~\ref{fig:hist_number}, we show the histograms as a function of colour index for the various classes of pulsators. The distribution peaks around
0.5\,mag for the observed $\gamma$\,Dor IS. In this region, about 50\% of stars show g-mode pulsations.

The upper right panel of Fig.~\ref{fig:variable_on_CMD} depicts the
distribution of p-mode pulsators. We observe that p-mode pulsators are
present within the $\bprp$ range of approximately 0.2\,mag to 0.6\,mag. These
stars are classified as $\delta$\,Sct stars, as their temperatures do
not reach the range of $\beta$\,Cep stars. The observed
IS of $\delta$\,Sct stars in the cluster is broader than that of the
$\gamma$\,Dor stars. We find that the observed $\delta$\,Sct IS by \cite{Murphy2019} matches the observation, while the theoretical IS by \cite{Dupret2005A&A} is still slightly redder than the observations. Examining Fig.~\ref{fig:hist_number}, we note that
$\delta$\,Sct stars dominate the IS. Approximately 60\% to 80\% of the
stars within the IS are identified as $\delta$\,Sct stars. This
fraction is consistent with the findings of a previous study conducted
on the Pleiades cluster \citep{Bedding2023ApJ}. The fraction slightly exceeds the value found for the \textit{Kepler} field \citep{Murphy2019}, which may be due to the age difference.

In the lower-left panel of Fig.~\ref{fig:variable_on_CMD}, we observe
the positions of eclipsing binaries on the CMD. The majority of these
binaries appear to lie near the main-sequence turn-off. This
observation is consistent with the nature that high-mass stars tend to
exhibit a higher binary fraction \citep{Moe2017ApJS}. In table~\ref{tab:EB}, we listed the TIC number, \textit{Gaia} colour index, and the orbital period for each eclipsing binary. The orbital period was measured using the procedure described in \citep{Li2020_gdor_in_EB}, which is mainly based on the $\mathrm{O}-\mathrm{C}$ (O minus C) method \citep{Sterken2005}. We find that TIC\,382529041 shows a variation in eclipse depths. Its $\mathrm{O}-\mathrm{C}$ diagram reveals a long-period third-component gravity perturbation with an estimated very high eccentricity.

\begin{table}[]
    \centering
        \caption{\EBnumber eclipsing binaries in NGC\,2516. }
    \label{tab:EB}
    \begin{tabular}{lll}
    \hline
    TIC & $\bprp$ & $P_\mathrm{orb}$ (d) \\
    \hline
341793209 & 0.458 & 0.28233919(23)\\ 
372913337 & 0.17 & 1.786037(8)\\ 
372913472 & 0.161 & 0.30749301(13)\\ 
382529041 & 0.115 & 0.3475920(13)\\ 
410451083 & 0.094 & 0.8593202(12)\\ 
\hline
    \end{tabular} 
    \tablefoot{We list TIC number, \textit{Gaia} colour index $\bprp$, and the orbital period $P_\mathrm{orb}$.}
\end{table}

The stars exhibiting surface modulation are situated in the lower
right panel of Fig.~\ref{fig:variable_on_CMD}. It is evident that
there exists a distinct gap wherein very few stars with surface
modulations are observed, ranging from $\bprp \sim 0.2$\,mag to
approximately 0.6\,mag. This is the region predominantly occupied by
$\delta$\,Sct and $\gamma$\,Dor stars. This gap is also evident in the
green histogram displayed in Fig.\ref{fig:hist_number}. We observe a
considerably higher fraction of cool stars displaying surface
modulation with $\bprp > 0.6$\,mag. This occurrence is attributed to
low-mass stars with a convective envelope, resulting in surface
activity. Furthermore, some stars located at the MSTO also show
surface modulation in the presence of fast rotation close to the
critical rate.

\subsection{Gravity-mode pulsators}

\begin{table}[]
    \centering
        \caption{\textcolor{black}{TAR fitting results of the \gdornumberwithcleargmodepattern g-mode pulsators in NGC\,2516. } }
    \label{tab:TAR}
    \begin{tabular}{lll}
    \hline
    TIC & $f_\mathrm{rot}$ ($\mathrm{d^{-1}}$) & $\Pi_0$ (s) \\
    \hline
281582674 & $2.953_{-0.006}^{+0.006}$ & $4840_{-60}^{+60}$  \\ 
308307454 & $0.97_{-0.04}^{+0.03}$ & $4340_{-160}^{+150}$  \\ 
308992761 & $2.965_{-0.007}^{+0.006}$ & $4900_{-70}^{+60}$  \\ 
341043961 & $3.011_{-0.012}^{+0.011}$ & $5040_{-110}^{+90}$  \\ 
358466708 & $1.265_{-0.03}^{+0.027}$ & $7500_{-700}^{+800}$  \\ 
358466729 & $2.1191_{-0.0029}^{+0.003}$ & $4730_{-90}^{+80}$  \\ 
364398040 & $2.947_{-0.014}^{+0.014}$ & $4870_{-120}^{+120}$  \\ 
372912679 & $3.005_{-0.013}^{+0.011}$ & $4820_{-110}^{+100}$  \\ 
372913043 & $2.974_{-0.006}^{+0.007}$ & $4890_{-60}^{+60}$  \\ 
410451583 & $2.953_{-0.008}^{+0.008}$ & $5020_{-70}^{+80}$  \\ 
410452218 & $2.80_{-0.010}^{+0.010}$ & $4800_{-90}^{+90}$  \\ 
\hline
    \end{tabular}
    \tablefoot{We give the TIC numbers, near-core rotation rates $f_\mathrm{rot}$, and asymptotic spacings $\Pi_0$. }
\end{table}

\begin{figure*}
    \centering
    \includegraphics{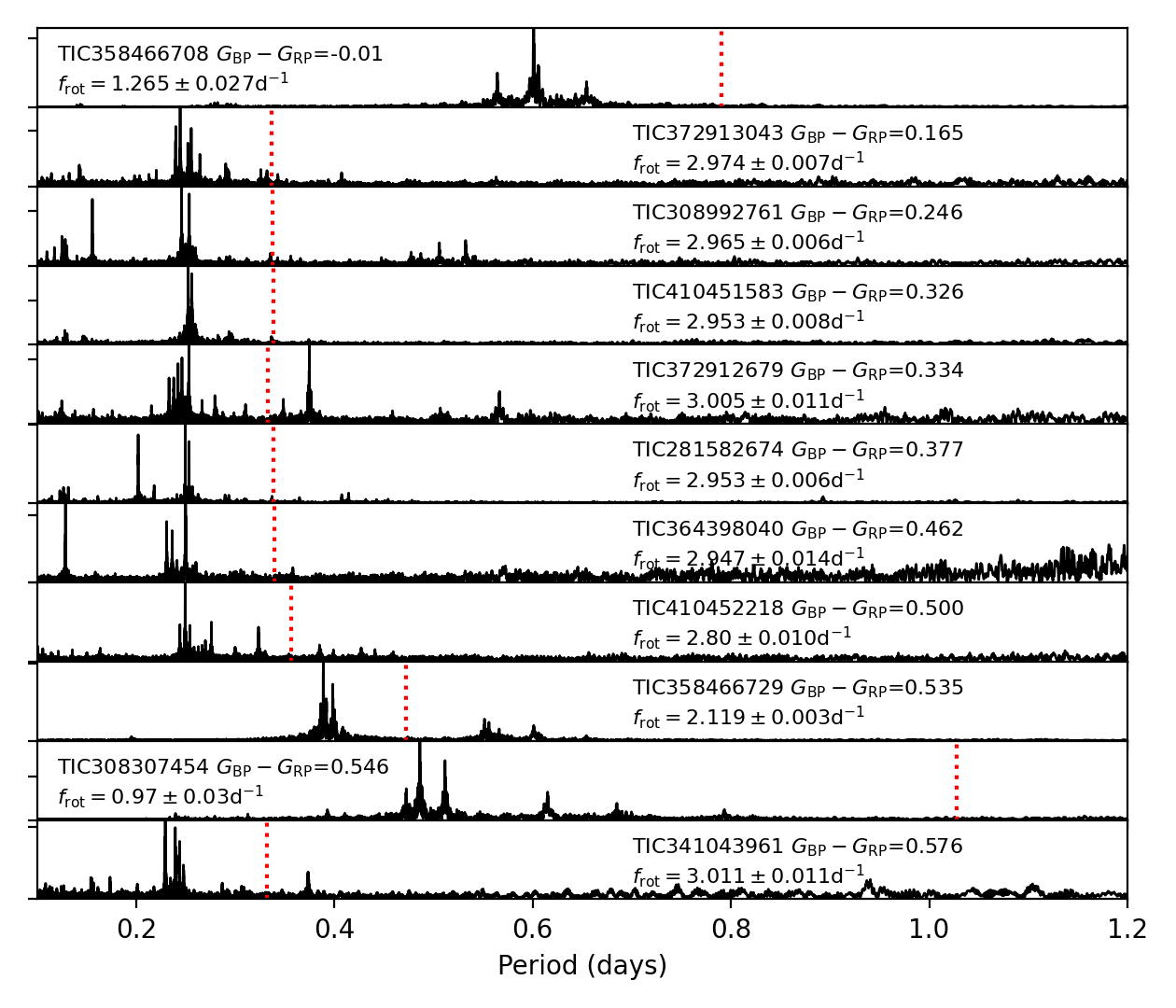}
    \caption{\gdornumberwithcleargmodepattern g-mode pulsators with
      clear period spacing patterns, sorted in descending order by
      their \textit{Gaia} colour index $\bprp$ from top to bottom. The red
      vertical lines mark their near-core rotation periods derived
      from the modes by adopting the TAR. }
    \label{fig:all_gdor}
\end{figure*}

\begin{figure*}
    \centering
    \includegraphics[width=\linewidth]{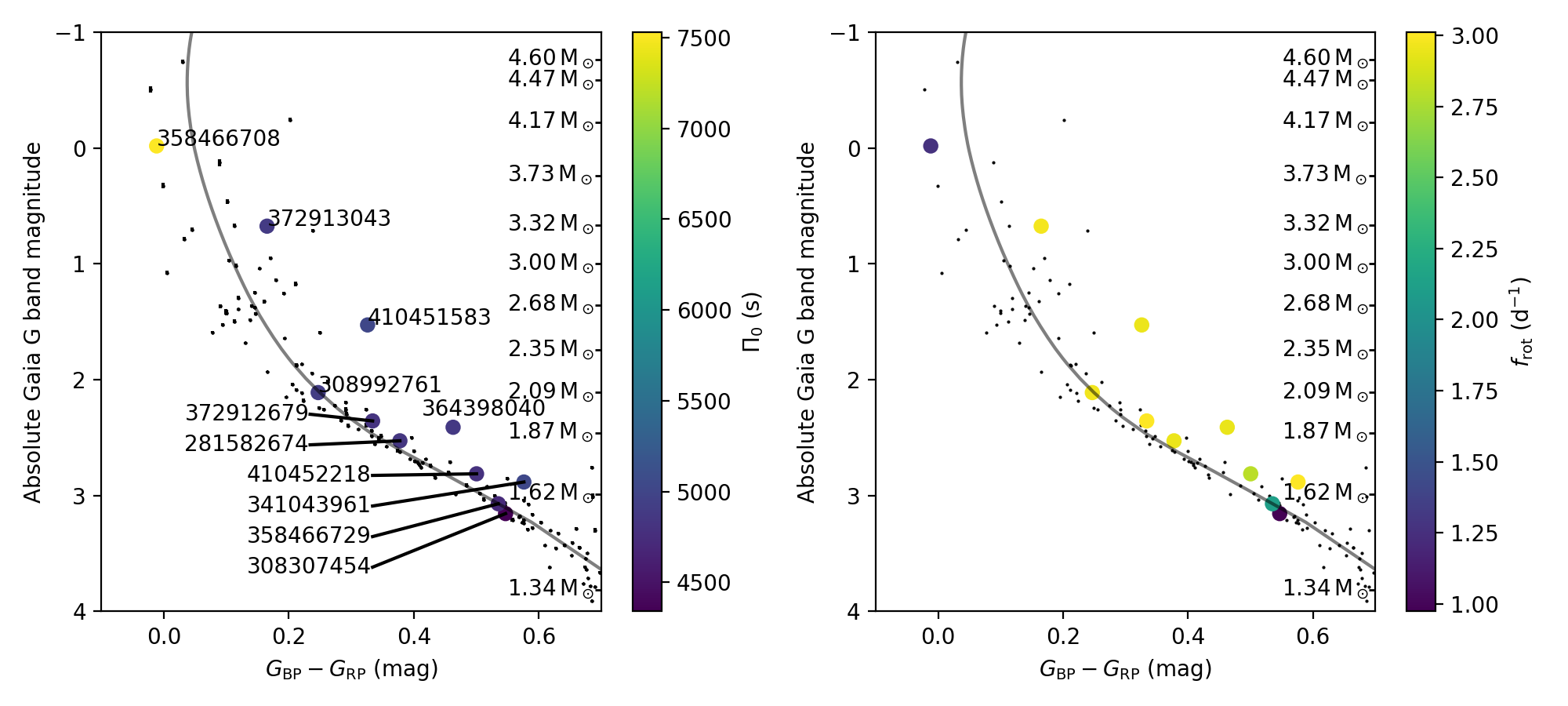}
    \caption{Location of $\gamma$\,Dor stars on the CMD with the
      best-fitting isochrone. The large circles show the $\gamma$\,Dor
      stars with clear period spacings, colour-coded by their buoyancy
      travel time (left panel) or near-core rotation rates (right
      panel). The black dots are the other cluster members and the grey
      solid line is the best-fitting isochrone. We
      also indicate the masses of the models on the right edges of each panel. In the left panel, we show the TIC numbers of the stars. 
% @GangLi: please explain why you list the 4 TIC numbers on the left panel
    }
    \label{fig:gdor_on_CMD}
\end{figure*}

\textcolor{black}{We first briefly recall the analysis procedure to detect period-spacing patterns and deduce mode identification as designed by \cite{Li2019_splitting_gdor}. After obtaining the light curves, we applied a detrending algorithm to eliminate the long-time trends. This involved calculating the median flux for each data point within a 5-day moving window and normalising the flux by this median value. The window size used for TESS data is narrower than that applied to Kepler data \citep[which is 10 days by][]{Li2020_gdor_in_EB} because the TESS data exhibits more gaps and instrumental effects. }

\textcolor{black}{An iterative prewhitening method was used to extract frequencies from the light curves. During each iteration, the frequency with the highest peak was identified and a sine function was fitted to the light curve to model this peak. This process yielded the optimised frequency, amplitude, and phase for the peak. Following each iteration, the residual from the sine function fitting was used for the subsequent iteration. The prewhitening process was terminated once the signal-to-noise ratio (S/N) of the highest peak fell below four. }

Not all of the \gdornumber g-mode pulsators allow for mode
identification, because only some of them show clear period spacing
patterns. There are various period-spacing search algorithms for $\gamma$\,Dor stars, such as \cite{Van_Reeth2015_gdor_detection_method} and
\cite{Christophe2018}. Here, we identified such spacing
patterns following the \textcolor{black}{observation-oriented} method by \cite{Li2019_splitting_gdor}. 
\textcolor{black}{This method constructed a template given the first period, the first period spacing value, and the slope (the changing rate between period spacing and period). A cross-correlation factor between the template and the real observed amplitude spectrum was calculated to reveal if a period spacing pattern is present in the data. }
This method yielded hundreds of $\gamma$\,Dor stars with identified modes in the {\it
  Kepler\/} field, including $\gamma\,$Dor stars in eclipsing binaries
\citep{Li2020MNRAS_611, Li2019_r_mode, Li2020_gdor_in_EB}. A similar
method to search for the period spacing patterns was developed by
\cite{Garcia2022} and yielded tens of $\gamma$\,Dor field stars in the
TESS continuous viewing zone \cite{Garcia2022_60_gdor}.

We find \gdornumberwithcleargmodepattern g-mode 
\ca{pulsators in NGC\,2516} to 
exhibit clear period spacing patterns. We measured their asymptotic spacings and
near-core rotation rates following the framework by
\cite{VanReeth2016_TAR}, which relies on the traditional approximation
of rotation (TAR).
\textcolor{black}{In this approximation, it is assumed that only the radial component of the rotation vector matters, while the horizontal component is neglected \citep{Townsend2003}. The TAR is an excellent approximation for $\gamma\,$Dor and SPB pulsators, because the displacement vectors of their g modes are dominantly horizontal 
\citep[][for observed values]{DeCat2002A&A},
such that the horizontal component of the rotation vector can be ignored in the frequency regime of the gravito-inertial modes \citep[e.g.,][]{Townsend2003,Rui2024}.}

Given the influence of rotation, the period spacings for g modes are no
longer constant \citep{Bouabid2013}. Rather, the g-mode periods in the co-rotating frame
can be rewritten as
\begin{equation}
    P_{nlm,\mathrm{co}} = \frac{\Pi_0}{\sqrt{\lambda_{l, m, s}}}\left(n+\alpha_\mathrm{g}\right).
\end{equation}
In this equation, $\Pi_0$ is the asymptotic spacing
representing the travel time of the waves within their mode cavity
determined by the buoyancy frequency and is hence called the buoyancy
travel time \citep{Aerts2021RvMP},
$s=2f_\mathrm{rot}/f_\mathrm{co}$ is the spin parameter as a function
of the rotation frequency $f_\mathrm{rot}$ and pulsation frequency in
the co-rotating frame $f_\mathrm{co} = 1/P_{nlm,\mathrm{co}}$, $\lambda$ is the eigenvalue of
the Laplace tidal equation, which can be computed with the
\textsc{gyre} code \citep{Townsend2013GYRE,Townsend2018GYRE},
$\alpha_\mathrm{g}$ is a phase term, and $n,l,m$ are the quantum
numbers of the mode. \textcolor{black}{We use the convention that positive $m$ represents the prograde modes. } The periods in the co-rotating frame can be transformed to the inertial frame $P_{nlm, \mathrm{in}}$ by
\begin{equation}
    1/P_{nlm, \mathrm{in}} = 1/P_{nlm,\mathrm{co}} + m f_\mathrm{rot}.
\end{equation}

\textcolor{black}{Mode identification, which involves allocating $l$ and $m$ values, is an essential step in the TAR fitting. For $\gamma$\,Dor stars that rotate slowly, the $l=1$ and $l=2$ peaks are roughly equally spaced in period, and their period spacings exhibit a ratio of $\sqrt{3}$, facilitating their identification in the period \'{e}chelle diagram \citep[e.g.][]{Bedding2015gdor}. The prograde ($m>0$) and the zonal ($m=0$) modes show decreasing period spacings as a function of increasing pulsation period, while the retrograde $m<0$ modes show increasing period spacings 
\citep[e.g.][]{Bouabid2013,VanReeth2016_TAR,Ouazzani2017}. For $\gamma$\,Dor stars that rotate rapidly, amplitude spectra often display characteristic frequency peak groups. The peak group with the highest amplitude corresponds to $l=1,~m=1$ modes, namely the prograde sectoral modes, while $l=2,~m=2$ modes exhibit periods that are half that of $l=1$ modes \cite[e.g.][]{Saio2018}. It is very rare to detect non-sectoral ($l \neq m$) modes in rapidly rotating $\gamma$\,Dor stars \citep{Saio2018, Li2020MNRAS_611}. Rossby modes are seen at periods slightly lower than the twice that of $l=1$ modes with increasing period spacing as a function of pulsation period, and are often characterised by a rapidly increasing and then slowly decreasing amplitude profile for each mode as a function of increasing period \citep{VanReeth2016_TAR, Saio2018_r_modes, Li2019_r_mode}. For Rossby modes, the quantum number $l$ is not used; instead, the letter $k$ is used, which falls within the range $k \leq -1$ \citep{Saio2018_r_modes}. The most frequently observed Rossby modes have $k=-2,~m=-1$ \citep{Li2019_r_mode}. Conversely, for g modes, $k \geq 0$, with $k = l - |m|$ \citep{Saio2018_r_modes}. }

\textcolor{black}{For our application to the NGC\,2516 $\gamma\,$Dor stars, a Markov chain Monte Carlo (MCMC) optimisation algorithm was applied to obtain the best-fitting $f_\mathrm{rot}$ and $\Pi_0$ by comparing the observed and calculated period spacings \citep{VanReeth2016_TAR, Li2020MNRAS_611}. }
All the period spacing patterns and the TAR fitting results are shown in Sect.~\ref{appandix_sec:all_gdor_figures}.

% @GangLi: check if it is a nicer plot when sorted according to frot (see my
% email) and if yes adapt the text below accordingly. This would be
% the most logical explanation in terms of physics...
Figure~\ref{fig:all_gdor} displays all the
\gdornumberwithcleargmodepattern g-mode pulsators with clear period
spacing patterns. We observe a distinct orderliness in the amplitude
spectra of these stars within the same star cluster after sorting them
by \textit{Gaia} colour index. We find that with colour index between 0.5\,mag and 0.165\,mag, those stars show similar near-core rotation rates, which is about 3\,$\mathrm{d^{-1}}$. 
The dominant mode frequencies at about 0.25\,d are identified as $l=1$, $m=1$ g modes subject to the Coriolis acceleration and projected into the line-of-sight with value $f_{\mathrm{rot}}$. Additionally, another group of frequencies is observed at a period of approximately 0.13\,d, corresponding to $l=2$, $m=2$ g modes shifted by $2 f_{\mathrm{rot}}$ towards
the observer's reference frame.

% @GangLi: adapt notation of colour bar (see my email)
We present the results of asymptotic spacing $\Pi_0$ and near-core
rotation frequency $f_{\mathrm{rot}}$ in Fig.~\ref{fig:gdor_on_CMD} and Table~\ref{tab:TAR}. In
the left panel of Fig.~\ref{fig:gdor_on_CMD}, we observe that the
majority of our g-mode pulsators have $\Pi_0$ around $\sim$4900\,s.
This value is typical for $\gamma$\,Dor stars, although it is somewhat
higher than the average value of the $\gamma$\,Dor field stars derived
from {\it Kepler\/} and TESS data \citep[which is 4000\,s, as reported
  in][]{Li2020MNRAS_611, Garcia2022_60_gdor}. The value of $\Pi_0$ is correlated with the
central hydrogen abundance, along with other key quantities describing
the internal mixing, as well as the initial chemical composition
\citep{Mombarg2019, Ouazzani2019A&A}. Hence it provides an estimation of the stellar
age, provided that the level of internal mixing can be deduced and
that the metallicity of the star is known. Here, we assume all the cluster members were born with the same initial metallicity. Consequently, the higher
value of $\Pi_0$ can be attributed to the young age of the cluster. It
is worth noting that $\Pi_0$ remains relatively consistent for all
pulsators within the $\bprp$ range of 0.2\,mag to 0.6\,mag, indicating a limited
sensitivity to (core) mass. TIC\,358466708 exhibits a value of $\Pi_0\approx
7500$\,s, which is significantly higher than the typical value for
$\gamma$\,Dor stars but is typical for SPB stars
\citep{Pedersen2021}. Additionally, the
high temperature of this star ($12070\pm170\,\mathrm{K}$) confirms its SPB nature. TIC\,372913043 is also a SPB star with temperature of $11510\pm240\,\mathrm{K}$. Their temperatures were measured using the high-resolution spectra, as discussed in Sect.~\ref{sec:spectra}.
% @GangLi: check if the other star with Teff>11000K in Fig. 11 is also an SPB, given its high Teff
% from spectroscopy it does not have a correct colour index in Gaia.
% Unless the Teff from spectroscopy is wrong... or unless it is
% another star where you find Teff>11000K in Fig. 11 (I cannot
% identify the TIC number...).
% If Fig.11 is correct then you have another SPB, and then it must
% be a fast rotator that looks redder...
% so adapt the paragraph above and below accordingly by
% also discussing this star here. for now, I left it unchanged in the text.

We show the near-core rotation rates in the right panel of
Fig.~\ref{fig:gdor_on_CMD} and find a clear correlation with colour
index. The near-core rotation rates increase from about 1\,$\mathrm{d^{-1}}$ at $\bprp\approx0.6\,\mathrm{mag}$ to
3\,$\mathrm{d^{-1}}$. However, the SPB star TIC\,358466708 shows a
relatively slow rotation, back to $1\,\mathrm{d^{-1}}$. This is in
agreement with the findings for {\it Kepler\/} fields stars, where the
SPB stars are on average slower rotators than the $\gamma\,$Dor stars
\citep[][see their Fig.\,6]{Aerts2021RvMP}. The rotation
rates of the $\gamma$\,Dor stars in NGC\,2516 are significantly higher
than those obtained from the {\it Kepler\/} and TESS data
\citep{Li2020MNRAS_611, Garcia2022_60_gdor}, where the median rotation rate is
approximately $1\,\mathrm{d^{-1}}$. The faster rotation that we find
here for the cluster stars is due to their younger age compared to those
of field stars \citep{Mombarg2021}.

Additionally, we identify two
g-mode pulsators (TIC\,358466708 and TIC\,372913043) located at the MSTO, which are categorised as SPB
stars. The effective temperatures measured by the spectra in Sect.~\ref{sec:spectra} further confirm the classification of SPB stars. 
TIC\,372913043 is somewhat suspicious, as it exhibits a small asymptotic spacing value ($4890^{+60}_{-80}\,\mathrm{s}$, Table~\ref{tab:TAR}), which should be within the typical range for $\gamma$\,Dor stars. We investigated the possibility of contamination and found the contamination level to be 1.8\%, indicating that 1.8\% of the flux in the aperture comes from other sources. There is only one star that could potentially contaminate the target, which is also a cluster member (TIC\,372913044, with apparent \textit{Gaia} G band magnitude of 12.84\,mag). As a comparison, TIC\,372913043 has a magnitude of 8.74\,mag, 4.1\,mag brighter than the neighbouring star. However, given its low colour index ($\bprp=0.92\,\mathrm{mag}$, corresponding to $T_\mathrm{eff}<5830\,\mathrm{K}$), the neighbouring star TIC\,372913044 is unlikely to exhibit g-mode period spacings. Therefore, we ruled out contamination as a cause. The reason for the small asymptotic spacing in this SPB star is likely due to evolution: as seen in Fig.~\ref{fig:gdor_on_CMD}, TIC\,372913043 is located in the \textcolor{black}{low-temperature side of the eMSTO} and is departing the main sequence, which means its asymptotic spacing should be decreasing rapidly.

\textcolor{black}{The mode identifications for two g-mode stars need verification. For TIC\,358466708, Figures.~\ref{appendix_fig:TIC358466708_period_spacing} and \ref{appendix_fig:TIC358466708_period_spacing_TAR} reveal that the modes are $l=2,~m=2$ with periods around 0.28\,d. As illustrated in the top panel of Fig.~\ref{fig:all_gdor}, TIC\,358466708 exhibits $l=1$ modes at periods  of approximately 0.56\,d; however, the $l=1$ period spacings could not be identified. Consequently, we only used the $l=2, m=2$ modes for the fitting with the TAR to determine the near-core rotation rate and the asymptotic spacing. The other star, TIC\,358466729, exhibits identifiable Rossby modes around 0.55\,d, as indicated in Figs \ref{appendix_fig:TIC358466729_period_spacing} and \ref{appendix_fig:TIC358466729_period_spacing_TAR}. Nevertheless, we were unable to identify any period spacing within its $l=1$ g-mode region, which should occur at approximately 0.4\,d, as shown in the third panel counting from the bottom in Fig.~\ref{fig:all_gdor}.  }

We also identify some g-mode pulsators situated
between the SPB and $\gamma$\,Dor strips. This is fully in line with
the findings by \citet{DeRidder2023A&A} from Gaia DR3 light curves and
points to fast rotators with g modes \citet{Aerts2023-DR3}.

\subsection{Pressure-mode pulsators}
\begin{figure*}
    \centering
    \includegraphics[width=0.9\linewidth]{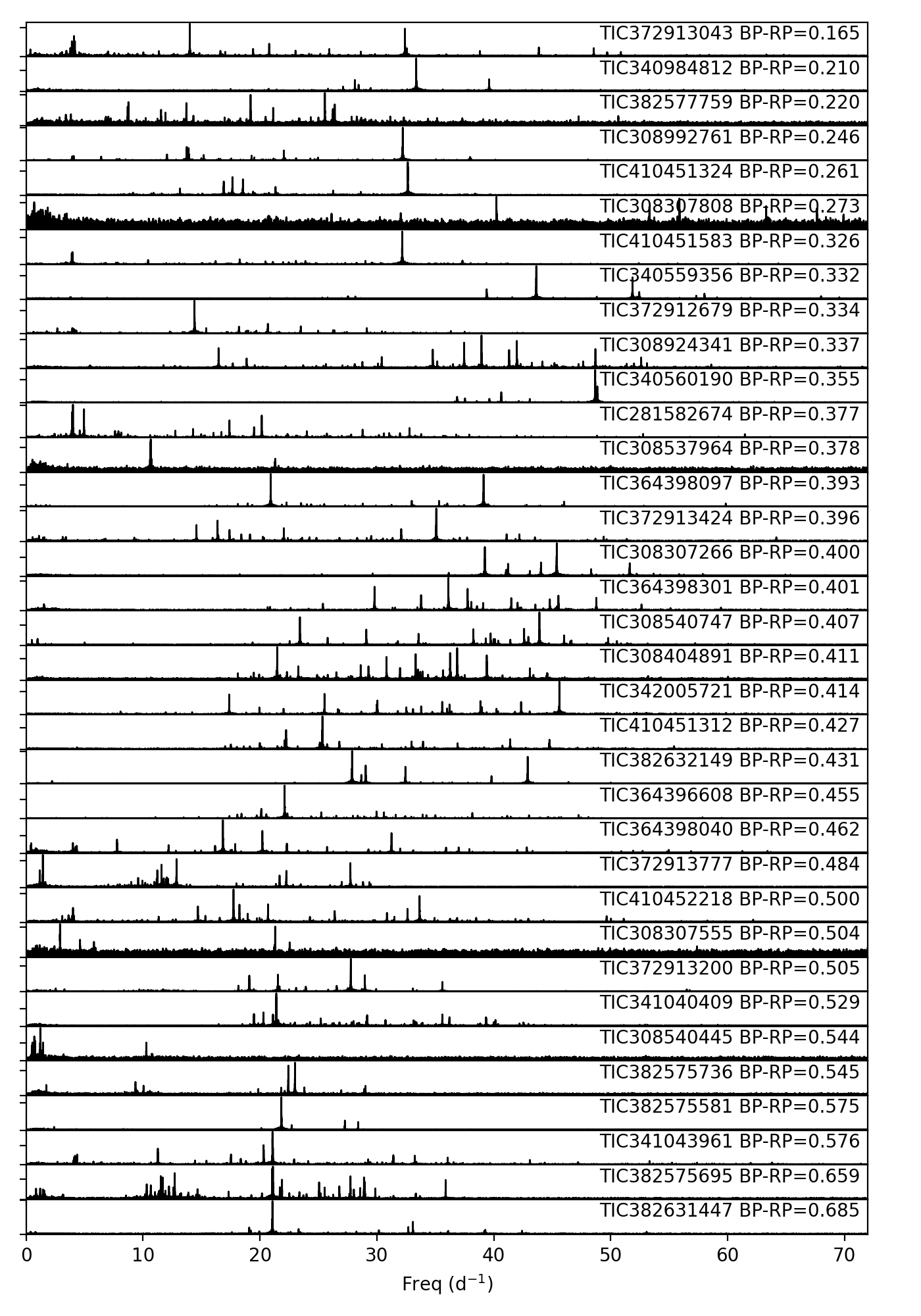}
    \caption{Amplitude spectra of the $\delta$\,Sct stars in
      NGC\,2516, sorted by colour index. The top panel has the
      star with the smallest colour index hence the highest effective
      temperature.}
    \label{fig:all_dsct}
\end{figure*}

\begin{figure}
    \centering
    \includegraphics[width=\linewidth]{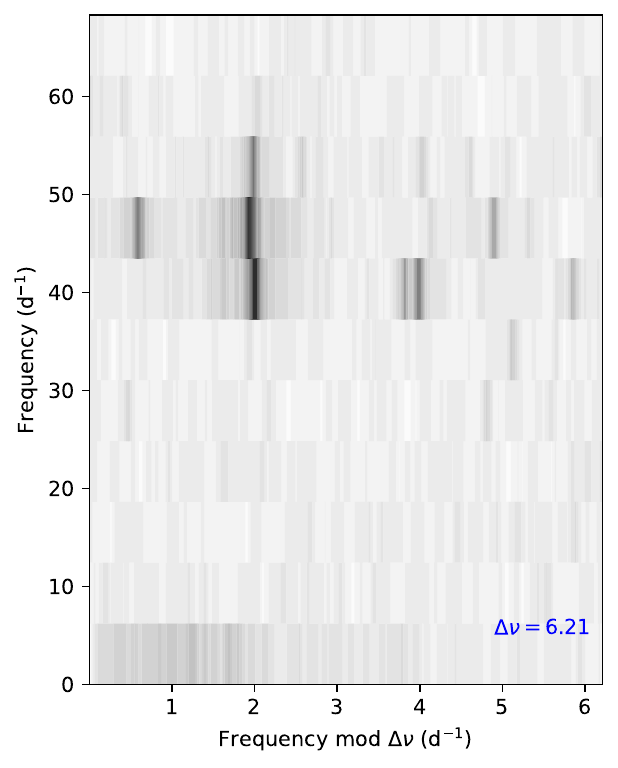}
    \caption{\Echelle\ diagram of the $\delta$\,Sct star TIC\,308307266. A frequency separation is seen with $\Delta \nu \approx 6.21\,\mathrm{d^{-1}}$.}
    \label{fig:dsct_freq_separation}
\end{figure}

The mode identification of $\delta$\,Sct stars is a long-standing
question, and leads to difficulty in asteroseismic exploitation of
their internal physical properties \citep{Bedding2020}.
We show all the amplitude spectra of the $\delta$\,Sct
stars in NGC\,2516, sorted by their colour index, in
Fig.~\ref{fig:all_dsct}. We find that their amplitude spectra are
well-ordered as a function of colour index. For
the low-temperature stars ($\bprp>0.411$\,mag, the bottom part of
Fig.~\ref{fig:all_dsct}), a series of strong frequency peaks appear near
$21\,\mathrm{d^{-1}}$, which is identified as the radial
fundamental mode frequency \citep{Bedding2020}. In the $\bprp$ region from 0.5\,mag to 0.4\,mag (middle
part of Fig.~\ref{fig:all_dsct}), we find a relation between the mean
pulsation frequency and temperature. With increasing temperature, the
envelope of the pulsation frequencies moves
to higher frequency \citep{Balona+Dziembowski2011, Barcelo-Forteza++2018, Bowman+Kurtz2018, Barcelo-Forteza++2020,Hasanzadeh++2021}. Finally, for the hotter stars
(with $\bprp\leq0.396$\,mag), another series of strong peaks aligns at
frequency around $33\,\mathrm{d^{-1}}$, which we interpret as the
signature of second overtone radial modes given their factor $\sim\! 0.64$ shorter period
than those of the fundamental modes \citep[e.g.][]{Netzel2022}.

We attempted to identify regular frequency separations among these
$\delta$\,Sct stars. In Fig.~\ref{fig:dsct_freq_separation}, the \echelle\ diagram of TIC\,308307266 displays a ridge corresponding to
$l=1$ modes with four distinct peaks separated by a large frequency
spacing of $\Delta \nu \approx 6.21\,\mathrm{d^{-1}}$, roughly in
agreement with mean stellar densities of $\delta\,$Sct stars if we keep
in mind that $\Delta \nu$ is affected by fast rotation.  We
did not observe a corresponding ridge for its $l=0$ modes, but this is
not so surprising given that such ridges due to radial modes tend to
be strongly blended while those of the dipole modes are not, making the
latter easier to find \citep{Bedding2020}.  No other $\delta\,$Sct
star in the cluster was found to reveal a clear large frequency
separation.

\subsection{Oscillations of the cluster giants}\label{subsec:solar-like}

\begin{figure}
    \centering
    \includegraphics[width=\linewidth]{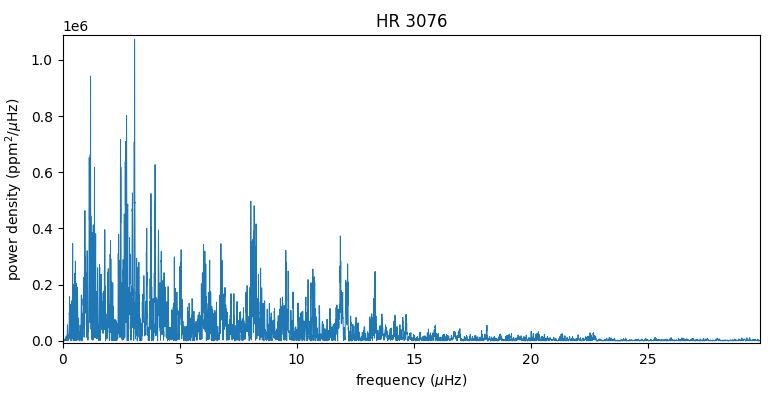}
    \caption{Power density spectra of the red giant HR\,3076.  }
    \label{fig:solar_like}
\end{figure}

% \begin{figure}
%     \centering
%     \includegraphics[width=\linewidth]{figures/all_solar_like_in_NGC2516_amplitude.png}
%     \caption{Power spectra of the three post-main-sequence stars. The
%       x-axis shows the frequency in $\mu$Hz, while the y-axis shows
%       the amplitude in ppm.  The vertical dashed lines mark the locations
%       of $\nu_\mathrm{max}$ for solar-like oscillation, as predicted
%       from the observed radii, effective temperatures, and isochrone
%       masses. }
% % @GangLi: either you give the units and values for the y-axis as is,
% % or you replace with an amplitude spectrum with y-axis in ppm
%     \label{fig:solar_like}
% \end{figure}

There are three red giants in NGC\,2516: HR\,3076 (TIC\,382510863; HD\,64320; $G=6.34$), SAO\,250043 (TIC\,372913375; $G=6.30$) and HD\,65662 (TIC\,358466601; $G=5.18$). 
%Their bright apparent magnitudes mean the TESS pixels become saturated, but the pixel data reveal that their bleed columns are not excessively long and that they still fall within the circular region around the stars. Therefore, we employed large circular apertures that encompass all the pixels with signal belonging to the star to construct their light curves, following a similar strategy developed in previous studies by \cite{White2017} and \cite{Handberg2021}.
We used the SPOC 2-minute light curves to search for solar-like oscillations.
To measure \numax, we used the {\tt nuSYD} method described by \textcolor{black}{\citet{Sreenivas2024}}. In HR\,3076, we found a very clear detection (Fig.~\ref{fig:solar_like}) with $\numax = 10.8\pm0.3\,\muhz$ \textcolor{black}{(measured using {\tt nuSYD})} and $\Dnu=1.27 \pm 0.02\,\muhz$ \textcolor{black}{(measured by constructing an \echelle\ diagram).} In the other two stars, the oscillations were not so clear but the likely values of \numax\ are $12.5 \pm 0.5\,\muhz$ for SAO\,250043 and $2.6\pm 0.3\,\muhz$ for HD\,65662.  We were not able to identify \Dnu\ for these two stars.

For HR\,3076, we have both \numax\ and \Dnu\ and so we used the standard asteroseismic scaling relations \citep[e.g.,][]{Jackiewicz_2021} to calculate the mass. \textcolor{black}{We remind the reader that the {\tt nuSYD} method has an optional correction that we did not make because it requires sufficient SNR to measure the width of the oscillation envelope reasonably accurately. Therefore, in the scaling relations we used the uncorrected solar value for \numax\ of $3154\pm30$\,\muhz\ \citep{Sreenivas2024}.}  Using $\Teff = 4704 \pm 122$\,K from \citet{Stassun++2019} gives a mass of $3.8 \pm 0.5\,\Msun$. 

For stars without a value of \Dnu, we can still use \numax\ to estimate a mass by using the Gaia luminosity \textcolor{black}{\citep[e.g.,][]{Miglio2012, Hon2021}}.
%
% % @GangLi: please provide the amplitudes on the y-axis
% % and discuss if the amplitudes are compatible with the scalings
% % predicted in Kallinger et al. (2014)
% Figure~\ref{fig:solar_like} illustrates the power spectra of the three
% post-main-sequence stars in the cluster. From the best-fitting isochrone, we know that their
% masses are about 5\,M$_\odot$. We observed that all three stars exhibit
% power excesses with quasi-periodicities in the low-frequency region up
% to several $\mu$Hz. These features occur at timescales consistent with
% the convective turnover timescale in the envelope of evolved stars for
% the appropriate mass in this part of the HRD. According to the
% theoretical predictions by \citet{Charbonnel2017} magnetic activity
% may occur, causing granulation and accompanying stochastic variability
% to occur in the measured frequency range. Such granulation is
% predicted to occur at frequencies that scale with $\nu_\mathrm{max}$
% of solar-like oscillators, given the universality of the scaling laws
% found by \citet{Kallinger2014}.
% We cannot easily measure a value of $\nu_\mathrm{max}$ from the observed amplitude spectra
% in Fig.~\ref{fig:solar_like}. To obtain an estimate of
% $\nu_\mathrm{max}$, 
%
We calculated the luminosities of the stars using the
equation\footnote{\url{https://gea.esac.esa.int/archive/documentation/GDR2/Data_analysis/chap_cu8par/sec_cu8par_process/ssec_cu8par_process_flame.html}}
\begin{equation}
    -2.5\log L = G+BC -M_\mathrm{bol, \odot}-A_G,
\end{equation}
where $G$ is the absolute magnitude in the \textit{Gaia} G band, $BC$
is the temperature-only dependent bolometric correction,
$M_\mathrm{bol, \odot}=4.74$ is the solar absolute bolometric
magnitude, and $A_G$ is the extinction in the \textit{Gaia} G band
given by the isochrone fitting in
Sect.~\ref{sec:isochrone_fitting}. The bolometric correction is
calculated as
\begin{equation}
    BC = \sum_{i=0}^4 b_i \left(T_\mathrm{eff}-T_\mathrm{eff,\odot}\right)^i,
\end{equation}
with the coefficients $b_i$ given by \cite{Andrae2018}. We used the
effective temperatures $T_\mathrm{eff}$ given by \textit{Gaia} to
calculate the luminosities, and subsequently relied on the Stefan–Boltzmann law to estimate the radii $R$, namely
\begin{equation}
    L \propto R^2 T_\mathrm{eff}^4.\label{equ:stefan-boltzmann}
\end{equation}

We obtained luminosities of 
$755\pm47\,\Lsun$ for HR\,3076, 
$740\pm42\,\Lsun$ for SAO\,250043, and
$2500\pm240\,\Lsun$ for HD\,65662.
We used effective temperatures of $4660 \pm 122$\,K for SAO\,250043 and $4281\pm 122$ for HD\,65662. Combining with our \numax\ measurements gave masses of 
$5.3 \pm 0.6$ for HR\,3076,
$6.2 \pm 0.7$ for SAO\,250043 and
$5.9 \pm 1.1$ for HD\,65662. Meanwhile, the best-fitting isochrone gives masses of these three red giant stars around 5.1\,$\mathrm{M_\odot}$. The isochrone-derived masses show large spread because the 
\ca{isochrones at that region overlap. 
These values should therefore be treated with caution, pending a full seismic analysis that includes fitting of individual mode frequencies.} In particular, the two mass estimates for HD\,3076 differ by a few sigma.

% We subsequently estimated the masses from the closest models of the
% best-fitting isochrone and find them to range from 5.07 to
% 5.13\,$\mathrm{M_\odot}$. Finally, we predict $\nu_\mathrm{max}$
% using the standard scaling relation \citep[][]{Brown++1991, Kjeldsen+Bedding1995}:
% \begin{equation}
%     \frac{\nu_\mathrm{max}}{\nu_\mathrm{max, \odot}} = \left(\frac{M}{M_\odot}\right) \left(\frac{R}{R_\odot}\right)^{-2}\left(\frac{T_\mathrm{eff}}{T_\mathrm{eff, \odot}}\right)^{-1/2},
% \end{equation}
% with $\nu_\mathrm{max, \odot} = 3090\,\mathrm{\mu Hz}$.  The vertical
% dashed lines in Fig.~\ref{fig:solar_like} mark the locations of the
% estimated $\nu_\mathrm{max}$. We find that our rough estimates for
% $\nu_\mathrm{max}$ are somewhat higher than the frequencies at maximum
% power but overall the agreement is satisfactory, given that the used
% isochrones to estimate the mass is for one chosen type of input
% physics, while uncertainties in such a choice are large \citep[as discussed in ][]{Charbonnel2017}. Future spectroscopy or
% spectropolarimetry of these cluster members would be extremely valuable
% to exploit their variability in more detail. 

\subsection{Rotation rate as a function of colour index}
\begin{figure*}
    \centering
    \includegraphics[width=0.9\linewidth]{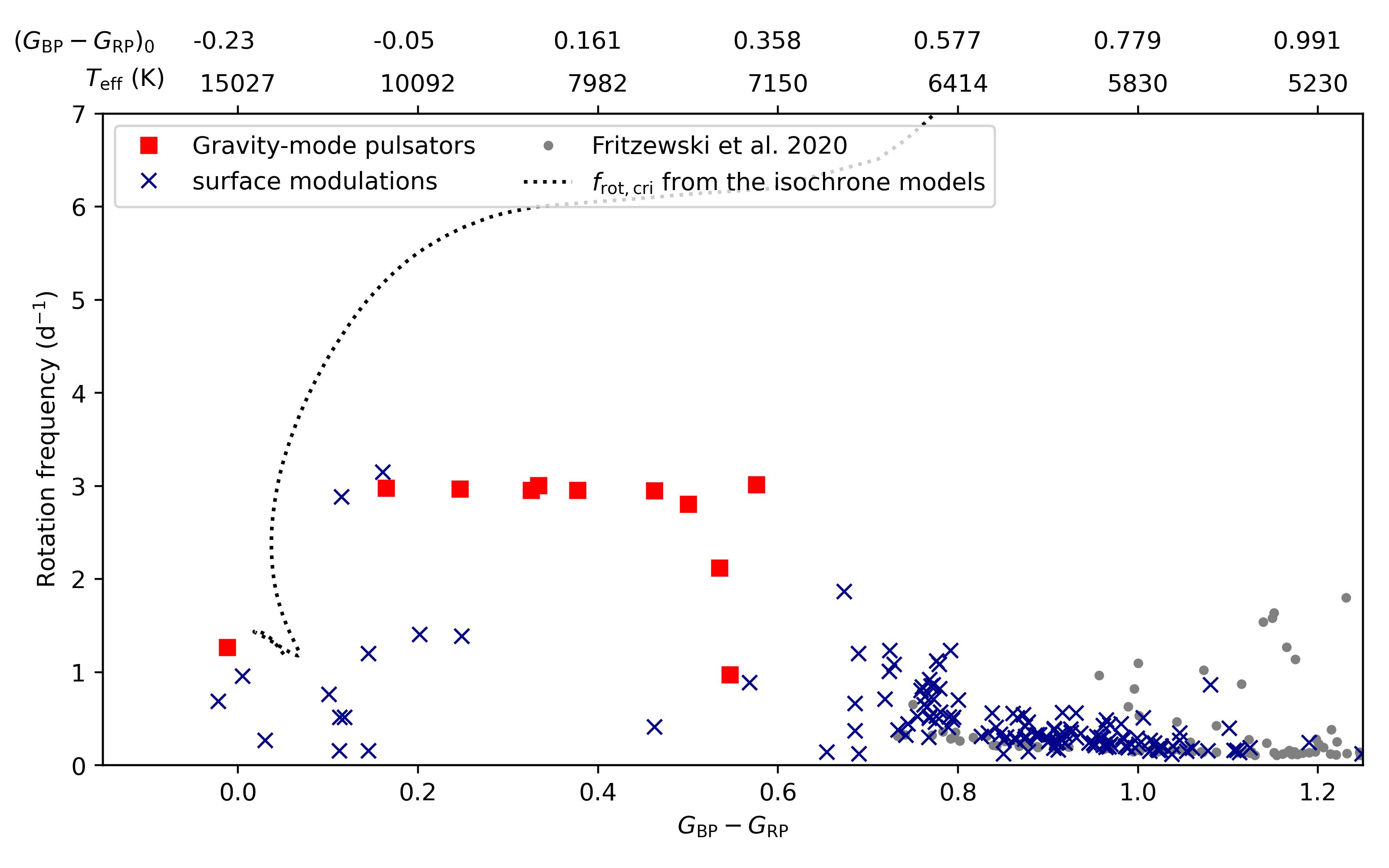}
    \caption{Observed rotation rates for cluster members from surface
      modulations or g-mode pulsations as a function of colour
      index. The y-axis is the rotation frequency in $\mathrm{d}^{-1}$. The bottom x-axis shows the observed \textit{Gaia} colour index, while the top x-axis shows both the effective temperature and the intrinsic \textit{Gaia} colour index from the best-fitting isochrone. The red squares show cluster members with near-core
      rotation rates measured from their identified g modes, while the dark blue crosses are stars with surface rotation measurements from modulations in their light curves. Rotation uncertainties are typically smaller than the marker sizes. The grey dots are the surface
      rotation rates for the late-type stars used for gyrochronology
      in the cluster reported by \cite{Fritzewski2020A&A}. The black dotted
      line represents the Keplerian critical rotation rate $f_\mathrm{rot, cri}$ from the best-fitting
      isochrone model.  }
    \label{fig:rotation_vs_bprp}
\end{figure*}

In this section, we want to give the relation between stellar rotation rates and their colour index $\bprp$. The peaks of the fundamental frequencies in the cluster stars with rotational
modulation directly give the surface rotation rates, while the g-mode
period spacings deliver the near-core rotation rates. We can compare them together because the radial differential rotations in main-sequence stars are not strong \citep[e.g.][]{Li2020MNRAS_611}.

We want to measure the surface rotation of stars for two reasons: firstly, to slightly extend the samples of rotation beyond the red edge of the instability strip, and secondly, to compare with the near-core rotation measured through asteroseismology in order to determine the mild radial differential rotation of stars. 
We use the stars with these measurements to investigate how the rotation rates change
as a function of colour index, which represents the effective temperature. 

To properly determine the frequency and uncertainty of a surface modulation signal, we adopt a strategy akin to that used for solar-like oscillators. The surface activity, excited and damped over time, performs similar to the stochastically excited oscillations observed in solar-like oscillators. Consequently, a critically sampled power spectrum of a surface modulation signal is expected to exhibit a Lorentzian profile, with the data following a $\chi^2$ distribution with two degrees of freedom \citep{Anderson1990}. The likelihood function, which quantifies the probability of observed data $D$ given a parameter $\theta$ \citep[see][]{Anderson1990}, is given by the equation:
\begin{equation}
    \ln p(D|\theta) = -\sum\left( \ln M_i(\theta) + \frac{D_i}{M_i(\theta)}  \right),\label{equ:solar_like_Lorentzian_likelihood}
\end{equation}
where $D_i$ represents the $i^\mathrm{th}$ data point, and $M_i(\theta)$ denotes the Lorentzian profile. The Lorentzian profile's parameters, $\theta$, include the central frequency (which is the surface rotation frequency we want to measure), amplitude, Full Width Half Maximum (FWHM), and background noise. We used the \textsc{emcee} package \citep{2013PASP..125..306F} to optimise the likelihood function in Eq.~\ref{equ:solar_like_Lorentzian_likelihood}, which is an implementation of the affine-invariant ensemble sampler of \cite{Goodman2010}. In the MCMC algorithm, we used 30 parallel chains and 5000 steps, and discarded the first 1000 steps for the final posterior distributions. We visually inspected all the posterior distributions to ensure all the chains converged. The determined surface rotation frequencies and their uncertainties are listed in Table~\ref{appendix_tab:surface_modulation}, and all the surface modulation signals are illustrated in Fig.~\ref{appendix_fig:all_surface_modulation}. The uncertainties of these frequencies are on the order of $10^{-3}$ to $10^{-4}\,\mathrm{d^{-1}}$, roughly akin to the frequency resolution of the 4-year light curve ($\sim 0.0007\,\mathrm{d^{-1}}$). In contrast, the uncertainties of stable oscillation signals (where amplitude, frequency, and phase remain constant) are significantly smaller. These are influenced by the signal-to-noise ratio and are typically less than one-tenth of the frequency resolution \citep{Montgomery1999}. Figure~\ref{appendix_fig:comparison_SM} compares the surface modulation periods in this work and those from \cite{Fritzewski2020A&A}, \cite{Bouma2021}, and \cite{Healy2020}, and a general consistency is seen.

In Fig.~\ref{fig:rotation_vs_bprp}, we present the
measured rotation rates as a function of colour index. For the cool
stars (with $\bprp > 0.6$\,mag), their rotation rates increase with increasing temperature. The rotation
rate reaches approximately $1\,\mathrm{d^{-1}}$ at $\bprp\approx0.7$\,mag,
indicating the diminishing effect of magnetic braking. The stars with
g-mode pulsations occur in the $\bprp$ range of 0.2\,mag to 0.6\,mag, where
rotational modulation is less common. Given the weak radial
differential rotation found in hundreds of intermediate-mass stars
\citep{Li2020MNRAS_611}, it is reasonable to compare the rotation
rates for all cluster stars having measurements for either the
near-core region or the surface. The near-core rotation rates measured
by g modes are approximately $3\,\mathrm{d^{-1}}$, significantly
faster than those derived for the cooler stars with magnetic
braking. For stars with $\bprp < 0.2$\,mag, both the rotation rates
measured by surface modulation and g-mode pulsations decrease
dramatically, displaying a large scatter of
approximately $1.5\,\mathrm{d^{-1}}$, which might be suppressed by the critical rotation rates, as discussed below.

In addition, we have calculated the critical rotation rates, which we
define as the rate at which the centrifugal and gravitational
accelerations at the
equator on an isobar are equal \citep[see the definition in ][]{Paxton2019ApJS}. To simplify the
calculation, we ignored the ellipsoidal distortion, that is
\begin{equation}
    \Omega_\mathrm{crit} = 2\pi f_\mathrm{rot, crit} = \sqrt{\frac{\mathrm{G}M}{R^3}},
\end{equation}
where $\mathrm{G}$ is the gravitational constant, $M$ is the stellar
mass, and $R$ is the radius. We used the theoretical radii and masses obtained from the best-fitting isochrone stellar model (dotted line in Fig.~\ref{fig:rotation_vs_bprp}). 

%Second, we estimated the masses
%and radii by using the observed luminosity $L$, temperature
%$T_\mathrm{eff}$, and surface gravity $\log g$ values from
%\textit{Gaia} by solving the equation $g = \mathrm{G}M/R^2$ along with
%Eq.~\ref{equ:stefan-boltzmann}.% we represent these results by the plus symbols in Figure~\ref{fig:rotation_vs_bprp}. 

The critical rotation
rates obtained by best-fitting isochrone reveal three distinct
regimes in Fig.~\ref{fig:rotation_vs_bprp}. First, a decreasing trend
is observed when $\bprp$ exceeds 0.6\,mag, declining to approximately
$6\,\mathrm{d^{-1}}$, which represents the transition region from
late-type to early-type stars; second, a relatively constant value is
maintained between 5 and $6\,\mathrm{d^{-1}}$ within the $\bprp$ range
of 0.2\,mag to 0.6\,mag, where $\gamma$\,Dor and $\delta$\,Sct stars occur;
third, a rapid decline is observed for $\bprp<0.2$\,mag, with the critical
rotation rate dropping to approximately $2\,\mathrm{d^{-1}}$, which
coincides with the main-sequence turn-off of the cluster, where a few surface
modulation stars and one SPB star appear.
% @GangLi: check if the 2nd leftmost square is also an SPB - cf. my
% earlier question/remark about this. Adapt if needed.

Within the region of $\bprp$ between 0.2\,mag and 0.6\,mag, the rotation rates measured for the g-mode pulsators correspond approximately to 50\% of
the critical rotation rates. As pointed out by \cite{Mombarg2021} and \cite{Henneco2021}, the TAR is still valid for $\Omega/\Omega_\mathrm{crit, Roche}$ up to 0.8, where $\Omega_\mathrm{crit, Roche} = \sqrt{8/27}\Omega_\mathrm{crit}$.
\cite{Mombarg2023calibrating_AM} found that the initial rotation frequencies of six slowly rotating field $\gamma\,$Dor stars near the
terminal age main sequence must have originated from stars rotating below 10\% of the initial critical frequency at the zero-age main sequence. Our values of $\Omega/\Omega_\mathrm{crit}\approx0.5$ show
that the $\gamma\,$Dor stars in NGC\,2516 are born with five times higher rotation rate. At NGC\,2516's turn-off, the rotation rates closely approach the critical rate, which is consistent with the
findings by \citet{Aerts2021RvMP} that all rotation rates, from zero to critical, are observed between stellar birth and the end of the main
sequence. Our findings are also consistent with some
spectropolarimetric observations of Be stars \citep[e.g. Regulus,
 Achernar, and $\alpha$ Arae, see][]{McAlister2005,
 Domiciano_de_Souza2003, Meilland2007}.

\textcolor{black}{We also calculated the observed critical rotation rates, which were calculated by the observed stellar radii and masses from the \textit{Gaia} photometry. }The theoretical critical rotation rates derived from isochrone models
are approximately 1\,$\mathrm{d^{-1}}$ higher than the observed
critical rotation rates, revealing a systematic deviation. This
deviation is easily understood in terms of a variety of choices for
the input physics of the models, notably the occurrence of major
differences in internal mixing \citep{Pedersen2021}. Moreover, part of
this systematic shift may arise from the neglected gravity darkening or centrifugal distortion of the
fast rotation of the cluster stars, making it challenging to precisely
define their $T_\mathrm{eff}$, $\log g$, and radius. 

In Sect.~\ref{sec:isochrone_fitting}, we determined that the best-fitting isochrones have $v / v_\mathrm{crit}\leq 0.4$, a finding that is incompatible with the value derived from g-mode pulsators (which is $\sim0.5$). This discrepancy is likely due to the inaccurate determination of $v_\mathrm{crit}$. In the meantime, \cite{Brogaard2023} reported that for NGC\,6866, the best-fitting isochrone indicates a lower rotation rate than the rate measured through asteroseismology. This phenomenon might suggest that the actual impact of rotation on stellar evolution is less significant than what is predicted by current 1-D models.

\subsection{No circumstellar disk around the fast-rotating B stars}

Fast-rotating stars may experience disk formation events due to
outbursts that eject surface material from their equator into a
decretion disk. 
This phenomenon, commonly known as the Be phenomenon, has been observed for the Be star HD\,49330 \citep{Huat2009}. Such decretion events can be triggered by the beating of g-mode pulsations \citep{Kurtz2015}. 
Notably, \citet{Neiner2020AA} also showed that the beating of the
stochastically-excited gravito-inertial waves in HD\,49330 is efficient for transporting the angular momentum from the core to the surface, causing the Be phenomenon.
If similar Be phenomena are observed in the rapid rotators within our sample, the asteroseismology data might offer further insights into the mechanisms triggering these occurrences.

Outburst akin to the one detected in the optical CoRoT light curve of HD\,49330 can also be detected at mid-infrared wavelengths, such as the Wide-field Infrared Survey Explorer (WISE; \citealt{Wright2010}) photometry bands $W1$ and $W2$. 
This wavelength range notably contains the primary emissions from the decretion disk.
\citet{Granada2018AJ} found that around half of the B-type stars in four young clusters have $W1-W2$ colours redder than $0.05$, while the intrinsic colours of early-type stars are around $-0.05$.  
\citet{Jian2023} confirmed this conclusion, identifying 916 early-type stars that have shown outbursts associated with the Be phenomenon over the past 13 years.
Given the presence of fast rotators in NGC\,2516, it is plausible that some of them might experience Be phenomena, potentially be captured in their WISE photometry.

\begin{figure}
    \centering
    \includegraphics[width=\linewidth]{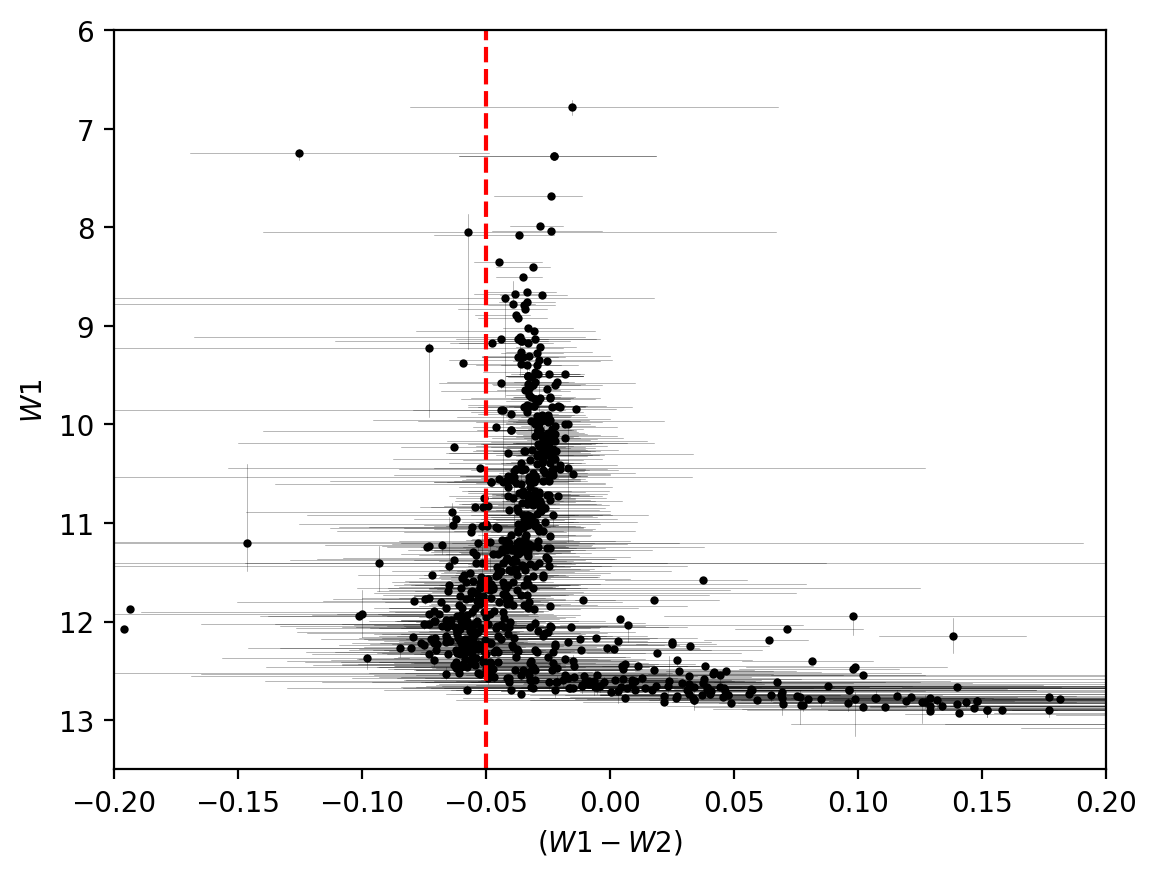}
    \caption{WISE colour-magnitude diagram of the sample stars, with the median value of $(W1-W2)$ and $W1$ plotted as black dots, and their extreme values plotted as horizontal and vertical errorbars. The red dashed-vertical line indicates the intrinsic colour ($-0.05$) of the stars with $T_\mathrm{eff} > 6000$\,K.}
    \label{fig:cmd_wise}
\end{figure}

Following the method introduced by \citet{Jian2023}, we
extracted the WISE epoch photometry for all our sample stars.
Figure~\ref{fig:cmd_wise} presents the colour-magnitude diagram with the median and extreme values derived from the WISE photometry for our sample of cluster members.
Stars exhibiting the Be phenomena typically display a "redder and brighter" variation in their photometry.
Their median $(W1-W2)$ will then range from \num{0} to \SI{0.3}{mag}, and the vertical bar in Figure~\ref{fig:cmd_wise} can extend up to \SI{1.5}{mag} (see Figure~9 in \citealt{Jian2023} for comparison).
Our analysis revealed that none of the stars within NGC\,2516 demonstrated such behaviour. 
It is important to note that the larger $(W1-W2)$ observed at $W1 > 12\,$mag corresponds to stars with lower effective temperatures, where their redder colours stem from the presence of molecular lines in the $W1$ band rather than from the Be phenomenon.

The reason for the absence of the Be phenomenon in NGC\,2516 is still uncertain, but there are potential reasons worth considering. 
One possibility might be the relatively old age of NGC\,2516. 
There are 35 Be stars reported in \citet{Jian2023} that are associated with open clusters, and most of these clusters' ages fall below 45 million years, which is less than half the age of NGC\,2516.
Another possible reason might be the rotation rate. 
The rotation rates we found for the A-type stars of NGC\,2516 are roughly at about 50\% of the critical rate. 
However, as shown in Fig.~\ref{fig:rotation_vs_bprp}, the B-type stars (with $\bprp < 0.2\,\mathrm{mag}$) exhibit much larger $v/v_\mathrm{crit}$ due to the decrease in critical rotation rates at the MSTO. 
The critical rotation rate at the MSTO is hard to determine, because of the rapid change in stellar physics. Therefore, it is still unclear if the rotation rates of our B-type stars are sufficient to induce the formation of decretion disks. 
It is noteworthy that some field SPB stars observed by {\it Kepler} display minor outbursts in their light curve despite the absence of disks \citep{VanBeeck2021}, indicating that additional factors beyond rotation may contribute to the occurrence or absence of the Be phenomenon \citep[see earlier comment on ][]{Kurtz2015}.

\section{Spectroscopic observations}\label{sec:spectra}

To accomplish the goal of calibrating stellar models through detailed
asteroseismic modelling of the g-mode pulsators in NGC\,2516 with
methods as in \citet{Aerts2018ApJS}, relying solely on photometric and
Gaia observations are sub-optimal. High-resolution spectroscopic
observations, delivering better constraints on $T_{\rm eff}$, $\log
g$, and metallicity offer a powerful addition as observables. Accurate
measurements of parameters such as $T_\mathrm{eff}$, metallicity, and
$\log g$ not only narrow down the range of possible models but also
help resolve degeneracies in the high-dimensional parameter space.
Additionally, measurements of surface element abundances play a
crucial role in constraining the internal mixing processes within
radiative envelopes, as demonstrated in previous studies
\citep{Pedersen2018,Mombarg2020,Mombarg2022,Pedersen2021,Michielsen2021}.

For these reasons, we aimed to collect high-resolution spectra of
the most interesting cluster members. Given their high potential for
cluster modelling, we focused on the g-mode pulsators exhibiting clear
period spacing patterns in this cluster and reported the data assembled
so far. We conducted spectroscopic follow-up observations using
the Fibre-fed Extended Range Optical Spectrograph (FEROS)
\citep{Kaufer1999}, which is mounted on the ESO/MPG 2.2-m telescope at
La Silla, Chile. FEROS is a high-resolution spectrograph with an
approximate resolution ($R$) of 48,000 and a wavelength coverage from
360 to 920\,nm. The data reduction was performed using the publicly
available pipeline Collection of Elemental Routines for \Echelle\
Spectra (CERES)
\citep{Brahm2017}\footnote{\url{https://github.com/rabrahm/ceres}}. All the spectra and the best-fitting results are shown in Sect.~\ref{appandix_sec:all_gdor_figures}. 

We derived the global parameters, including radial velocity,
$T_{\mathrm{eff}}$, $\log g$, projected equatorial rotation velocity
($v\sin i$), microturbulent velocity in the atmosphere($\xi$), and
metallicity ([M/H]), using the spectrum analysis algorithm
\textsc{zeta-Payne} \citep{Straumit2022}. This algorithm is fully
automated and machine-learning-based, specifically optimised for
intermediate- to high-mass stars with spectral types O, B, A, and F
and optimally suited to exploit FEROS spectroscopy of g-mode pulsators
\citep{Gebruers2022}.
The observational details and parameter results are summarised in
Table~\ref{tab:FEROS_results}. These observations were conducted
between 12th and 16th of March, 2023, with varying exposure times of
600, 900, and 1800\,s, depending on the brightness of the target
star. We aimed for a high signal-to-noise ratio (S/N) of 200 to ensure
accurate abundance determinations. In cases where achieving S/N=200
with a single exposure time was not feasible for faint stars, we
combined multiple shorter exposures with S/N=150 to reach the desired
S/N=200. However, it is worth noting that we still did not
achieve a high S/N for some stars, such that follow-up measurements
are still needed to get precise values of the stellar parameters.

\begin{figure}
    \centering
    \includegraphics[width=\linewidth]{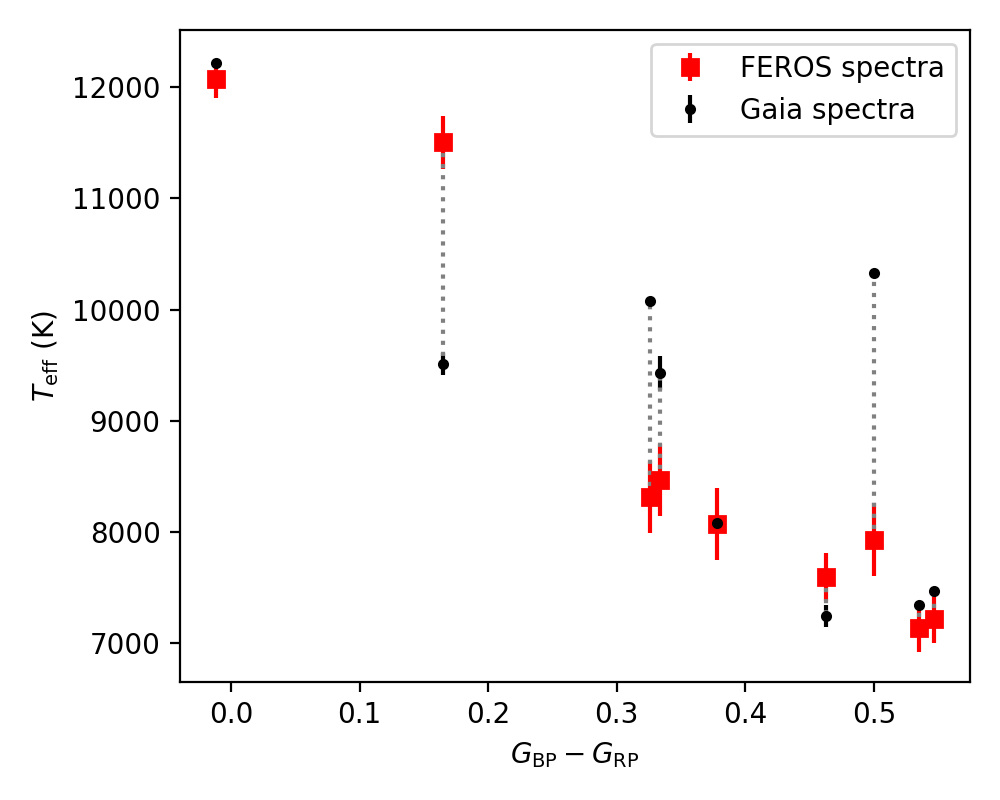} 
    \caption{Effective temperature of the g-mode stars as a
      function of their \textit{Gaia} colour index. We show the
      effective temperatures obtained by the FEROS spectra and
      \textit{Gaia} GSP-Phot connected by vertical dotted line.
% @GangLi: please redo the plot as requested
    }
    \label{fig:Teff_comparison_between_FEROS_and_Gaia}
\end{figure}

As demonstrated by \cite{Gebruers2022}, internal uncertainties occur
in the spectrum analysis, in addition to statistical uncertainties,
even when the S/N is high. Therefore, we calculated the uncertainties
in surface parameters and radial velocity by quadratic summation of
the observed and internal uncertainties.
Figure~\ref{fig:Teff_comparison_between_FEROS_and_Gaia} contains a
comparison between the effective temperatures obtained from the FEROS
spectra and those derived using the \textit{Gaia} General Stellar
Parametrizer from Photometry (GSP-Phot). Our analysis reveals that the
temperatures derived from the spectra exhibit a smoother relation with
the \textit{Gaia} colour index, whereas the GSP-Phot temperatures
display considerable scatter. In some cases, this scatter is as high
as approximately $2000\,\mathrm{K}$. Furthermore, we determined
the metallicities of these stars and found them to align with solar
metallicity within the uncertainties. The radial velocities (RV) of
the stars also exhibit consistency, averaging around $\sim 25\,\,\pm \sim
7\,\mathrm{km/s}$, consistent with \cite{Gonzalez2000AJ}. Notably, multiple observations of the same target
stars do not reveal significant RV variations, excluding
short-period binarity among the monitored stars.

\begin{table*}
    \tiny
    \centering
    \caption{Results of the spectroscopic observations of the g-mode
      pulsators in NGC\,2516. Some stars were observed more than once. }
    \label{tab:FEROS_results}
    
    \begin{tabular}{ccrrrrlrrrrr}
    \hline
TIC & DATE-OBS & EXP & $m_\mathrm{g}$ & BP-RP & S/N & RV (km/s) & $T_\mathrm{eff}$ (K) & $\log g$ (dex) & $v\sin i$ (km/s) & $\xi$ (km/s) & [M/H] (dex) \\ 
    \hline
281582674 & 2023-03-12 & 1800 & 10.36 & 0.38 & 101 & $26\pm20$ & $8100\pm300$ & $3.97\pm0.21$ & $274\pm24$ & $3.1\pm2.0$ & $0.02\pm0.20$\\
281582674 & 2023-03-16 & 1800 & 10.36 & 0.38 & 121 & $25\pm12$ & $8100\pm300$ & $3.95\pm0.20$ & $274\pm23$ & $3.2\pm1.9$ & $-0.01\pm0.20$\\
308307454 & 2023-03-13 & 1800 & 11.20 & 0.55 & 56 & $24\pm4$ & $7220\pm220$ & $3.8\pm0.4$\phantom{$0$} & $56\pm12$ & $2.8\pm1.4$ & $-0.27\pm0.21$\\
308307454 & 2023-03-14 & 1800 & 11.20 & 0.55 & 77 & $23\pm3$ & $7250\pm220$ & $3.7\pm0.3$\phantom{$0$} & $56\pm12$ & $3.0\pm1.4$ & $-0.32\pm0.21$\\
308307454 & 2023-03-16 & 1800 & 11.20 & 0.55 & 67 & $25\pm3$ & $7250\pm220$ & $3.7\pm0.4$\phantom{$0$} & $55\pm12$ & $2.9\pm1.4$ & $-0.31\pm0.21$\\
358466708 & 2023-03-12 & 600 & 8.05 & -0.01 & 191 & $27\pm5$ & $12070\pm170$ & $3.83\pm0.06$ & $181\pm16$ & $0.6\pm2.0$ & $0.08\pm0.16$\\
358466729 & 2023-03-13 & 1800 & 11.15 & 0.54 & 81 & $24\pm6$ & $7140\pm220$ & $4.1\pm0.4$\phantom{$0$} & $129\pm14$ & $2.3\pm1.4$ & $-0.11\pm0.21$\\
358466729 & 2023-03-15 & 1800 & 11.15 & 0.54 & 91 & $22\pm4$ & $7160\pm220$ & $4.1\pm0.3$\phantom{$0$} & $130\pm13$ & $2.4\pm1.4$ & $-0.11\pm0.21$\\
364398040 & 2023-03-13 & 1800 & 10.50 & 0.46 & 92 & $22\pm8$ & $7590\pm220$ & $3.9\pm0.4$\phantom{$0$} & $271\pm18$ & $3.9\pm1.4$ & $-0.12\pm0.20$\\
364398040 & 2023-03-15 & 1800 & 10.50 & 0.46 & 140 & $18\pm4$ & $7610\pm220$ & $3.9\pm0.3$\phantom{$0$} & $274\pm15$ & $4.0\pm1.3$ & $-0.09\pm0.20$\\
372912679 & 2023-03-14 & 1800 & 10.37 & 0.33 & 130 & $24\pm7$ & $8500\pm300$ & $4.19\pm0.19$ & $134\pm21$ & $3.2\pm2.0$ & $-0.11\pm0.19$\\
372912679 & 2023-03-15 & 1800 & 10.37 & 0.33 & 156 & $24\pm7$ & $8500\pm300$ & $4.21\pm0.19$ & $134\pm20$ & $3.3\pm2.0$ & $-0.13\pm0.19$\\
372913043 & 2023-03-12 & 900 & 8.74 & 0.17 & 144 & $22\pm4$ & $11510\pm240$ & $4.12\pm0.11$ & $96\pm19$ & $0.9\pm2.3$ & $0.09\pm0.22$\\
410451583 & 2023-03-14 & 1800 & 9.58 & 0.33 & 218 & $26\pm6$ & $8300\pm300$ & $3.91\pm0.18$ & $304\pm20$ & $3.2\pm1.9$ & $-0.21\pm0.18$\\
410452218 & 2023-03-14 & 1800 & 10.86 & 0.50 & 83 & $23\pm12$ & $7900\pm300$ & $3.92\pm0.23$ & $272\pm27$ & $2.4\pm2.0$ & $0.12\pm0.22$\\
410452218 & 2023-03-16 & 1800 & 10.86 & 0.50 & 86 & $24\pm13$ & $7900\pm220$ & $3.9\pm0.3$\phantom{$0$} & $268\pm18$ & $2.3\pm1.4$ & $0.16\pm0.21$\\
\hline
    \end{tabular}
    \tablefoot{We list TIC
      number (TIC), observation date (DATE-OBS), exposure time (EXP),
      \textit{Gaia} g-band magnitude $m_\mathrm{g}$, \textit{Gaia}
      colour index (BP-RP), radial velocity (RV), effective
      temperature ($T_\mathrm{eff}$), surface gravity $\log g$,
      projected rotational velocity ($v\sin i$), microturbulent
      velocity ($\xi$), and metallicity ([M/H]).  }
\end{table*}

\begin{figure}
    \centering
    \includegraphics[width=\linewidth]{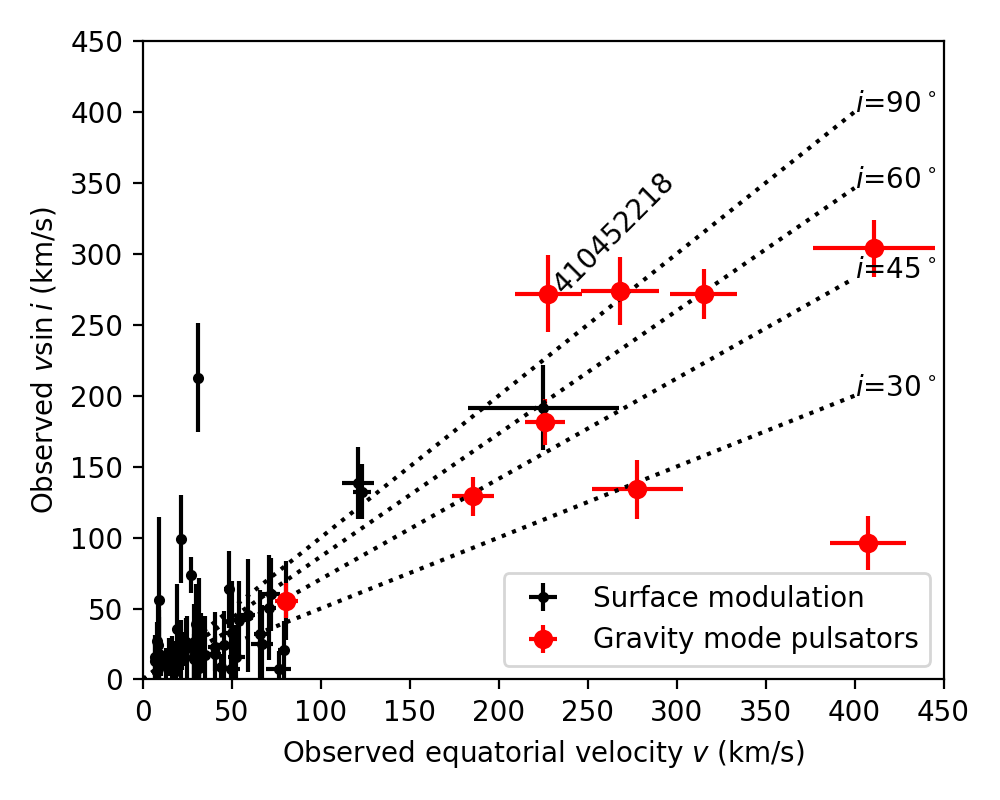}
    \caption{Comparison between the observed equatorial velocity and the observed $v\sin i$ by \textit{Gaia}. For the surface modulation stars (black dot symbols), the $v\sin i$ comes from the \textit{Gaia}'s \texttt{vbroad} parameter, and the $v$ is calculated using the rotation frequency and stellar radius. For the g-mode pulsators (red circle symbols), the $v \sin i$ are measured by the FEROS spectra, and their $v$ is calculated by the near-core rotation rate and stellar radius. The dotted lines mark the locations with various inclinations. }
    \label{fig:v_vsini_inclination}
\end{figure}

Figure~\ref{fig:v_vsini_inclination} illustrates the comparison
between various estimates of the projected equatorial rotation
velocity ($v \sin i$) and the true equatorial velocity for several of
the cluster members. This provides a direct estimate of the
inclination of the rotation axis of the stars in the
line of sight. Various remarks are in order. For stars exhibiting
surface modulations, we have a direct determination of the surface rotation
period. We computed their equatorial velocities by multiplying with
the radii as determined in
Section~\ref{subsec:solar-like}. Additionally, we obtained their
projected equatorial velocities using the \texttt{vbroad} parameter
provided by the \textit{Gaia} Radial Velocity Spectrometer (RVS)
\citep{Fremat2023}. For stars displaying period spacing patterns with
identified g-mode pulsations, we have the near-core rotation rates and
the $v \sin i$ values were measured from FEROS spectra. For these
stars, we recalculated their radii using the more accurate effective
temperatures derived from the FEROS spectra. We then determined their
equatorial velocities by multiplying their radii with the near-core
rotation rates, assuming these stars have rigid rotation.
Figure~\ref{fig:v_vsini_inclination} reveals that most stars with
surface modulations have low $v \sin i$ values not significantly
different from zero, indicative of slow rotation. When the equatorial
velocity surpasses approximately $100\,\mathrm{km/s}$, the stars have
$v \sin i$ values significantly above zero. We find that these stars
cover a range of inclinations, spanning from approximately $30^\circ$
to $90^\circ$. Although it concerns relatively a few cluster members,
most of the stars with inclination estimates have values $i \gtrsim
45^\circ$, resembling a distribution akin to random orientations. It is worth noting that TIC 410452218 is the only star for which the projected equatorial velocity ($v \sin i$) is larger than the equatorial velocity. This discrepancy might be due to poor consideration of rotational distortion or gravity darkening.

We stress that the determination of the inclination angles comes with
a significant degree of uncertainty. First, it is important to note
that in the case of rapidly rotating stars, there are variations in
both radius and temperature between the pole and equator. Therefore,
the radius determined through the Stefan-Boltzmann law should be
considered as an averaged value that is also influenced by the
effective temperature and inclination. Secondly, differential rotation
may occur while we have assumed rigid rotation for the g-mode
pulsators. Observations have indicated that the difference between the
near-core and surface rotation rates typically does not exceed 10\%
for $\gamma\,$Dor stars from the \textit{Kepler} field \citep{Li2020MNRAS_611}, but these stars are much older than those in NGC\,2516. However, recent
two-dimensional models suggest a more complex internal rotation
profile and show that the core might rotate approximately 50\% faster
than the envelope \citep{Bouchaud2020, Mombarg2023}. This level of
differentiality in the rotation adds to the complexity and uncertainty
in determining the inclination angles in these cases.

\section{Conclusions}\label{sec:conclusions}
In this study, we conduct the first asteroseismic study of the
young open cluster NGC\,2516, \ca{from an observational perspective.} Our findings reveal a variety of
different types of variable stars in this cluster. These stars, along
with the cluster itself, can help us calibrate the physical processes
inside stars to unprecedented precision and answer long-standing questions about star clusters.

Firstly, we applied the rotating MIST isochrones to fit the color-magnitude
diagram (CMD) of the cluster. We find that isochrones with \textcolor{black}{$v /
v_\mathrm{crit} \leq 0.4$} closely matched the observational data
and yielded consistent results for age and extinction. However, those
with \textcolor{black}{$v / v_\mathrm{crit} \geq 0.5$} did not provide a good
fit to the data. By combining the fitting results obtained from
isochrones with \textcolor{black}{$\Omega / \Omega_\mathrm{crit} \leq 0.4$}, we report an
age of $102\pm15\,\mathrm{Myr}$ and an extinction value at 550\,nm of
$A_0 = 0.53\pm0.04\,\mathrm{mag}$. Our newly determined age shows that
NGC\,2516 is younger than the Pleiades, while we provide a
slightly larger extinction and reddening than the previous study.

We used the TESS data in cycle 1 and cycle 3 to conduct photometry and obtained the light curves of the member stars. The almost continuous
TESS light curves provided excellent data for asteroseismology. We find
\gdornumber g-mode pulsators, \dsctnumber $\delta$\,Sct pulsators,
\surfacemodulationnumber stars with surface modulations, \EBnumber
eclipsing binaries, and three post-main-sequence stars.

Among the g-mode pulsators, there are \gdornumberwithcleargmodepattern
with clear period spacing patterns, which allow us to determine their
near-core rotation rates and asymptotic spacings. We identified most
of these stars as $\gamma$\,Dor stars based on their asymptotic spacings (around $\sim 4900\,\mathrm{s}$) and effective temperatures (between 7000\,K and 10000\,K). Their asymptotic spacings are
larger than those $\gamma$\,Dor stars in the {\it Kepler\/}
field because they are young. Additionally, there are two stars with even higher effective temperatures ($> 10000\,\mathrm{K}$). Considering the position of
these stars at the top of the main sequence, we classified them as SPB stars. Our findings reveal that the g-mode stars in NGC\,2516 exhibit rapid near-core rotation rates, some reaching values up to $3\,\mathrm{d^{-1}}$. This is significantly faster than the average value for those observed for
single pulsators in both the {\it Kepler\/} and TESS fields. The combination of a large asymptotic period spacing and high near-core rotation rates
aligns with the nature that these stars are very young.

For the $\delta$\,Sct stars, we observe that their amplitude spectra
are well ordered when sorted by colour. In cooler stars, we find a
series of frequencies at approximately $21.07\,\mathrm{d^{-1}}$, while
the hotter p-mode pulsators display another series of frequency peaks around
$33\,\mathrm{d^{-1}}$. These frequencies are identified as the
fundamental frequency and second overtone, respectively. Notably, the
mean pulsation frequency increases as the temperature rises,
indicating a frequency -- temperature relation in $\delta$\,Sct
stars. We cannot identify reliable frequency separations
among these $\delta$\,Sct stars, except for one cluster member.

For the three post-main-sequence stars, we observe a low-frequency
power excess in their power density spectra in agreement with
theoretical predictions for granulation resulting from magnetic
activity. Based on the closest model from the best-fitting isochrone,
these evolved stars have masses larger than $5\,\mathrm{M_\odot}$, which is greater than the most massive star from the \textit{Kepler} sample \citep{Yu2018}.

The surface modulation signals observed in \surfacemodulationnumber
stars offer direct measurements of surface rotation rates. Combined
with near-core rotation rates determined from g-mode pulsations, we
unveil a rotation\,--\,temperature relationship among the
main-sequence stars in NGC\,2516. Our findings show that as
temperature increases, the typical rotation rate follows a rising
trend until it reaches a plateau (whose red edge is at $\bprp=0.6\,\mathrm{mag}$ or $T_\mathrm{eff} = 7150\,\mathrm{K}$), where it stabilises at approximately
50\% of the critical rotation rates. Furthermore, we calculate the
critical rotation rates using both the best-fitting stellar models and
observations. We observe a rapid drop in the critical rotation rate at
the main-sequence turn-off, in agreement with the observed stellar
rotation rates. It is noteworthy that the rotation rate in B-type
stars may approach nearly $\sim 90\%$ of the critical rotation
rate. However, we did not find any evidence of circumstellar disks in
these cluster stars.

Finally, we carried out high-resolution spectroscopic observations for
the g-mode pulsators displaying period spacing patterns. Our spectra,
with high resolution and high S/N, enabled precise measurements of the
global stellar parameters and radial velocity. Our analysis confirms
that the cluster has solar metallicity and exhibits a radial velocity
of approximately $\sim 25\,\mathrm{km/s}$. These accurate
spectroscopic observations will provide invaluable information for our
planned detailed ensemble asteroseismic modelling of the g-mode
pulsators under the constraint of equal initial chemical composition
at birth, reducing computational complexity and resolving parameter
degeneracies.

NGC\,2516 revealed itself as an optimal cluster for
combined detailed asteroseismic modelling of its g-mode pulsators.  So
far such type of asteroseismic cluster modelling based on g-mode
pulsators has not yet been done. The only young open cluster that was scrutinised by g-mode asteroseismology so far is UBC1, which
has only one g-mode pulsator with identified modes 
\ca{and this cluster member has a moderate rotation -- see the study by \citet{Fritzewski2024}. 
Adopting the same methodology as developed by these authors, we encountered the limitation of the currently available asteroseismic model grids for $\gamma\,$Dor stars, in the sense that none of the models predict values of $\Pi_0$ comparable to the observed ones. We ascribe this mismatch to the fast rotation of the member stars in NGC\,2516, 
along with the use of non-rotating equilibrium models in the existing $\gamma\,$Dor grids. Proper asteroseismic modelling of the identified modes of the $\gamma\,$Dor members of NGC\,2516 likely requires stellar evolution models with rotational deformation to achieve meaningful comparisons between measured and predicted asteroseismic observables. This will be taken up in a follow-up study. 
Since the addition of just one g-mode pulsator 
already implied a drastic reduction in the uncertainty of the cluster age for UBC1, our future modelling of the eleven g-mode pulsators and of the $\delta\,$Sct overtone pulsators in NGC\,2516 promises to be extremely rewarding to increase our knowledge of the
interior physics of intermediate-mass stars and of young open clusters, including proper age-dating.}

\begin{acknowledgements}
The authors are grateful to Professor Aaron Dotter for his
contribution to the construction of the rotating isochrones used in
this work. GL also thanks Doctor Sarah Gebruers for her help in the
exploitation of the FEROS spectroscopy.
\newline
The research leading to these results has received funding
from the KU\,Leuven Research Council (grant C16/18/005: PARADISE). 
GL acknowledges the Research Foundation Flanders (FWO) Grant for a long stay abroad 
\ca{(grant V422323N)} and the Dick Hunstead Fund for Astrophysics for his 2-month stay at the University of Sydney. GL also acknowledges the travel support from the National Natural Science Foundation of China (NSFC) through grant 12273002. \textcolor{black}{JSGM acknowledges funding the French Agence Nationale de la Recherche (ANR), under grant MASSIF (ANR-21-CE31-0018-02). }
CA acknowledges financial support from the FWO under grant K802922N (Sabbatical leave)
and from the European Research Council (ERC) under the Horizon Europe
programme (Synergy Grant agreement N$^\circ$101071505: 4D-STAR). 
\ca{While partially funded by the European Union, views and opinions expressed are however those of the authors only and do not necessarily reflect those of the European Union or the European Research Council. Neither the European Union nor the granting authority can be held responsible for them. }

\newline
This research has made use of NASA's Astrophysics Data System
Bibliographic Services and of the SIMBAD database and the VizieR
catalogue access tool, operated at CDS, Strasbourg, France.
This publication makes use of data products from the Wide-field
Infrared Survey Explorer, which is a joint project of the University
of California, Los Angeles, and the Jet Propulsion
Laboratory/California Institute of Technology, and NEOWISE, which is a
project of the Jet Propulsion Laboratory/California Institute of
Technology. WISE and NEOWISE are funded by the National Aeronautics
and Space Administration.  This work has made use of data from the
European Space Agency (ESA) mission \emph{Gaia}
(\url{https://www.cosmos.esa.int/gaia}), processed by the \emph{Gaia}
Data Processing and Analysis Consortium (DPAC,
\url{https://www.cosmos.esa.int/web/gaia/dpac/consortium}). Funding
for the DPAC has been provided by national institutions, in particular
the institutions participating in the \emph{Gaia} Multilateral
Agreement. This paper includes data collected by the TESS mission,
which are publicly available from the Mikulski Archive for Space
Telescopes (MAST).

\newline
%    \textbf{Software:}
%%%%% the bibfile
This research made use of \texttt{Astropy}, a community-developed
core \texttt{Python} package for Astronomy \citep{2013A&A...558A..33A} and \texttt{Lightkurve}, a
\texttt{Python} package for Kepler and TESS data analysis \citep{2018ascl.soft12013L}. 
%    This work made use of \texttt{ColorBrewer2} \url{http://www.ColorBrewer2.org}.
This research also made use of the following \texttt{Python} packages:
\texttt{astroquery} \citep{2019AJ....157...98G}, \textsc{emcee} \citep{2013PASP..125..306F}, 
\texttt{corner} \citep{2016JOSS....1...24F},
%    \texttt{IPython} \citep{ipython};
\texttt{MatPlotLib} \citep{2005ASPC..347...91B},
\texttt{NumPy} \citep{harris2020array}, and \texttt{Pandas} \citep{reback2020pandas}.
%    \texttt{Scikit-learn} \citep{scikit-learn}
%    \texttt{SciPy} \citep{scipy};
%    \texttt{seaborn} \citep{Waskom2021}
 
\end{acknowledgements}

   \bibliographystyle{aa} % style aa.bst
   \bibliography{V2-ligangreference} % your references Yourfile.bib

\begin{thebibliography}{219}
\expandafter\ifx\csname natexlab\endcsname\relax\def\natexlab#1{#1}\fi

\bibitem[{{Abt}(1979)}]{Abt1979ApJ}
{Abt}, H.~A. 1979, \apj, 230, 485

\bibitem[{{Abt} {et~al.}(1969){Abt}, {Clements}, {Doose}, \&
  {Harris}}]{Abt1969AJ_rotational_velocities}
{Abt}, H.~A., {Clements}, A.~E., {Doose}, L.~R., \& {Harris}, D.~H. 1969, \aj,
  74, 1153

\bibitem[{{Abt} \& {Levy}(1972)}]{Abt1972}
{Abt}, H.~A. \& {Levy}, S.~G. 1972, \apj, 172, 355

\bibitem[{{Aerts}(2021)}]{Aerts2021RvMP}
{Aerts}, C. 2021, Reviews of Modern Physics, 93, 015001

\bibitem[{{Aerts} {et~al.}(2010){Aerts}, {Christensen-Dalsgaard}, \&
  {Kurtz}}]{Aerts2010book}
{Aerts}, C., {Christensen-Dalsgaard}, J., \& {Kurtz}, D.~W. 2010,
  {Asteroseismology, Springer-Verlag, Heidelberg}

\bibitem[{{Aerts} \& {De Cat}(2003)}]{Aerts2003}
{Aerts}, C. \& {De Cat}, P. 2003, \ssr, 105, 453

\bibitem[{{Aerts} \& {Mathis}(2023)}]{Aerts2023A&A}
{Aerts}, C. \& {Mathis}, S. 2023, \aap, 677, A68

\bibitem[{{Aerts} {et~al.}(2019){Aerts}, {Mathis}, \&
  {Rogers}}]{Aerts2019ARA&A}
{Aerts}, C., {Mathis}, S., \& {Rogers}, T.~M. 2019, \araa, 57, 35

\bibitem[{{Aerts} {et~al.}(2023){Aerts}, {Molenberghs}, \& {De
  Ridder}}]{Aerts2023-DR3}
{Aerts}, C., {Molenberghs}, G., \& {De Ridder}, J. 2023, \aap, 672, A183

\bibitem[{{Aerts} {et~al.}(2018){Aerts}, {Molenberghs}, {Michielsen},
  {Pedersen}, {Bj{\"o}rklund}, {Johnston}, {Mombarg}, {Bowman}, {Buysschaert},
  {P{\'a}pics}, {Sekaran}, {Sundqvist}, {Tkachenko}, {Truyaert}, {Van Reeth},
  \& {Vermeyen}}]{Aerts2018ApJS}
{Aerts}, C., {Molenberghs}, G., {Michielsen}, M., {et~al.} 2018, \apjs, 237, 15

\bibitem[{{Aerts} {et~al.}(2003){Aerts}, {Thoul}, {Daszy{\'n}ska}, {Scuflaire},
  {Waelkens}, {Dupret}, {Niemczura}, \& {Noels}}]{Aerts2003Sci}
{Aerts}, C., {Thoul}, A., {Daszy{\'n}ska}, J., {et~al.} 2003, Science, 300,
  1926

\bibitem[{{Anderson} {et~al.}(1990){Anderson}, {Duvall}, \&
  {Jefferies}}]{Anderson1990}
{Anderson}, E.~R., {Duvall}, Thomas~L., J., \& {Jefferies}, S.~M. 1990, \apj,
  364, 699

\bibitem[{{Andrae} {et~al.}(2018){Andrae}, {Fouesneau}, {Creevey}, {Ordenovic},
  {Mary}, {Burlacu}, {Chaoul}, {Jean-Antoine-Piccolo}, {Kordopatis}, {Korn},
  {Lebreton}, {Panem}, {Pichon}, {Th{\'e}venin}, {Walmsley}, \&
  {Bailer-Jones}}]{Andrae2018}
{Andrae}, R., {Fouesneau}, M., {Creevey}, O., {et~al.} 2018, \aap, 616, A8

\bibitem[{{Antoci} {et~al.}(2014){Antoci}, {Cunha}, {Houdek}, {Kjeldsen},
  {Trampedach}, {Handler}, {L{\"u}ftinger}, {Arentoft}, \&
  {Murphy}}]{antocietal2014}
{Antoci}, V., {Cunha}, M., {Houdek}, G., {et~al.} 2014, \apj, 796, 118

\bibitem[{{Antonello} \& {Mantegazza}(1986)}]{Antonello1986A&A}
{Antonello}, E. \& {Mantegazza}, L. 1986, \aap, 164, 40

\bibitem[{{Asplund} {et~al.}(2009){Asplund}, {Grevesse}, {Sauval}, \&
  {Scott}}]{Asplund2009}
{Asplund}, M., {Grevesse}, N., {Sauval}, A.~J., \& {Scott}, P. 2009, \araa, 47,
  481

\bibitem[{{Astropy Collaboration} {et~al.}(2013){Astropy Collaboration},
  {Robitaille}, {Tollerud}, {Greenfield}, {Droettboom}, {Bray}, {Aldcroft},
  {Davis}, {Ginsburg}, {Price-Whelan}, {Kerzendorf}, {Conley}, {Crighton},
  {Barbary}, {Muna}, {Ferguson}, {Grollier}, {Parikh}, {Nair}, {Unther},
  {Deil}, {Woillez}, {Conseil}, {Kramer}, {Turner}, {Singer}, {Fox}, {Weaver},
  {Zabalza}, {Edwards}, {Azalee Bostroem}, {Burke}, {Casey}, {Crawford},
  {Dencheva}, {Ely}, {Jenness}, {Labrie}, {Lim}, {Pierfederici}, {Pontzen},
  {Ptak}, {Refsdal}, {Servillat}, \& {Streicher}}]{2013A&A...558A..33A}
{Astropy Collaboration}, {Robitaille}, T.~P., {Tollerud}, E.~J., {et~al.} 2013,
  \aap, 558, A33

\bibitem[{{Bagnulo} {et~al.}(2003){Bagnulo}, {Landstreet}, {Lo Curto},
  {Szeifert}, \& {Wade}}]{Bagnulo2003}
{Bagnulo}, S., {Landstreet}, J.~D., {Lo Curto}, G., {Szeifert}, T., \& {Wade},
  G.~A. 2003, \aap, 403, 645

\bibitem[{{Bailey} {et~al.}(2018){Bailey}, {Mateo}, {White}, {Shectman}, \&
  {Crane}}]{Bailey2018}
{Bailey}, J.~I., {Mateo}, M., {White}, R.~J., {Shectman}, S.~A., \& {Crane},
  J.~D. 2018, \mnras, 475, 1609

\bibitem[{{Balona} \& {Dziembowski}(2011)}]{Balona+Dziembowski2011}
{Balona}, L.~A. \& {Dziembowski}, W.~A. 2011, \mnras, 417, 591

\bibitem[{{Balona} {et~al.}(1994){Balona}, {Krisciunas}, \&
  {Cousins}}]{Balona1994}
{Balona}, L.~A., {Krisciunas}, K., \& {Cousins}, A.~W.~J. 1994, \mnras, 270,
  905

\bibitem[{{Barcel{\'o} Forteza} {et~al.}(2020){Barcel{\'o} Forteza}, {Moya},
  {Barrado}, {Solano}, {Mart{\'\i}n-Ruiz}, {Su{\'a}rez}, \& {Garc{\'\i}a
  Hern{\'a}ndez}}]{Barcelo-Forteza++2020}
{Barcel{\'o} Forteza}, S., {Moya}, A., {Barrado}, D., {et~al.} 2020, \aap, 638,
  A59

\bibitem[{{Barcel{\'o} Forteza} {et~al.}(2018){Barcel{\'o} Forteza}, {Roca
  Cort{\'e}s}, \& {Garc{\'{\i}}a}}]{Barcelo-Forteza++2018}
{Barcel{\'o} Forteza}, S., {Roca Cort{\'e}s}, T., \& {Garc{\'{\i}}a}, R.~A.
  2018, \aap, 614, A46

\bibitem[{{Barnes}(2003)}]{Barnes2003}
{Barnes}, S.~A. 2003, \apj, 586, 464

\bibitem[{{Barnes}(2007)}]{Barnes2007}
{Barnes}, S.~A. 2007, \apj, 669, 1167

\bibitem[{{Barrett} {et~al.}(2005){Barrett}, {Hunter}, {Miller}, {Hsu}, \&
  {Greenfield}}]{2005ASPC..347...91B}
{Barrett}, P., {Hunter}, J., {Miller}, J.~T., {Hsu}, J.~C., \& {Greenfield}, P.
  2005, in Astronomical Society of the Pacific Conference Series, Vol. 347,
  Astronomical Data Analysis Software and Systems XIV, ed. P.~{Shopbell},
  M.~{Britton}, \& R.~{Ebert}, 91

\bibitem[{{Bastian} \& {de Mink}(2009)}]{Bastian2009MNRAS}
{Bastian}, N. \& {de Mink}, S.~E. 2009, \mnras, 398, L11

\bibitem[{{Bastian} {et~al.}(2018){Bastian}, {Kamann}, {Cabrera-Ziri},
  {Georgy}, {Ekstr{\"o}m}, {Charbonnel}, {de Juan Ovelar}, \&
  {Usher}}]{Bastian2018MNRAS}
{Bastian}, N., {Kamann}, S., {Cabrera-Ziri}, I., {et~al.} 2018, \mnras, 480,
  3739

\bibitem[{{Bastian} {et~al.}(2016){Bastian}, {Niederhofer},
  {Kozhurina-Platais}, {Salaris}, {Larsen}, {Cabrera-Ziri}, {Cordero},
  {Ekstr{\"o}m}, {Geisler}, {Georgy}, {Hilker}, {Kacharov}, {Li}, {Mackey},
  {Mucciarelli}, \& {Platais}}]{Bastian2016MNRAS}
{Bastian}, N., {Niederhofer}, F., {Kozhurina-Platais}, V., {et~al.} 2016,
  \mnras, 460, L20

\bibitem[{{Basu} \& {Chaplin}(2017)}]{Basu2017asda.book}
{Basu}, S. \& {Chaplin}, W.~J. 2017, {Asteroseismic Data Analysis: Foundations
  and Techniques}

\bibitem[{{Basu} {et~al.}(2011){Basu}, {Grundahl}, {Stello}, {Kallinger},
  {Hekker}, {Mosser}, {Garc{\'\i}a}, {Mathur}, {Brogaard}, {Bruntt}, {Chaplin},
  {Gai}, {Elsworth}, {Esch}, {Ballot}, {Bedding}, {Gruberbauer}, {Huber},
  {Miglio}, {Yildiz}, {Kjeldsen}, {Christensen-Dalsgaard}, {Gilliland},
  {Fanelli}, {Ibrahim}, \& {Smith}}]{Basu2011}
{Basu}, S., {Grundahl}, F., {Stello}, D., {et~al.} 2011, \apjl, 729, L10

\bibitem[{{Bedding} {et~al.}(2015){Bedding}, {Murphy}, {Colman}, \&
  {Kurtz}}]{Bedding2015gdor}
{Bedding}, T.~R., {Murphy}, S.~J., {Colman}, I.~L., \& {Kurtz}, D.~W. 2015, in
  European Physical Journal Web of Conferences, Vol. 101, European Physical
  Journal Web of Conferences, 01005

\bibitem[{{Bedding} {et~al.}(2023){Bedding}, {Murphy}, {Crawford}, {Hey},
  {Huber}, {Kjeldsen}, {Li}, {Mann}, {Torres}, {White}, \&
  {Zhou}}]{Bedding2023ApJ}
{Bedding}, T.~R., {Murphy}, S.~J., {Crawford}, C., {et~al.} 2023, \apjl, 946,
  L10

\bibitem[{{Bedding} {et~al.}(2020){Bedding}, {Murphy}, {Hey}, {Huber}, {Li},
  {Smalley}, {Stello}, {White}, {Ball}, {Chaplin}, {Colman}, {Fuller},
  {Gaidos}, {Harbeck}, {Hermes}, {Holdsworth}, {Li}, {Li}, {Mann}, {Reese},
  {Sekaran}, {Yu}, {Antoci}, {Bergmann}, {Brown}, {Howard}, {Ireland},
  {Isaacson}, {Jenkins}, {Kjeldsen}, {McCully}, {Rabus}, {Rains}, {Ricker},
  {Tinney}, \& {Vanderspek}}]{Bedding2020}
{Bedding}, T.~R., {Murphy}, S.~J., {Hey}, D.~R., {et~al.} 2020, \nat, 581, 147

\bibitem[{{Bertelli} {et~al.}(2003){Bertelli}, {Nasi}, {Girardi}, {Chiosi},
  {Zoccali}, \& {Gallart}}]{Bertelli2003AJ}
{Bertelli}, G., {Nasi}, E., {Girardi}, L., {et~al.} 2003, \aj, 125, 770

\bibitem[{{Bouabid} {et~al.}(2013){Bouabid}, {Dupret}, {Salmon},
  {Montalb{\'a}n}, {Miglio}, \& {Noels}}]{Bouabid2013}
{Bouabid}, M.~P., {Dupret}, M.~A., {Salmon}, S., {et~al.} 2013, \mnras, 429,
  2500

\bibitem[{{Bouchaud} {et~al.}(2020){Bouchaud}, {Domiciano de Souza},
  {Rieutord}, {Reese}, \& {Kervella}}]{Bouchaud2020}
{Bouchaud}, K., {Domiciano de Souza}, A., {Rieutord}, M., {Reese}, D.~R., \&
  {Kervella}, P. 2020, \aap, 633, A78

\bibitem[{{Bouma} {et~al.}(2021){Bouma}, {Curtis}, {Hartman}, {Winn}, \&
  {Bakos}}]{Bouma2021}
{Bouma}, L.~G., {Curtis}, J.~L., {Hartman}, J.~D., {Winn}, J.~N., \& {Bakos},
  G.~{\'A}. 2021, \aj, 162, 197

\bibitem[{{Bowman} \& {Kurtz}(2018)}]{Bowman+Kurtz2018}
{Bowman}, D.~M. \& {Kurtz}, D.~W. 2018, \mnras, 476, 3169

\bibitem[{{Brahm} {et~al.}(2017){Brahm}, {Jord{\'a}n}, \&
  {Espinoza}}]{Brahm2017}
{Brahm}, R., {Jord{\'a}n}, A., \& {Espinoza}, N. 2017, \pasp, 129, 034002

\bibitem[{{Brandt} \& {Huang}(2015)}]{Brandt2015ApJ}
{Brandt}, T.~D. \& {Huang}, C.~X. 2015, \apj, 807, 25

\bibitem[{{Brasseur} {et~al.}(2019){Brasseur}, {Phillip}, {Fleming},
  {Mullally}, \& {White}}]{Brasseur2019}
{Brasseur}, C.~E., {Phillip}, C., {Fleming}, S.~W., {Mullally}, S.~E., \&
  {White}, R.~L. 2019, {Astrocut: Tools for creating cutouts of TESS images},
  Astrophysics Source Code Library, record ascl:1905.007

\bibitem[{{Brogaard} {et~al.}(2023){Brogaard}, {Arentoft}, {Miglio}, {Casali},
  {Thomsen}, {Tailo}, {Montalb{\'a}n}, {Grisoni}, {Willett}, {Stokholm},
  {Grundahl}, {Stello}, \& {Sandquist}}]{Brogaard2023}
{Brogaard}, K., {Arentoft}, T., {Miglio}, A., {et~al.} 2023, \aap, 679, A23

\bibitem[{{Cantat-Gaudin} {et~al.}(2018){Cantat-Gaudin}, {Jordi}, {Vallenari},
  {Bragaglia}, {Balaguer-N{\'u}{\~n}ez}, {Soubiran}, {Bossini}, {Moitinho},
  {Castro-Ginard}, {Krone-Martins}, {Casamiquela}, {Sordo}, \&
  {Carrera}}]{Cantat-Gaudin2018}
{Cantat-Gaudin}, T., {Jordi}, C., {Vallenari}, A., {et~al.} 2018, \aap, 618,
  A93

\bibitem[{{Casagrande} \& {VandenBerg}(2018)}]{Casagrande2018}
{Casagrande}, L. \& {VandenBerg}, D.~A. 2018, \mnras, 479, L102

\bibitem[{{Choi} {et~al.}(2016){Choi}, {Dotter}, {Conroy}, {Cantiello},
  {Paxton}, \& {Johnson}}]{Choi2016ApJ_MIST}
{Choi}, J., {Dotter}, A., {Conroy}, C., {et~al.} 2016, \apj, 823, 102

\bibitem[{{Christophe} {et~al.}(2018){Christophe}, {Ballot}, {Ouazzani},
  {Antoci}, \& {Salmon}}]{Christophe2018}
{Christophe}, S., {Ballot}, J., {Ouazzani}, R.~M., {Antoci}, V., \& {Salmon},
  S.~J.~A.~J. 2018, \aap, 618, A47

\bibitem[{{Corbard} {et~al.}(1999){Corbard}, {Blanc-F{\'e}raud}, {Berthomieu},
  \& {Provost}}]{Corbard1999}
{Corbard}, T., {Blanc-F{\'e}raud}, L., {Berthomieu}, G., \& {Provost}, J. 1999,
  \aap, 344, 696

\bibitem[{{Correnti} {et~al.}(2015){Correnti}, {Goudfrooij}, {Puzia}, \& {de
  Mink}}]{Correnti2015}
{Correnti}, M., {Goudfrooij}, P., {Puzia}, T.~H., \& {de Mink}, S.~E. 2015,
  \mnras, 450, 3054

\bibitem[{{Cox}(1955)}]{Cox1955ApJ}
{Cox}, A.~N. 1955, \apj, 121, 628

\bibitem[{{Danielski} {et~al.}(2018){Danielski}, {Babusiaux}, {Ruiz-Dern},
  {Sartoretti}, \& {Arenou}}]{Danielski2018}
{Danielski}, C., {Babusiaux}, C., {Ruiz-Dern}, L., {Sartoretti}, P., \&
  {Arenou}, F. 2018, \aap, 614, A19

\bibitem[{{D'Antona} {et~al.}(2015){D'Antona}, {Di Criscienzo}, {Decressin},
  {Milone}, {Vesperini}, \& {Ventura}}]{DAntona2015MNRAS}
{D'Antona}, F., {Di Criscienzo}, M., {Decressin}, T., {et~al.} 2015, \mnras,
  453, 2637

\bibitem[{{De Cat} \& {Aerts}(2002)}]{DeCat2002A&A}
{De Cat}, P. \& {Aerts}, C. 2002, \aap, 393, 965

\bibitem[{{Deheuvels} {et~al.}(2015){Deheuvels}, {Ballot}, {Beck}, {Mosser},
  {{\O}stensen}, {Garc{\'\i}a}, \& {Goupil}}]{Deheuvels2015}
{Deheuvels}, S., {Ballot}, J., {Beck}, P.~G., {et~al.} 2015, \aap, 580, A96

\bibitem[{{Deheuvels} {et~al.}(2020){Deheuvels}, {Ballot}, {Eggenberger},
  {Spada}, {Noll}, \& {den Hartogh}}]{Deheuvels2020}
{Deheuvels}, S., {Ballot}, J., {Eggenberger}, P., {et~al.} 2020, \aap, 641,
  A117

\bibitem[{{Deheuvels} {et~al.}(2014){Deheuvels}, {Do{\u{g}}an}, {Goupil},
  {Appourchaux}, {Benomar}, {Bruntt}, {Campante}, {Casagrande}, {Ceillier},
  {Davies}, {De Cat}, {Fu}, {Garc{\'\i}a}, {Lobel}, {Mosser}, {Reese},
  {Regulo}, {Schou}, {Stahn}, {Thygesen}, {Yang}, {Chaplin},
  {Christensen-Dalsgaard}, {Eggenberger}, {Gizon}, {Mathis},
  {Molenda-{\.Z}akowicz}, \& {Pinsonneault}}]{Deheuvels2014}
{Deheuvels}, S., {Do{\u{g}}an}, G., {Goupil}, M.~J., {et~al.} 2014, \aap, 564,
  A27

\bibitem[{{Deubner} {et~al.}(1979){Deubner}, {Ulrich}, \&
  {Rhodes}}]{Deubner1979}
{Deubner}, F.~L., {Ulrich}, R.~K., \& {Rhodes}, E.~J., J. 1979, \aap, 72, 177

\bibitem[{{Di Mauro} {et~al.}(2016){Di Mauro}, {Ventura}, {Cardini}, {Stello},
  {Christensen-Dalsgaard}, {Dziembowski}, {Patern{\`o}}, {Beck}, {Bloemen},
  {Davies}, {De Smedt}, {Elsworth}, {Garc{\'\i}a}, {Hekker}, {Mosser}, \&
  {Tkachenko}}]{DiMauro2016}
{Di Mauro}, M.~P., {Ventura}, R., {Cardini}, D., {et~al.} 2016, \apj, 817, 65

\bibitem[{{Domiciano de Souza} {et~al.}(2003){Domiciano de Souza}, {Kervella},
  {Jankov}, {Abe}, {Vakili}, {di Folco}, \& {Paresce}}]{Domiciano_de_Souza2003}
{Domiciano de Souza}, A., {Kervella}, P., {Jankov}, S., {et~al.} 2003, \aap,
  407, L47

\bibitem[{{Dotter}(2016)}]{Dotter2016ApJS_MIST}
{Dotter}, A. 2016, \apjs, 222, 8

\bibitem[{{Dupret} {et~al.}(2005{\natexlab{a}}){Dupret}, {Grigahc{\`e}ne},
  {Garrido}, {Gabriel}, \& {Scuflaire}}]{Dupret2005A&A}
{Dupret}, M.~A., {Grigahc{\`e}ne}, A., {Garrido}, R., {Gabriel}, M., \&
  {Scuflaire}, R. 2005{\natexlab{a}}, \aap, 435, 927

\bibitem[{{Dupret} {et~al.}(2005{\natexlab{b}}){Dupret}, {Grigahc{\`e}ne},
  {Garrido}, {Gabriel}, \& {Scuflaire}}]{Dupret2005}
{Dupret}, M.~A., {Grigahc{\`e}ne}, A., {Garrido}, R., {Gabriel}, M., \&
  {Scuflaire}, R. 2005{\natexlab{b}}, \aap, 435, 927

\bibitem[{{Eggen}(1964)}]{Eggen1964IAUS}
{Eggen}, O.~J. 1964, in The Galaxy and the Magellanic Clouds, ed. F.~J. {Kerr},
  Vol.~20, 10

\bibitem[{{Endal} \& {Sofia}(1978)}]{Endal1978}
{Endal}, A.~S. \& {Sofia}, S. 1978, \apj, 220, 279

\bibitem[{{Espinosa Lara} \& {Rieutord}(2011)}]{Espinosa_Lara2011}
{Espinosa Lara}, F. \& {Rieutord}, M. 2011, \aap, 533, A43

\bibitem[{{Feinstein} {et~al.}(1973){Feinstein}, {Marraco}, \&
  {Mirabel}}]{Feinstein1973}
{Feinstein}, A., {Marraco}, H.~G., \& {Mirabel}, I. 1973, \aaps, 9, 233

\bibitem[{{Foreman-Mackey}(2016)}]{2016JOSS....1...24F}
{Foreman-Mackey}, D. 2016, The Journal of Open Source Software, 1, 24

\bibitem[{{Foreman-Mackey} {et~al.}(2013){Foreman-Mackey}, {Hogg}, {Lang}, \&
  {Goodman}}]{2013PASP..125..306F}
{Foreman-Mackey}, D., {Hogg}, D.~W., {Lang}, D., \& {Goodman}, J. 2013, \pasp,
  125, 306

\bibitem[{{Fr{\'e}mat} {et~al.}(2023{\natexlab{a}}){Fr{\'e}mat}, {Royer},
  {Marchal}, {Blomme}, {Sartoretti}, {Guerrier}, {Panuzzo}, {Katz}, {Seabroke},
  {Th{\'e}venin}, {Cropper}, {Benson}, {Damerdji}, {Haigron}, {Lobel}, {Smith},
  {Baker}, {Chemin}, {David}, {Dolding}, {Gosset}, {Jan{\ss}en}, {Jasniewicz},
  {Plum}, {Samaras}, {Snaith}, {Soubiran}, {Vanel}, {Zorec}, {Zwitter},
  {Brouillet}, {Caffau}, {Crifo}, {Fabre}, {Fragkoudi}, {Huckle}, {Lasne},
  {Leclerc}, {Mastrobuono-Battisti}, {Jean-Antoine Piccolo}, \&
  {Viala}}]{Fremat2023A&A}
{Fr{\'e}mat}, Y., {Royer}, F., {Marchal}, O., {et~al.} 2023{\natexlab{a}},
  \aap, 674, A8

\bibitem[{{Fr{\'e}mat} {et~al.}(2023{\natexlab{b}}){Fr{\'e}mat}, {Royer},
  {Marchal}, {Blomme}, {Sartoretti}, {Guerrier}, {Panuzzo}, {Katz}, {Seabroke},
  {Th{\'e}venin}, {Cropper}, {Benson}, {Damerdji}, {Haigron}, {Lobel}, {Smith},
  {Baker}, {Chemin}, {David}, {Dolding}, {Gosset}, {Jan{\ss}en}, {Jasniewicz},
  {Plum}, {Samaras}, {Snaith}, {Soubiran}, {Vanel}, {Zorec}, {Zwitter},
  {Brouillet}, {Caffau}, {Crifo}, {Fabre}, {Fragkoudi}, {Huckle}, {Lasne},
  {Leclerc}, {Mastrobuono-Battisti}, {Jean-Antoine Piccolo}, \&
  {Viala}}]{Fremat2023}
{Fr{\'e}mat}, Y., {Royer}, F., {Marchal}, O., {et~al.} 2023{\natexlab{b}},
  \aap, 674, A8

\bibitem[{{Fritzewski} {et~al.}(2020){Fritzewski}, {Barnes}, {James}, \&
  {Strassmeier}}]{Fritzewski2020A&A}
{Fritzewski}, D.~J., {Barnes}, S.~A., {James}, D.~J., \& {Strassmeier}, K.~G.
  2020, \aap, 641, A51

\bibitem[{{Fritzewski} {et~al.}(2024){Fritzewski}, {Van Reeth}, {Aerts}, {Van
  Beeck}, {Gossage}, \& {Li}}]{Fritzewski2024}
{Fritzewski}, D.~J., {Van Reeth}, T., {Aerts}, C., {et~al.} 2024, \aap, 681,
  A13

\bibitem[{{Gaia Collaboration} {et~al.}(2023{\natexlab{a}}){Gaia
  Collaboration}, {De Ridder}, {Ripepi}, \& {Aerts}}]{DeRidder2023A&A}
{Gaia Collaboration}, {De Ridder}, J., {Ripepi}, V., \& {Aerts}, C.
  2023{\natexlab{a}}, \aap, 674, A36

\bibitem[{{Gaia Collaboration} {et~al.}(2023{\natexlab{b}}){Gaia
  Collaboration}, {Vallenari}, {Brown}, \& {Prusti}}]{GaiaColl2023}
{Gaia Collaboration}, {Vallenari}, A., {Brown}, A.~G.~A., \& {Prusti}, T.
  2023{\natexlab{b}}, \aap, 674, A1

\bibitem[{{Garcia} {et~al.}(2022{\natexlab{a}}){Garcia}, {Van Reeth}, {De
  Ridder}, \& {Aerts}}]{Garcia2022_60_gdor}
{Garcia}, S., {Van Reeth}, T., {De Ridder}, J., \& {Aerts}, C.
  2022{\natexlab{a}}, \aap, 668, A137

\bibitem[{{Garcia} {et~al.}(2022{\natexlab{b}}){Garcia}, {Van Reeth}, {De
  Ridder}, {Tkachenko}, {IJspeert}, \& {Aerts}}]{Garcia2022}
{Garcia}, S., {Van Reeth}, T., {De Ridder}, J., {et~al.} 2022{\natexlab{b}},
  \aap, 662, A82

\bibitem[{{Gebruers} {et~al.}(2022){Gebruers}, {Tkachenko}, {Bowman}, {Van
  Reeth}, {Burssens}, {IJspeert}, {Mahy}, {Straumit}, {Xiang}, {Rix}, \&
  {Aerts}}]{Gebruers2022}
{Gebruers}, S., {Tkachenko}, A., {Bowman}, D.~M., {et~al.} 2022, \aap, 665, A36

\bibitem[{{Gehan} {et~al.}(2018){Gehan}, {Mosser}, {Michel}, {Samadi}, \&
  {Kallinger}}]{Gehan2018}
{Gehan}, C., {Mosser}, B., {Michel}, E., {Samadi}, R., \& {Kallinger}, T. 2018,
  \aap, 616, A24

\bibitem[{{Gieseking} \& {Karimie}(1982)}]{Gieseking1982}
{Gieseking}, F. \& {Karimie}, M.~T. 1982, \aaps, 49, 497

\bibitem[{{Ginsburg} {et~al.}(2019){Ginsburg}, {Sip{\H{o}}cz}, {Brasseur},
  {Cowperthwaite}, {Craig}, {Deil}, {Guillochon}, {Guzman}, {Liedtke}, {Lian
  Lim}, {Lockhart}, {Mommert}, {Morris}, {Norman}, {Parikh}, {Persson},
  {Robitaille}, {Segovia}, {Singer}, {Tollerud}, {de Val-Borro}, {Valtchanov},
  {Woillez}, {Astroquery Collaboration}, \& {a subset of astropy
  Collaboration}}]{2019AJ....157...98G}
{Ginsburg}, A., {Sip{\H{o}}cz}, B.~M., {Brasseur}, C.~E., {et~al.} 2019, \aj,
  157, 98

\bibitem[{{Girardi} {et~al.}(2013){Girardi}, {Goudfrooij}, {Kalirai}, {Kerber},
  {Kozhurina-Platais}, {Rubele}, {Bressan}, {Chandar}, {Marigo}, {Platais}, \&
  {Puzia}}]{Girardi2013MNRAS}
{Girardi}, L., {Goudfrooij}, P., {Kalirai}, J.~S., {et~al.} 2013, \mnras, 431,
  3501

\bibitem[{{Glatt} {et~al.}(2009){Glatt}, {Grebel}, {Gallagher}, {Nota},
  {Sabbi}, {Sirianni}, {Clementini}, {Da Costa}, {Tosi}, {Harbeck}, {Koch}, \&
  {Kayser}}]{Glatt2009}
{Glatt}, K., {Grebel}, E.~K., {Gallagher}, John~S., I., {et~al.} 2009, \aj,
  138, 1403

\bibitem[{{Gonz{\'a}lez} \& {Lapasset}(2000)}]{Gonzalez2000AJ}
{Gonz{\'a}lez}, J.~F. \& {Lapasset}, E. 2000, \aj, 119, 2296

\bibitem[{{Goodman} \& {Weare}(2010)}]{Goodman2010}
{Goodman}, J. \& {Weare}, J. 2010, Communications in Applied Mathematics and
  Computational Science, 5, 65

\bibitem[{{Gossage} {et~al.}(2019){Gossage}, {Conroy}, {Dotter},
  {Cabrera-Ziri}, {Dolphin}, {Bastian}, {Dalcanton}, {Goudfrooij}, {Johnson},
  {Williams}, {Rosenfield}, {Kalirai}, \& {Fouesneau}}]{Gossage2019ApJ}
{Gossage}, S., {Conroy}, C., {Dotter}, A., {et~al.} 2019, \apj, 887, 199

\bibitem[{{Goupil} {et~al.}(2005){Goupil}, {Dupret}, {Samadi}, {Boehm},
  {Alecian}, {Suarez}, {Lebreton}, \& {Catala}}]{Goupil2005}
{Goupil}, M.~J., {Dupret}, M.~A., {Samadi}, R., {et~al.} 2005, Journal of
  Astrophysics and Astronomy, 26, 249

\bibitem[{{Granada} {et~al.}(2018){Granada}, {Jones}, {Sigut}, {Semaan},
  {Georgy}, {Meynet}, \& {Ekstr{\"o}m}}]{Granada2018AJ}
{Granada}, A., {Jones}, C.~E., {Sigut}, T.~A.~A., {et~al.} 2018, \aj, 155, 50

\bibitem[{{Grassitelli} {et~al.}(2015){Grassitelli}, {Fossati}, {Langer},
  {Miglio}, {Istrate}, \& {Sanyal}}]{Grassitelli2015}
{Grassitelli}, L., {Fossati}, L., {Langer}, N., {et~al.} 2015, \aap, 584, L2

\bibitem[{{Gratton} {et~al.}(2012){Gratton}, {Carretta}, \&
  {Bragaglia}}]{Gratton2012}
{Gratton}, R.~G., {Carretta}, E., \& {Bragaglia}, A. 2012, \aapr, 20, 50

\bibitem[{{Guzik} {et~al.}(2000){Guzik}, {Kaye}, {Bradley}, {Cox}, \&
  {Neuforge}}]{Guzik2000}
{Guzik}, J.~A., {Kaye}, A.~B., {Bradley}, P.~A., {Cox}, A.~N., \& {Neuforge},
  C. 2000, \apjl, 542, L57

\bibitem[{{Handler}(2009)}]{Handler2009}
{Handler}, G. 2009, in American Institute of Physics Conference Series, Vol.
  1170, Stellar Pulsation: Challenges for Theory and Observation, ed. J.~A.
  {Guzik} \& P.~A. {Bradley}, 403--409

\bibitem[{Harris {et~al.}(2020)Harris, Millman, van~der Walt, Gommers,
  Virtanen, Cournapeau, Wieser, Taylor, Berg, Smith, Kern, Picus, Hoyer, van
  Kerkwijk, Brett, Haldane, del R{\'{i}}o, Wiebe, Peterson,
  G{\'{e}}rard-Marchant, Sheppard, Reddy, Weckesser, Abbasi, Gohlke, \&
  Oliphant}]{harris2020array}
Harris, C.~R., Millman, K.~J., van~der Walt, S.~J., {et~al.} 2020, Nature, 585,
  357

\bibitem[{{Hasanzadeh} {et~al.}(2021){Hasanzadeh}, {Safari}, \&
  {Ghasemi}}]{Hasanzadeh++2021}
{Hasanzadeh}, A., {Safari}, H., \& {Ghasemi}, H. 2021, \mnras, 505, 1476

\bibitem[{{Healy} \& {McCullough}(2020)}]{Healy2020}
{Healy}, B.~F. \& {McCullough}, P.~R. 2020, \apj, 903, 99

\bibitem[{{Hekker} {et~al.}(2011){Hekker}, {Basu}, {Stello}, {Kallinger},
  {Grundahl}, {Mathur}, {Garc{\'\i}a}, {Mosser}, {Huber}, {Bedding},
  {Szab{\'o}}, {De Ridder}, {Chaplin}, {Elsworth}, {Hale},
  {Christensen-Dalsgaard}, {Gilliland}, {Still}, {McCauliff}, \&
  {Quintana}}]{Hekker2011}
{Hekker}, S., {Basu}, S., {Stello}, D., {et~al.} 2011, \aap, 530, A100

\bibitem[{{Henneco} {et~al.}(2021){Henneco}, {Van Reeth}, {Prat}, {Mathis},
  {Mombarg}, \& {Aerts}}]{Henneco2021}
{Henneco}, J., {Van Reeth}, T., {Prat}, V., {et~al.} 2021, \aap, 648, A97

\bibitem[{{Hon} {et~al.}(2021){Hon}, {Huber}, {Kuszlewicz}, {Stello}, {Sharma},
  {Tayar}, {Zinn}, {Vrard}, \& {Pinsonneault}}]{Hon2021}
{Hon}, M., {Huber}, D., {Kuszlewicz}, J.~S., {et~al.} 2021, \apj, 919, 131

\bibitem[{{Houdek}(2000)}]{houdek2000}
{Houdek}, G. 2000, in Astronomical Society of the Pacific Conference Series,
  Vol. 210, Delta Scuti and Related Stars, ed. M.~{Breger} \& M.~{Montgomery},
  454

\bibitem[{{Huat} {et~al.}(2009){Huat}, {Hubert}, {Baudin}, {Floquet}, {Neiner},
  {Fr{\'e}mat}, {Guti{\'e}rrez-Soto}, {Andrade}, {de Batz}, {Diago}, {Emilio},
  {Espinosa Lara}, {Fabregat}, {Janot-Pacheco}, {Leroy}, {Martayan}, {Semaan},
  {Suso}, {Auvergne}, {Catala}, {Michel}, \& {Samadi}}]{Huat2009}
{Huat}, A.~L., {Hubert}, A.~M., {Baudin}, F., {et~al.} 2009, \aap, 506, 95

\bibitem[{{Irwin} {et~al.}(2009){Irwin}, {Aigrain}, {Bouvier}, {Hebb},
  {Hodgkin}, {Irwin}, \& {Moraux}}]{Irwin2009}
{Irwin}, J., {Aigrain}, S., {Bouvier}, J., {et~al.} 2009, \mnras, 392, 1456

\bibitem[{{Jackiewicz}(2021)}]{Jackiewicz_2021}
{Jackiewicz}, J. 2021, Frontiers in Astronomy and Space Sciences, 7, 102

\bibitem[{{Jian} {et~al.}(2023){Jian}, {Matsunaga}, {Jiang}, {Yuan}, \&
  {Zhang}}]{Jian2023}
{Jian}, M., {Matsunaga}, N., {Jiang}, B., {Yuan}, H., \& {Zhang}, R. 2023,
  arXiv e-prints, arXiv:2311.08395

\bibitem[{{Johnston} {et~al.}(2019){Johnston}, {Aerts}, {Pedersen}, \&
  {Bastian}}]{Johnston2019A&A}
{Johnston}, C., {Aerts}, C., {Pedersen}, M.~G., \& {Bastian}, N. 2019, \aap,
  632, A74

\bibitem[{{Kaufer} {et~al.}(1999){Kaufer}, {Stahl}, {Tubbesing},
  {N{\o}rregaard}, {Avila}, {Francois}, {Pasquini}, \& {Pizzella}}]{Kaufer1999}
{Kaufer}, A., {Stahl}, O., {Tubbesing}, S., {et~al.} 1999, The Messenger, 95, 8

\bibitem[{{Kaye} {et~al.}(1999){Kaye}, {Handler}, {Krisciunas}, {Poretti}, \&
  {Zerbi}}]{Kaye1999}
{Kaye}, A.~B., {Handler}, G., {Krisciunas}, K., {Poretti}, E., \& {Zerbi},
  F.~M. 1999, \pasp, 111, 840

\bibitem[{{Kippenhahn} \& {Thomas}(1970)}]{Kippenhahn1970}
{Kippenhahn}, R. \& {Thomas}, H.~C. 1970, in IAU Colloq. 4: Stellar Rotation,
  ed. A.~{Slettebak}, 20

\bibitem[{{Koester} \& {Reimers}(1996)}]{Koester1996}
{Koester}, D. \& {Reimers}, D. 1996, \aap, 313, 810

\bibitem[{{Kraft}(1967)}]{Kraft1967}
{Kraft}, R.~P. 1967, \apj, 150, 551

\bibitem[{{Kurtz} {et~al.}(2023){Kurtz}, {Jayaraman}, {Sowicka}, {Handler},
  {Saio}, {Labadie-Bartz}, \& {Lee}}]{Kurtz2023}
{Kurtz}, D.~W., {Jayaraman}, R., {Sowicka}, P., {et~al.} 2023, \mnras, 521,
  4765

\bibitem[{{Kurtz} {et~al.}(2014){Kurtz}, {Saio}, {Takata}, {Shibahashi},
  {Murphy}, \& {Sekii}}]{Kurtz2014}
{Kurtz}, D.~W., {Saio}, H., {Takata}, M., {et~al.} 2014, \mnras, 444, 102

\bibitem[{{Kurtz} {et~al.}(2015){Kurtz}, {Shibahashi}, {Murphy}, {Bedding}, \&
  {Bowman}}]{Kurtz2015}
{Kurtz}, D.~W., {Shibahashi}, H., {Murphy}, S.~J., {Bedding}, T.~R., \&
  {Bowman}, D.~M. 2015, \mnras, 450, 3015

\bibitem[{{Ledoux}(1951)}]{Ledoux1951}
{Ledoux}, P. 1951, \apj, 114, 373

\bibitem[{{Lee} \& {Saio}(1987)}]{Lee1987}
{Lee}, U. \& {Saio}, H. 1987, \mnras, 224, 513

\bibitem[{{Lee} \& {Saio}(1997)}]{Lee1997}
{Lee}, U. \& {Saio}, H. 1997, \apj, 491, 839

\bibitem[{{Li} {et~al.}(2016){Li}, {de Grijs}, {Deng}, {Geller}, {Xin}, {Hu},
  \& {Faucher-Gigu{\`e}re}}]{LiChengYuan2016Natur}
{Li}, C., {de Grijs}, R., {Deng}, L., {et~al.} 2016, \nat, 529, 502

\bibitem[{{Li} {et~al.}(2019{\natexlab{a}}){Li}, {Sun}, {de Grijs}, {Deng},
  {Wang}, {Cordoni}, \& {Milone}}]{ChLi2019ApJ}
{Li}, C., {Sun}, W., {de Grijs}, R., {et~al.} 2019{\natexlab{a}}, \apj, 876, 65

\bibitem[{{Li} {et~al.}(2019{\natexlab{b}}){Li}, {Bedding}, {Murphy}, {Van
  Reeth}, {Antoci}, \& {Ouazzani}}]{Li2019_splitting_gdor}
{Li}, G., {Bedding}, T.~R., {Murphy}, S.~J., {et~al.} 2019{\natexlab{b}},
  \mnras, 482, 1757

\bibitem[{{Li} {et~al.}(2020{\natexlab{a}}){Li}, {Guo}, {Fuller}, {Bedding},
  {Murphy}, {Colman}, \& {Hey}}]{Li2020_gdor_in_EB}
{Li}, G., {Guo}, Z., {Fuller}, J., {et~al.} 2020{\natexlab{a}}, \mnras, 497,
  4363

\bibitem[{{Li} {et~al.}(2019{\natexlab{c}}){Li}, {Van Reeth}, {Bedding},
  {Murphy}, \& {Antoci}}]{Li2019_r_mode}
{Li}, G., {Van Reeth}, T., {Bedding}, T.~R., {Murphy}, S.~J., \& {Antoci}, V.
  2019{\natexlab{c}}, \mnras, 487, 782

\bibitem[{{Li} {et~al.}(2020{\natexlab{b}}){Li}, {Van Reeth}, {Bedding},
  {Murphy}, {Antoci}, {Ouazzani}, \& {Barbara}}]{Li2020MNRAS_611}
{Li}, G., {Van Reeth}, T., {Bedding}, T.~R., {et~al.} 2020{\natexlab{b}},
  \mnras, 491, 3586

\bibitem[{{Lightkurve Collaboration} {et~al.}(2018{\natexlab{a}}){Lightkurve
  Collaboration}, {Cardoso}, {Hedges}, {Gully-Santiago}, {Saunders}, {Cody},
  {Barclay}, {Hall}, {Sagear}, {Turtelboom}, {Zhang}, {Tzanidakis}, {Mighell},
  {Coughlin}, {Bell}, {Berta-Thompson}, {Williams}, {Dotson}, \&
  {Barentsen}}]{Lightkurve2018}
{Lightkurve Collaboration}, {Cardoso}, J. V. d.~M., {Hedges}, C., {et~al.}
  2018{\natexlab{a}}, {Lightkurve: Kepler and TESS time series analysis in
  Python}, Astrophysics Source Code Library, record ascl:1812.013

\bibitem[{{Lightkurve Collaboration} {et~al.}(2018{\natexlab{b}}){Lightkurve
  Collaboration}, {Cardoso}, {Hedges}, {Gully-Santiago}, {Saunders}, {Cody},
  {Barclay}, {Hall}, {Sagear}, {Turtelboom}, {Zhang}, {Tzanidakis}, {Mighell},
  {Coughlin}, {Bell}, {Berta-Thompson}, {Williams}, {Dotson}, \&
  {Barentsen}}]{2018ascl.soft12013L}
{Lightkurve Collaboration}, {Cardoso}, J. V. d.~M., {Hedges}, C., {et~al.}
  2018{\natexlab{b}}, {Lightkurve: Kepler and TESS time series analysis in
  Python}, Astrophysics Source Code Library, record ascl:1812.013

\bibitem[{{Lim} {et~al.}(2019){Lim}, {Rauw}, {Naz{\'e}}, {Sung}, {Hwang}, \&
  {Park}}]{Lim2019NatAs}
{Lim}, B., {Rauw}, G., {Naz{\'e}}, Y., {et~al.} 2019, Nature Astronomy, 3, 76

\bibitem[{{Lomb}(1976)}]{Lomb1976}
{Lomb}, N.~R. 1976, \apss, 39, 447

\bibitem[{{Lund} {et~al.}(2016){Lund}, {Basu}, {Silva Aguirre}, {Chaplin},
  {Serenelli}, {Garc{\'\i}a}, {Latham}, {Casagrande}, {Bieryla}, {Davies},
  {Viani}, {Buchhave}, {Miglio}, {Soderblom}, {Valenti}, {Stefanik}, \&
  {Handberg}}]{Lund2016}
{Lund}, M.~N., {Basu}, S., {Silva Aguirre}, V., {et~al.} 2016, \mnras, 463,
  2600

\bibitem[{{Maeder}(2009)}]{Maeder2009}
{Maeder}, A. 2009, {Physics, Formation and Evolution of Rotating Stars,
  Springer-Verlag, Heidelberg}

\bibitem[{{Maitzen} \& {Hensberge}(1981)}]{Maitzen1981A&A}
{Maitzen}, H.~M. \& {Hensberge}, H. 1981, \aap, 96, 151

\bibitem[{{Martins} \& {Palacios}(2013)}]{Martins2013}
{Martins}, F. \& {Palacios}, A. 2013, \aap, 560, A16

\bibitem[{{Mathis}(2009)}]{Mathis2009}
{Mathis}, S. 2009, \aap, 506, 811

\bibitem[{{Mathis}(2013)}]{Mathis2013LNP}
{Mathis}, S. 2013, in Lecture Notes in Physics, Berlin Springer Verlag, ed.
  M.~{Goupil}, K.~{Belkacem}, C.~{Neiner}, F.~{Ligni{\`e}res}, \& J.~J.
  {Green}, Vol. 865, 23

\bibitem[{{McAlister} {et~al.}(2005){McAlister}, {ten Brummelaar}, {Gies},
  {Huang}, {Bagnuolo}, {Shure}, {Sturmann}, {Sturmann}, {Turner}, {Taylor},
  {Berger}, {Baines}, {Grundstrom}, {Ogden}, {Ridgway}, \& {van
  Belle}}]{McAlister2005}
{McAlister}, H.~A., {ten Brummelaar}, T.~A., {Gies}, D.~R., {et~al.} 2005,
  \apj, 628, 439

\bibitem[{{McQuillan} {et~al.}(2014){McQuillan}, {Mazeh}, \&
  {Aigrain}}]{McQuillan2014ApJS}
{McQuillan}, A., {Mazeh}, T., \& {Aigrain}, S. 2014, \apjs, 211, 24

\bibitem[{{Meilland} {et~al.}(2007){Meilland}, {Stee}, {Vannier}, {Millour},
  {Domiciano de Souza}, {Malbet}, {Martayan}, {Paresce}, {Petrov}, {Richichi},
  \& {Spang}}]{Meilland2007}
{Meilland}, A., {Stee}, P., {Vannier}, M., {et~al.} 2007, \aap, 464, 59

\bibitem[{{Meingast} \& {Alves}(2019)}]{Meingast2019}
{Meingast}, S. \& {Alves}, J. 2019, \aap, 621, L3

\bibitem[{{Meingast} {et~al.}(2021){Meingast}, {Alves}, \&
  {Rottensteiner}}]{Meingast2021}
{Meingast}, S., {Alves}, J., \& {Rottensteiner}, A. 2021, \aap, 645, A84

\bibitem[{{Meynet} {et~al.}(1993){Meynet}, {Mermilliod}, \&
  {Maeder}}]{Meynet1993}
{Meynet}, G., {Mermilliod}, J.~C., \& {Maeder}, A. 1993, \aaps, 98, 477

\bibitem[{{Michielsen} {et~al.}(2021){Michielsen}, {Aerts}, \&
  {Bowman}}]{Michielsen2021}
{Michielsen}, M., {Aerts}, C., \& {Bowman}, D.~M. 2021, \aap, 650, A175

\bibitem[{{Miglio} {et~al.}(2012){Miglio}, {Brogaard}, {Stello}, {Chaplin},
  {D'Antona}, {Montalb{\'a}n}, {Basu}, {Bressan}, {Grundahl}, {Pinsonneault},
  {Serenelli}, {Elsworth}, {Hekker}, {Kallinger}, {Mosser}, {Ventura},
  {Bonanno}, {Noels}, {Silva Aguirre}, {Szabo}, {Li}, {McCauliff}, {Middour},
  \& {Kjeldsen}}]{Miglio2012}
{Miglio}, A., {Brogaard}, K., {Stello}, D., {et~al.} 2012, \mnras, 419, 2077

\bibitem[{{Moe} \& {Di Stefano}(2017)}]{Moe2017ApJS}
{Moe}, M. \& {Di Stefano}, R. 2017, \apjs, 230, 15

\bibitem[{{Mombarg}(2023)}]{Mombarg2023calibrating_AM}
{Mombarg}, J.~S.~G. 2023, \aap, 677, A63

\bibitem[{{Mombarg} {et~al.}(2022){Mombarg}, {Dotter}, {Rieutord},
  {Michielsen}, {Van Reeth}, \& {Aerts}}]{Mombarg2022}
{Mombarg}, J. S.~G., {Dotter}, A., {Rieutord}, M., {et~al.} 2022, \apj, 925,
  154

\bibitem[{{Mombarg} {et~al.}(2020){Mombarg}, {Dotter}, {Van Reeth},
  {Tkachenko}, {Gebruers}, \& {Aerts}}]{Mombarg2020}
{Mombarg}, J. S.~G., {Dotter}, A., {Van Reeth}, T., {et~al.} 2020, \apj, 895,
  51

\bibitem[{{Mombarg} {et~al.}(2023){Mombarg}, {Rieutord}, \& {Espinosa
  Lara}}]{Mombarg2023}
{Mombarg}, J.~S.~G., {Rieutord}, M., \& {Espinosa Lara}, F. 2023, \aap, 677, L5

\bibitem[{{Mombarg} {et~al.}(2021){Mombarg}, {Van Reeth}, \&
  {Aerts}}]{Mombarg2021}
{Mombarg}, J.~S.~G., {Van Reeth}, T., \& {Aerts}, C. 2021, \aap, 650, A58

\bibitem[{{Mombarg} {et~al.}(2019){Mombarg}, {Van Reeth}, {Pedersen},
  {Molenberghs}, {Bowman}, {Johnston}, {Tkachenko}, \& {Aerts}}]{Mombarg2019}
{Mombarg}, J.~S.~G., {Van Reeth}, T., {Pedersen}, M.~G., {et~al.} 2019, \mnras,
  485, 3248

\bibitem[{{Montgomery} \& {O'Donoghue}(1999)}]{Montgomery1999}
{Montgomery}, M.~H. \& {O'Donoghue}, D. 1999, Delta Scuti Star Newsletter, 13,
  28

\bibitem[{{Mosser} {et~al.}(2012){Mosser}, {Goupil}, {Belkacem}, {Marques},
  {Beck}, {Bloemen}, {De Ridder}, {Barban}, {Deheuvels}, {Elsworth}, {Hekker},
  {Kallinger}, {Ouazzani}, {Pinsonneault}, {Samadi}, {Stello}, {Garc{\'\i}a},
  {Klaus}, {Li}, {Mathur}, \& {Morris}}]{Mosser2012}
{Mosser}, B., {Goupil}, M.~J., {Belkacem}, K., {et~al.} 2012, \aap, 548, A10

\bibitem[{{Murphy} {et~al.}(2022){Murphy}, {Bedding}, {White}, {Li}, {Hey},
  {Reese}, \& {Joyce}}]{Murphy2022Pleiades}
{Murphy}, S.~J., {Bedding}, T.~R., {White}, T.~R., {et~al.} 2022, \mnras, 511,
  5718

\bibitem[{{Murphy} {et~al.}(2016){Murphy}, {Fossati}, {Bedding}, {Saio},
  {Kurtz}, {Grassitelli}, \& {Wang}}]{murphyetal2016a}
{Murphy}, S.~J., {Fossati}, L., {Bedding}, T.~R., {et~al.} 2016, \mnras, 459,
  1201

\bibitem[{{Murphy} {et~al.}(2019){Murphy}, {Hey}, {Van Reeth}, \&
  {Bedding}}]{Murphy2019}
{Murphy}, S.~J., {Hey}, D., {Van Reeth}, T., \& {Bedding}, T.~R. 2019, \mnras,
  485, 2380

\bibitem[{{Murphy} {et~al.}(2021){Murphy}, {Joyce}, {Bedding}, {White}, \&
  {Kama}}]{Murphy2021}
{Murphy}, S.~J., {Joyce}, M., {Bedding}, T.~R., {White}, T.~R., \& {Kama}, M.
  2021, \mnras, 502, 1633

\bibitem[{{Murphy} {et~al.}(2020){Murphy}, {Saio}, {Takada-Hidai}, {Kurtz},
  {Shibahashi}, {Takata}, \& {Hey}}]{murphyetal2020d}
{Murphy}, S.~J., {Saio}, H., {Takada-Hidai}, M., {et~al.} 2020, \mnras, 498,
  4272

\bibitem[{{Neiner} {et~al.}(2020){Neiner}, {Lee}, {Mathis}, {Saio}, {Lovekin},
  \& {Augustson}}]{Neiner2020AA}
{Neiner}, C., {Lee}, U., {Mathis}, S., {et~al.} 2020, \aap, 644, A9

\bibitem[{{Netzel} {et~al.}(2022){Netzel}, {Pietrukowicz}, {Soszy{\'n}ski}, \&
  {Wrona}}]{Netzel2022}
{Netzel}, H., {Pietrukowicz}, P., {Soszy{\'n}ski}, I., \& {Wrona}, M. 2022,
  \mnras, 510, 1748

\bibitem[{{Ouazzani} {et~al.}(2020){Ouazzani}, {Ligni{\`e}res}, {Dupret},
  {Salmon}, {Ballot}, {Christophe}, \& {Takata}}]{Ouazzani2020}
{Ouazzani}, R.~M., {Ligni{\`e}res}, F., {Dupret}, M.~A., {et~al.} 2020, \aap,
  640, A49

\bibitem[{{Ouazzani} {et~al.}(2019){Ouazzani}, {Marques}, {Goupil},
  {Christophe}, {Antoci}, {Salmon}, \& {Ballot}}]{Ouazzani2019A&A}
{Ouazzani}, R.~M., {Marques}, J.~P., {Goupil}, M.~J., {et~al.} 2019, \aap, 626,
  A121

\bibitem[{{Ouazzani} {et~al.}(2017){Ouazzani}, {Salmon}, {Antoci}, {Bedding},
  {Murphy}, \& {Roxburgh}}]{Ouazzani2017}
{Ouazzani}, R.-M., {Salmon}, S.~J.~A.~J., {Antoci}, V., {et~al.} 2017, \mnras,
  465, 2294

\bibitem[{{Palakkatharappil} \& {Creevey}(2023)}]{Palakkatharappil2023}
{Palakkatharappil}, D.~B. \& {Creevey}, O.~L. 2023, \aap, 674, A146

\bibitem[{{Pamos Ortega} {et~al.}(2023){Pamos Ortega}, {Mirouh}, {Garc{\'\i}a
  Hern{\'a}ndez}, {Su{\'a}rez Yanes}, \& {Barcel{\'o}
  Forteza}}]{Pamos-Ortega++2023}
{Pamos Ortega}, D., {Mirouh}, G.~M., {Garc{\'\i}a Hern{\'a}ndez}, A.,
  {Su{\'a}rez Yanes}, J.~C., \& {Barcel{\'o} Forteza}, S. 2023, \aap, 675, A167

\bibitem[{{Pamyatnykh}(1999)}]{Pamyatnykh1999AcA}
{Pamyatnykh}, A.~A. 1999, \actaa, 49, 119

\bibitem[{pandas~development team(2020)}]{reback2020pandas}
pandas~development team, T. 2020, pandas-dev/pandas: Pandas

\bibitem[{{P{\'a}pics} {et~al.}(2014){P{\'a}pics}, {Moravveji}, {Aerts},
  {Tkachenko}, {Triana}, {Bloemen}, \& {Southworth}}]{Papics2014A&A}
{P{\'a}pics}, P.~I., {Moravveji}, E., {Aerts}, C., {et~al.} 2014, \aap, 570, A8

\bibitem[{{Paxton} {et~al.}(2011){Paxton}, {Bildsten}, {Dotter}, {Herwig},
  {Lesaffre}, \& {Timmes}}]{Paxton2011ApJS}
{Paxton}, B., {Bildsten}, L., {Dotter}, A., {et~al.} 2011, \apjs, 192, 3

\bibitem[{{Paxton} {et~al.}(2013){Paxton}, {Cantiello}, {Arras}, {Bildsten},
  {Brown}, {Dotter}, {Mankovich}, {Montgomery}, {Stello}, {Timmes}, \&
  {Townsend}}]{Paxton2013ApJS}
{Paxton}, B., {Cantiello}, M., {Arras}, P., {et~al.} 2013, \apjs, 208, 4

\bibitem[{{Paxton} {et~al.}(2015){Paxton}, {Marchant}, {Schwab}, {Bauer},
  {Bildsten}, {Cantiello}, {Dessart}, {Farmer}, {Hu}, {Langer}, {Townsend},
  {Townsley}, \& {Timmes}}]{Paxton2015ApJS}
{Paxton}, B., {Marchant}, P., {Schwab}, J., {et~al.} 2015, \apjs, 220, 15

\bibitem[{{Paxton} {et~al.}(2018){Paxton}, {Schwab}, {Bauer}, {Bildsten},
  {Blinnikov}, {Duffell}, {Farmer}, {Goldberg}, {Marchant}, {Sorokina},
  {Thoul}, {Townsend}, \& {Timmes}}]{Paxton2018ApJS}
{Paxton}, B., {Schwab}, J., {Bauer}, E.~B., {et~al.} 2018, \apjs, 234, 34

\bibitem[{{Paxton} {et~al.}(2019){Paxton}, {Smolec}, {Schwab}, {Gautschy},
  {Bildsten}, {Cantiello}, {Dotter}, {Farmer}, {Goldberg}, {Jermyn}, {Kanbur},
  {Marchant}, {Thoul}, {Townsend}, {Wolf}, {Zhang}, \&
  {Timmes}}]{Paxton2019ApJS}
{Paxton}, B., {Smolec}, R., {Schwab}, J., {et~al.} 2019, \apjs, 243, 10

\bibitem[{{Pedersen} {et~al.}(2021){Pedersen}, {Aerts}, {P{\'a}pics},
  {Michielsen}, {Gebruers}, {Rogers}, {Molenberghs}, {Burssens}, {Garcia}, \&
  {Bowman}}]{Pedersen2021}
{Pedersen}, M.~G., {Aerts}, C., {P{\'a}pics}, P.~I., {et~al.} 2021, Nature
  Astronomy, 5, 715

\bibitem[{{Pedersen} {et~al.}(2018){Pedersen}, {Aerts}, {P{\'a}pics}, \&
  {Rogers}}]{Pedersen2018}
{Pedersen}, M.~G., {Aerts}, C., {P{\'a}pics}, P.~I., \& {Rogers}, T.~M. 2018,
  \aap, 614, A128

\bibitem[{{P{\'e}rez Hern{\'a}ndez} {et~al.}(1999){P{\'e}rez Hern{\'a}ndez},
  {Claret}, {Hern{\'a}ndez}, \& {Michel}}]{Perez_Hernandez_1999}
{P{\'e}rez Hern{\'a}ndez}, F., {Claret}, A., {Hern{\'a}ndez}, M.~M., \&
  {Michel}, E. 1999, \aap, 346, 586

\bibitem[{{Reed} {et~al.}(2000){Reed}, {Kilkenny}, {Kawaler}, {Mukadam},
  {Kleinman}, {Nitta-Kleinman}, {Provencal}, {Watson}, {Sullivan}, {Sullivan},
  {Shobbrook}, {Jiang}, {Ashoka}, {Seetha}, {Leibowitz}, {Ibbetson},
  {Mendelson}, {Meistas}, {Kalytis}, {Alisauskas}, {O'Donoghue}, {Martinez},
  {van Wyk}, {Stobie}, {Marang}, {Zola}, {Krzesinski}, {Ogloza}, {Moskalik},
  {Silvotti}, {Piccioni}, {Vauclair}, {Dolez}, {Rene-Fremy}, {Chevreton},
  {Ulla}, {Dreizler}, {Schuh}, {Deetjen}, {Solheim}, {Perez}, {Suarez},
  {Manteiga}, {Burleigh}, {Barstow}, {Kepler}, {Kanaan}, {Giovannini},
  {Metcalfe}, \& {Ostensen}}]{Reed2000BaltA}
{Reed}, M.~D., {Kilkenny}, D., {Kawaler}, S.~D., {et~al.} 2000, Baltic
  Astronomy, 9, 183

\bibitem[{{Reimers} \& {Koester}(1982)}]{Reimers1982}
{Reimers}, D. \& {Koester}, D. 1982, \aap, 116, 341

\bibitem[{{Ricker} {et~al.}(2015){Ricker}, {Winn}, {Vanderspek}, {Latham},
  {Bakos}, {Bean}, {Berta-Thompson}, {Brown}, {Buchhave}, {Butler}, {Butler},
  {Chaplin}, {Charbonneau}, {Christensen-Dalsgaard}, {Clampin}, {Deming},
  {Doty}, {De Lee}, {Dressing}, {Dunham}, {Endl}, {Fressin}, {Ge}, {Henning},
  {Holman}, {Howard}, {Ida}, {Jenkins}, {Jernigan}, {Johnson}, {Kaltenegger},
  {Kawai}, {Kjeldsen}, {Laughlin}, {Levine}, {Lin}, {Lissauer}, {MacQueen},
  {Marcy}, {McCullough}, {Morton}, {Narita}, {Paegert}, {Palle}, {Pepe},
  {Pepper}, {Quirrenbach}, {Rinehart}, {Sasselov}, {Sato}, {Seager},
  {Sozzetti}, {Stassun}, {Sullivan}, {Szentgyorgyi}, {Torres}, {Udry}, \&
  {Villasenor}}]{Ricker2015}
{Ricker}, G.~R., {Winn}, J.~N., {Vanderspek}, R., {et~al.} 2015, Journal of
  Astronomical Telescopes, Instruments, and Systems, 1, 014003

\bibitem[{{Riello} {et~al.}(2021){Riello}, {De Angeli}, {Evans}, {Montegriffo},
  {Carrasco}, {Busso}, {Palaversa}, {Burgess}, {Diener}, {Davidson}, {Rowell},
  {Fabricius}, {Jordi}, {Bellazzini}, {Pancino}, {Harrison}, {Cacciari}, {van
  Leeuwen}, {Hambly}, {Hodgkin}, {Osborne}, {Altavilla}, {Barstow}, {Brown},
  {Castellani}, {Cowell}, {De Luise}, {Gilmore}, {Giuffrida}, {Hidalgo},
  {Holland}, {Marinoni}, {Pagani}, {Piersimoni}, {Pulone}, {Ragaini}, {Rainer},
  {Richards}, {Sanna}, {Walton}, {Weiler}, \& {Yoldas}}]{Riello2021}
{Riello}, M., {De Angeli}, F., {Evans}, D.~W., {et~al.} 2021, \aap, 649, A3

\bibitem[{{Ripepi} {et~al.}(2015){Ripepi}, {Balona}, {Catanzaro}, {Marconi},
  {Palla}, \& {Giarrusso}}]{Ripepi2015}
{Ripepi}, V., {Balona}, L., {Catanzaro}, G., {et~al.} 2015, \mnras, 454, 2606

\bibitem[{{Royer} {et~al.}(2007){Royer}, {Zorec}, \& {G{\'o}mez}}]{Royer2007}
{Royer}, F., {Zorec}, J., \& {G{\'o}mez}, A.~E. 2007, \aap, 463, 671

\bibitem[{{Rui} {et~al.}(2024){Rui}, {Ong}, \& {Mathis}}]{Rui2024}
{Rui}, N.~Z., {Ong}, J.~M.~J., \& {Mathis}, S. 2024, \mnras, 527, 6346

\bibitem[{{Saio} {et~al.}(2018{\natexlab{a}}){Saio}, {Bedding}, {Kurtz},
  {Murphy}, {Antoci}, {Shibahashi}, {Li}, \& {Takata}}]{Saio2018}
{Saio}, H., {Bedding}, T.~R., {Kurtz}, D.~W., {et~al.} 2018{\natexlab{a}},
  \mnras, 477, 2183

\bibitem[{{Saio} {et~al.}(2018{\natexlab{b}}){Saio}, {Kurtz}, {Murphy},
  {Antoci}, \& {Lee}}]{Saio2018_r_modes}
{Saio}, H., {Kurtz}, D.~W., {Murphy}, S.~J., {Antoci}, V.~L., \& {Lee}, U.
  2018{\natexlab{b}}, \mnras, 474, 2774

\bibitem[{{Saio} {et~al.}(2015){Saio}, {Kurtz}, {Takata}, {Shibahashi},
  {Murphy}, {Sekii}, \& {Bedding}}]{Saio2015}
{Saio}, H., {Kurtz}, D.~W., {Takata}, M., {et~al.} 2015, \mnras, 447, 3264

\bibitem[{{Saio} {et~al.}(2021){Saio}, {Takata}, {Lee}, {Li}, \& {Van
  Reeth}}]{Saio2021}
{Saio}, H., {Takata}, M., {Lee}, U., {Li}, G., \& {Van Reeth}, T. 2021, \mnras,
  502, 5856

\bibitem[{{Sandquist} {et~al.}(2020){Sandquist}, {Stello}, {Arentoft},
  {Brogaard}, {Grundahl}, {Vanderburg}, {Hedlund}, {DeWitt}, {Ackerman},
  {Aguilar}, {Buckner}, {Juarez}, {Ortiz}, {Richarte}, {Rivera}, \&
  {Schlapfer}}]{Sandquist2020}
{Sandquist}, E.~L., {Stello}, D., {Arentoft}, T., {et~al.} 2020, \aj, 159, 96

\bibitem[{{Scargle}(1982)}]{Scargle1982}
{Scargle}, J.~D. 1982, \apj, 263, 835

\bibitem[{{Shajn} \& {Struve}(1929)}]{Shajn1929MNRAS}
{Shajn}, G. \& {Struve}, O. 1929, \mnras, 89, 222

\bibitem[{{Shibahashi}(1979)}]{Shibahashi1979}
{Shibahashi}, H. 1979, \pasj, 31, 87

\bibitem[{{Skumanich}(1972)}]{Skumanich1972ApJ}
{Skumanich}, A. 1972, \apj, 171, 565

\bibitem[{{Sreenivas} {et~al.}(2024){Sreenivas}, {Bedding}, {Li}, {Huber},
  {Stello}, {Crawford}, \& {Yu}}]{Sreenivas2024}
{Sreenivas}, K.~R., {Bedding}, T.~R., {Li}, Y., {et~al.} 2024, arXiv e-prints,
  arXiv:2401.17557

\bibitem[{{Stassun} {et~al.}(2019){Stassun}, {Oelkers}, {Paegert}, {Torres},
  {Pepper}, {De Lee}, {Collins}, {Latham}, {Muirhead}, {Chittidi},
  {Rojas-Ayala}, {Fleming}, {Rose}, {Tenenbaum}, {Ting}, {Kane}, {Barclay},
  {Bean}, {Brassuer}, {Charbonneau}, {Ge}, {Lissauer}, {Mann}, {McLean},
  {Mullally}, {Narita}, {Plavchan}, {Ricker}, {Sasselov}, {Seager}, {Sharma},
  {Shiao}, {Sozzetti}, {Stello}, {Vanderspek}, {Wallace}, \&
  {Winn}}]{Stassun++2019}
{Stassun}, K.~G., {Oelkers}, R.~J., {Paegert}, M., {et~al.} 2019, \aj, 158, 138

\bibitem[{{Stellingwerf}(1979)}]{stellingwerf1979}
{Stellingwerf}, R.~F. 1979, \apj, 227, 935

\bibitem[{{Stello} {et~al.}(2010){Stello}, {Basu}, {Bruntt}, {Mosser},
  {Stevens}, {Brown}, {Christensen-Dalsgaard}, {Gilliland}, {Kjeldsen},
  {Arentoft}, {Ballot}, {Barban}, {Bedding}, {Chaplin}, {Elsworth},
  {Garc{\'\i}a}, {Goupil}, {Hekker}, {Huber}, {Mathur}, {Meibom},
  {Sangaralingam}, {Baldner}, {Belkacem}, {Biazzo}, {Brogaard}, {Su{\'a}rez},
  {D'Antona}, {Demarque}, {Esch}, {Gai}, {Grundahl}, {Lebreton}, {Jiang},
  {Jevtic}, {Karoff}, {Miglio}, {Molenda-{\.Z}akowicz}, {Montalb{\'a}n},
  {Noels}, {Roca Cort{\'e}s}, {Roxburgh}, {Serenelli}, {Silva Aguirre},
  {Sterken}, {Stine}, {Szab{\'o}}, {Weiss}, {Borucki}, {Koch}, \&
  {Jenkins}}]{Stello2010}
{Stello}, D., {Basu}, S., {Bruntt}, H., {et~al.} 2010, \apjl, 713, L182

\bibitem[{{Sterken}(2005)}]{Sterken2005}
{Sterken}, C. 2005, in Astronomical Society of the Pacific Conference Series,
  Vol. 335, The Light-Time Effect in Astrophysics: Causes and cures of the O-C
  diagram, ed. C.~{Sterken}, 3

\bibitem[{{Sterken} \& {Jerzykiewicz}(1993)}]{Sterken1993}
{Sterken}, C. \& {Jerzykiewicz}, M. 1993, \ssr, 62, 95

\bibitem[{{Straumit} {et~al.}(2022){Straumit}, {Tkachenko}, {Gebruers},
  {Audenaert}, {Xiang}, {Zari}, {Aerts}, {Johnson}, {Kollmeier}, {Rix},
  {Beaton}, {Van Saders}, {Teske}, {Roman-Lopes}, {Ting}, \&
  {Rom{\'a}n-Z{\'u}{\~n}iga}}]{Straumit2022}
{Straumit}, I., {Tkachenko}, A., {Gebruers}, S., {et~al.} 2022, \aj, 163, 236

\bibitem[{{Sung} {et~al.}(2002){Sung}, {Bessell}, {Lee}, \& {Lee}}]{Sung2002AJ}
{Sung}, H., {Bessell}, M.~S., {Lee}, B.-W., \& {Lee}, S.-G. 2002, \aj, 123, 290

\bibitem[{{Szewczuk} \& {Daszy{\'n}ska-Daszkiewicz}(2018)}]{Szewczuk2018}
{Szewczuk}, W. \& {Daszy{\'n}ska-Daszkiewicz}, J. 2018, \mnras, 478, 2243

\bibitem[{{Tarricq} {et~al.}(2021){Tarricq}, {Soubiran}, {Casamiquela},
  {Cantat-Gaudin}, {Chemin}, {Anders}, {Antoja}, {Romero-G{\'o}mez},
  {Figueras}, {Jordi}, {Bragaglia}, {Balaguer-N{\'u}{\~n}ez}, {Carrera},
  {Castro-Ginard}, {Moitinho}, {Ramos}, \& {Bossini}}]{Tarricq2021}
{Tarricq}, Y., {Soubiran}, C., {Casamiquela}, L., {et~al.} 2021, \aap, 647, A19

\bibitem[{{Terndrup} {et~al.}(2002){Terndrup}, {Pinsonneault}, {Jeffries},
  {Ford}, {Stauffer}, \& {Sills}}]{Terndrup2002}
{Terndrup}, D.~M., {Pinsonneault}, M., {Jeffries}, R.~D., {et~al.} 2002, \apj,
  576, 950

\bibitem[{{Thompson} {et~al.}(1987){Thompson}, {Brown}, \&
  {Landstreet}}]{Thompson1987ApJS}
{Thompson}, I.~B., {Brown}, D.~N., \& {Landstreet}, J.~D. 1987, \apjs, 64, 219

\bibitem[{{Tokuno} \& {Takata}(2022)}]{Tokuno2022}
{Tokuno}, T. \& {Takata}, M. 2022, \mnras, 514, 4140

\bibitem[{{Townsend}(2003)}]{Townsend2003}
{Townsend}, R.~H.~D. 2003, \mnras, 340, 1020

\bibitem[{{Townsend} {et~al.}(2018){Townsend}, {Goldstein}, \&
  {Zweibel}}]{Townsend2018GYRE}
{Townsend}, R.~H.~D., {Goldstein}, J., \& {Zweibel}, E.~G. 2018, \mnras, 475,
  879

\bibitem[{{Townsend} \& {Teitler}(2013)}]{Townsend2013GYRE}
{Townsend}, R.~H.~D. \& {Teitler}, S.~A. 2013, \mnras, 435, 3406

\bibitem[{{Triana} {et~al.}(2017){Triana}, {Corsaro}, {De Ridder}, {Bonanno},
  {P{\'e}rez Hern{\'a}ndez}, \& {Garc{\'\i}a}}]{Triana2017}
{Triana}, S.~A., {Corsaro}, E., {De Ridder}, J., {et~al.} 2017, \aap, 602, A62

\bibitem[{{Triana} {et~al.}(2015){Triana}, {Moravveji}, {P{\'a}pics}, {Aerts},
  {Kawaler}, \& {Christensen-Dalsgaard}}]{Triana2015}
{Triana}, S.~A., {Moravveji}, E., {P{\'a}pics}, P.~I., {et~al.} 2015, \apj,
  810, 16

\bibitem[{{Unno} {et~al.}(1989){Unno}, {Osaki}, {Ando}, {Saio}, \&
  {Shibahashi}}]{Unno1989book}
{Unno}, W., {Osaki}, Y., {Ando}, H., {Saio}, H., \& {Shibahashi}, H. 1989,
  {Nonradial oscillations of stars}

\bibitem[{{Van Beeck} {et~al.}(2021){Van Beeck}, {Bowman}, {Pedersen}, {Van
  Reeth}, {Van Hoolst}, \& {Aerts}}]{VanBeeck2021}
{Van Beeck}, J., {Bowman}, D.~M., {Pedersen}, M.~G., {et~al.} 2021, \aap, 655,
  A59

\bibitem[{{Van Reeth} {et~al.}(2016){Van Reeth}, {Tkachenko}, \&
  {Aerts}}]{VanReeth2016_TAR}
{Van Reeth}, T., {Tkachenko}, A., \& {Aerts}, C. 2016, \aap, 593, A120

\bibitem[{{Van Reeth} {et~al.}(2015{\natexlab{a}}){Van Reeth}, {Tkachenko},
  {Aerts}, {P{\'a}pics}, {Degroote}, {Debosscher}, {Zwintz}, {Bloemen}, {De
  Smedt}, {Hrudkova}, {Raskin}, \& {Van Winckel}}]{VanReeth2015}
{Van Reeth}, T., {Tkachenko}, A., {Aerts}, C., {et~al.} 2015{\natexlab{a}},
  \aap, 574, A17

\bibitem[{{Van Reeth} {et~al.}(2015{\natexlab{b}}){Van Reeth}, {Tkachenko},
  {Aerts}, {P{\'a}pics}, {Degroote}, {Debosscher}, {Zwintz}, {Bloemen}, {De
  Smedt}, {Hrudkova}, {Raskin}, \& {Van
  Winckel}}]{Van_Reeth2015_gdor_detection_method}
{Van Reeth}, T., {Tkachenko}, A., {Aerts}, C., {et~al.} 2015{\natexlab{b}},
  \aap, 574, A17

\bibitem[{{Van Reeth} {et~al.}(2015{\natexlab{c}}){Van Reeth}, {Tkachenko},
  {Aerts}, {P{\'a}pics}, {Triana}, {Zwintz}, {Degroote}, {Debosscher},
  {Bloemen}, {Schmid}, {De Smedt}, {Fremat}, {Fuentes}, {Homan}, {Hrudkova},
  {Karjalainen}, {Lombaert}, {Nemeth}, {{\O}stensen}, {Van De Steene}, {Vos},
  {Raskin}, \& {Van Winckel}}]{VanReeth2015ApJS}
{Van Reeth}, T., {Tkachenko}, A., {Aerts}, C., {et~al.} 2015{\natexlab{c}},
  \apjs, 218, 27

\bibitem[{{von Zeipel}(1924)}]{von_Zeipel1924}
{von Zeipel}, H. 1924, \mnras, 84, 665

\bibitem[{{Waelkens}(1991)}]{Waelkens1991}
{Waelkens}, C. 1991, \aap, 246, 453

\bibitem[{{Wang} {et~al.}(2022){Wang}, {Langer}, {Schootemeijer}, {Milone},
  {Hastings}, {Xu}, {Bodensteiner}, {Sana}, {Castro}, {Lennon}, {Marchant}, {de
  Koter}, \& {de Mink}}]{WangChen2022}
{Wang}, C., {Langer}, N., {Schootemeijer}, A., {et~al.} 2022, Nature Astronomy,
  6, 480

\bibitem[{{Winget} {et~al.}(1991){Winget}, {Nather}, {Clemens}, {Provencal},
  {Kleinman}, {Bradley}, {Wood}, {Claver}, {Frueh}, {Grauer}, {Hine}, {Hansen},
  {Fontaine}, {Achilleos}, {Wickramasinghe}, {Marar}, {Seetha}, {Ashoka},
  {O'Donoghue}, {Warner}, {Kurtz}, {Buckley}, {Brickhill}, {Vauclair}, {Dolez},
  {Chevreton}, {Barstow}, {Solheim}, {Kanaan}, {Kepler}, {Henry}, \&
  {Kawaler}}]{Winget1991ApJ}
{Winget}, D.~E., {Nather}, R.~E., {Clemens}, J.~C., {et~al.} 1991, \apj, 378,
  326

\bibitem[{{Wright} {et~al.}(2010){Wright}, {Eisenhardt}, {Mainzer}, {Ressler},
  {Cutri}, {Jarrett}, {Kirkpatrick}, {Padgett}, {McMillan}, {Skrutskie},
  {Stanford}, {Cohen}, {Walker}, {Mather}, {Leisawitz}, {Gautier}, {McLean},
  {Benford}, {Lonsdale}, {Blain}, {Mendez}, {Irace}, {Duval}, {Liu}, {Royer},
  {Heinrichsen}, {Howard}, {Shannon}, {Kendall}, {Walsh}, {Larsen}, {Cardon},
  {Schick}, {Schwalm}, {Abid}, {Fabinsky}, {Naes}, \& {Tsai}}]{Wright2010}
{Wright}, E.~L., {Eisenhardt}, P. R.~M., {Mainzer}, A.~K., {et~al.} 2010, \aj,
  140, 1868

\bibitem[{{Xiong} {et~al.}(2016){Xiong}, {Deng}, {Zhang}, \&
  {Wang}}]{Xiong2016}
{Xiong}, D.~R., {Deng}, L., {Zhang}, C., \& {Wang}, K. 2016, \mnras, 457, 3163

\bibitem[{{Yu} {et~al.}(2018){Yu}, {Huber}, {Bedding}, {Stello}, {Hon},
  {Murphy}, \& {Khanna}}]{Yu2018}
{Yu}, J., {Huber}, D., {Bedding}, T.~R., {et~al.} 2018, \apjs, 236, 42

\bibitem[{{Zahn} {et~al.}(2010){Zahn}, {Ranc}, \& {Morel}}]{Zahn2010}
{Zahn}, J.~P., {Ranc}, C., \& {Morel}, P. 2010, \aap, 517, A7

\bibitem[{{Zerbi} {et~al.}(1998){Zerbi}, {Mantegazza}, {Campana}, \&
  {Antonello}}]{Zerbi1998PASP}
{Zerbi}, F.~M., {Mantegazza}, L., {Campana}, S., \& {Antonello}, E. 1998,
  \pasp, 110, 804

\end{thebibliography}

\begin{appendix}

\section{Surface modulations}

\begin{figure}
    \centering
    \includegraphics[width=\linewidth]{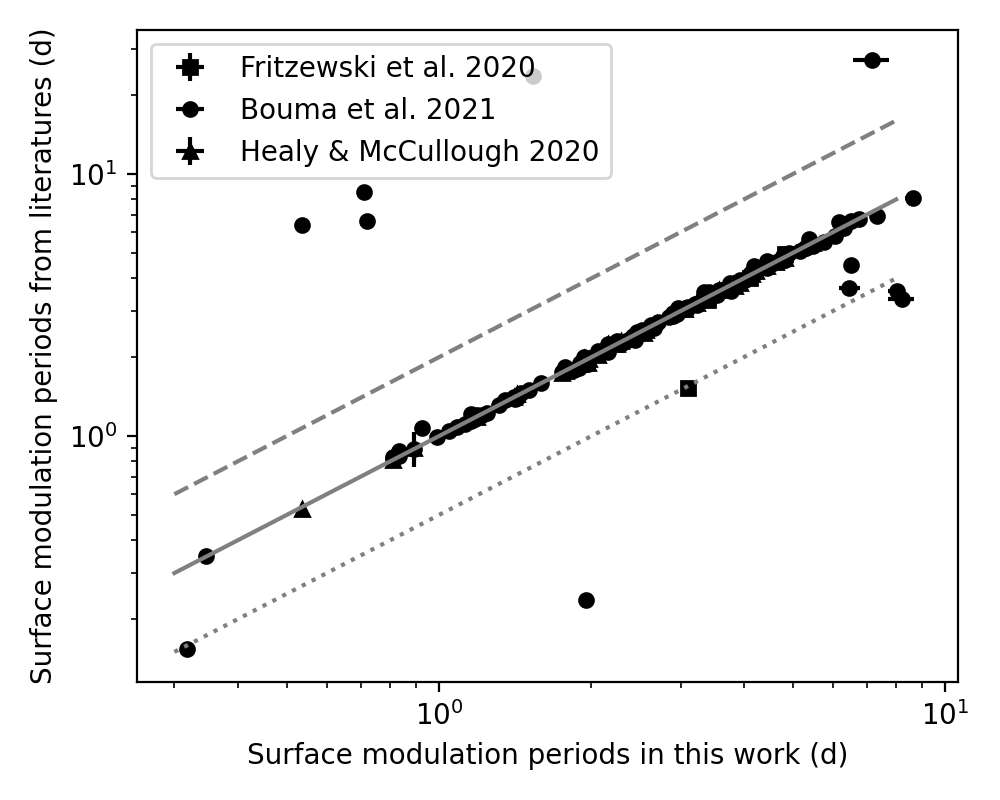}
    \caption{A comparison of the surface modulation periods in this work and from the previous literatures. We collected the results by \cite{Fritzewski2020A&A}, \cite{Bouma2021}, and \cite{Healy2020}. \textcolor{black}{The grey solid, dashed, and dotted lines show the 1:1, 2:1, and 1:2 relations.} }
    \label{appendix_fig:comparison_SM}
\end{figure}

\begin{figure*}
        \centering
        \includegraphics[width=0.9\linewidth]{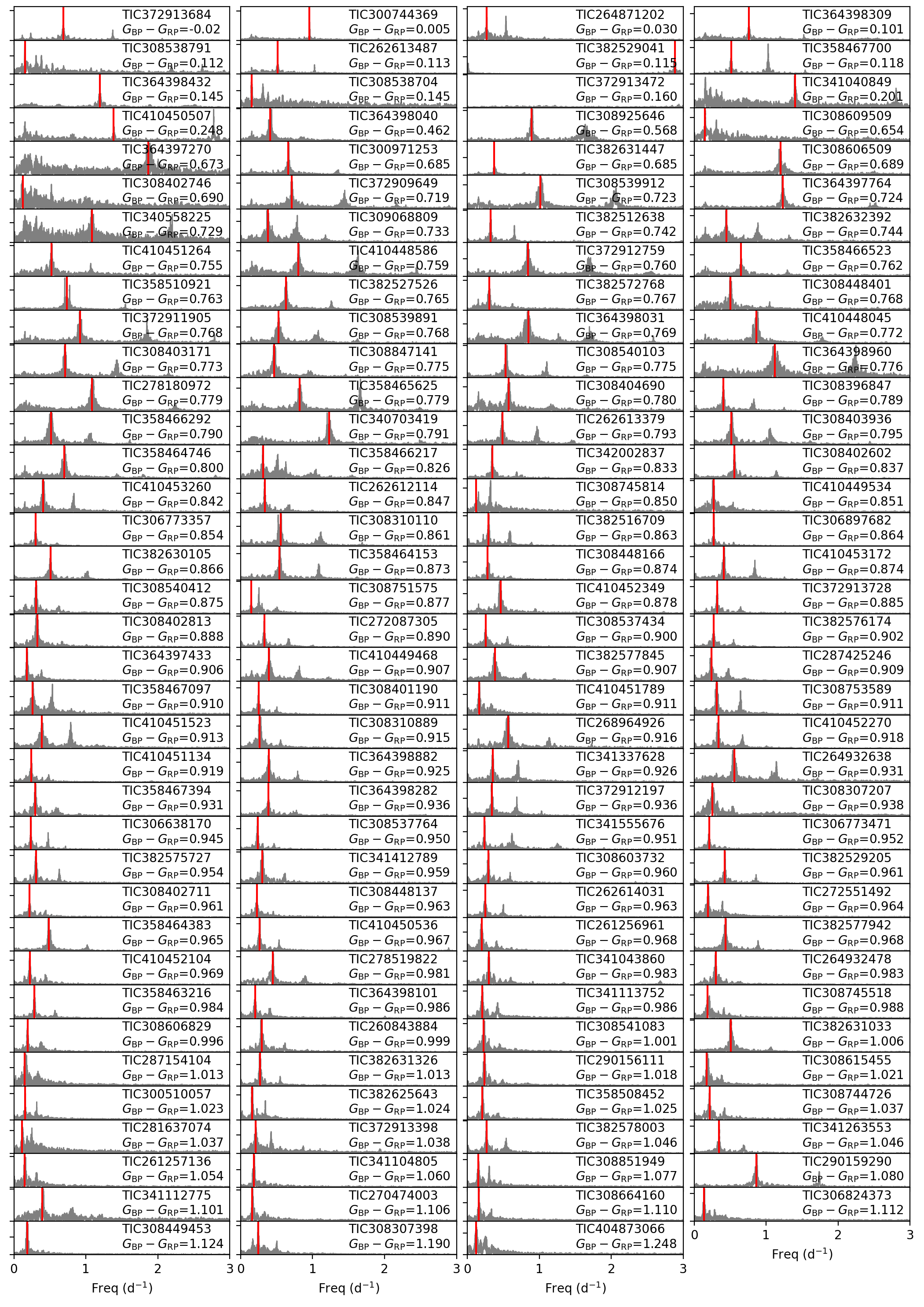}
        \caption{Surface modulation signals in the stars of NGC\,2516, sorted by \textit{Gaia} colour index. Stars with smaller colour indices are displayed at the top. Frequency (in $\mathrm{d^{-1}}$) is plotted along the x-axis. The y-axis, representing amplitude, is omitted for brevity. Each panel displays the amplitude spectrum in grey, with the rotation frequency indicated by a vertical red line. The TIC number and corresponding \textit{Gaia} colour index are noted in each panel. }
        \label{appendix_fig:all_surface_modulation}
    \end{figure*}

Figure~\ref{appendix_fig:comparison_SM} compares the surface modulation periods by this work and by the previous studies. We collected the results by \cite{Fritzewski2020A&A}, \cite{Bouma2021}, and \cite{Healy2020}. Our results show high consistency. Table~\ref{appendix_tab:surface_modulation} lists the periods of all the surface modulation stars. Figure~\ref{appendix_fig:all_surface_modulation} depicts all the surface modulation amplitude spectra.

\begin{sidewaystable*}[]
    \centering
    \tiny
    \caption{Surface modulation periods for the cluster member stars in this study. }\label{appendix_tab:surface_modulation}
    \begin{tabular}{llll|llll|llll|llll}
    \hline
    TIC & CI & Freq ($\mathrm{d^{-1}}$) & $P$ (d) &  TIC & CI & Freq ($\mathrm{d^{-1}}$)  & $P$ (d) &   TIC & CI & Freq ($\mathrm{d^{-1}}$)  & $P$ (d) &     TIC & CI & Freq ($\mathrm{d^{-1}}$)  & $P$ (d) \\
    \hline
372913684 & -0.022 & 0.6878(13) & 1.4538(28)  & 300744369 & 0.005 & 0.9567(5) & 1.0453(6)  & 264871202 & 0.030 & 0.2668(26) & 3.75(4)  & 364398309 & 0.101 & 0.7612(17) & 1.3137(29) \\
 308538791 & 0.113 & 0.155(7) & 6.5(3)  & 262613487 & 0.113 & 0.5131(6) & 1.9488(24)  & 382529041 & 0.115 & 2.8807(12) & 0.34714(14)  & 358467700 & 0.118 & 0.5154(15) & 1.940(6) \\
 364398432 & 0.145 & 1.198(3) & 0.8350(22)  & 308538704 & 0.145 & 0.1535(29) & 6.52(12)  & 372913472 & 0.161 & 3.1470(13) & 0.31777(13)  & 341040849 & 0.202 & 1.4039(23) & 0.7123(12) \\
 410450507 & 0.249 & 1.3869(22) & 0.7211(11)  & 364398040 & 0.463 & 0.410(4) & 2.440(23)  & 308925646 & 0.569 & 0.888(5) & 1.126(6)  & 308609509 & 0.654 & 0.140(11) & 7.2(6) \\
 364397270 & 0.674 & 1.869(7) & 0.5352(21)  & 300971253 & 0.686 & 0.6631(13) & 1.508(3)  & 382631447 & 0.686 & 0.3693(27) & 2.708(20)  & 308606509 & 0.690 & 1.199(3) & 0.8343(22) \\
 308402746 & 0.690 & 0.125(5) & 8.0(3)  & 372909649 & 0.719 & 0.7079(21) & 1.413(4)  & 308539912 & 0.724 & 1.010(5) & 0.990(5)  & 364397764 & 0.724 & 1.2315(22) & 0.8120(15) \\
 340558225 & 0.729 & 1.083(7) & 0.923(6)  & 309068809 & 0.733 & 0.377(6) & 2.65(4)  & 382512638 & 0.742 & 0.3231(12) & 3.095(11)  & 382632392 & 0.745 & 0.445(4) & 2.245(18) \\
 410451264 & 0.755 & 0.522(4) & 1.915(13)  & 410448586 & 0.759 & 0.803(3) & 1.246(5)  & 372912759 & 0.760 & 0.8429(22) & 1.186(3)  & 358466523 & 0.762 & 0.6512(18) & 1.536(4) \\
 358510921 & 0.763 & 0.7395(27) & 1.352(5)  & 382527526 & 0.766 & 0.6284(18) & 1.591(5)  & 382572768 & 0.768 & 0.3015(16) & 3.317(18)  & 308448401 & 0.768 & 0.5050(26) & 1.98(10) \\
 372911905 & 0.769 & 0.920(3) & 1.087(4)  & 308539891 & 0.769 & 0.526(6) & 1.903(23)  & 364398031 & 0.770 & 0.849(4) & 1.178(5)  & 410448045 & 0.772 & 0.8645(26) & 1.157(3) \\
 308403171 & 0.773 & 0.7111(24) & 1.406(5)  & 308847141 & 0.775 & 0.465(4) & 2.152(17)  & 308540103 & 0.776 & 0.528(3) & 1.892(11)  & 364398960 & 0.777 & 1.120(6) & 0.893(5) \\
 278180972 & 0.779 & 1.083(3) & 0.9231(29)  & 358465625 & 0.780 & 0.822(4) & 1.216(5)  & 308404690 & 0.781 & 0.571(3) & 1.752(9)  & 308396847 & 0.790 & 0.4061(15) & 2.463(9) \\
 358466292 & 0.790 & 0.518(4) & 1.931(15)  & 340703419 & 0.792 & 1.231(3) & 0.8127(20)  & 262613379 & 0.794 & 0.4849(16) & 2.062(7)  & 308403936 & 0.795 & 0.514(4) & 1.947(16) \\
 358464746 & 0.801 & 0.7024(23) & 1.424(5)  & 358466217 & 0.826 & 0.311(4) & 3.21(4)  & 342002837 & 0.834 & 0.3457(25) & 2.893(21)  & 308402602 & 0.838 & 0.5607(25) & 1.783(8) \\
 410453260 & 0.843 & 0.407(4) & 2.459(21)  & 262612114 & 0.847 & 0.3341(25) & 2.993(23)  & 308745814 & 0.851 & 0.122(7) & 8.2(5)  & 410449534 & 0.851 & 0.2689(25) & 3.72(3) \\
 306773357 & 0.854 & 0.3024(9) & 3.31(10)  & 308310110 & 0.861 & 0.5572(22) & 1.795(7)  & 382516709 & 0.864 & 0.2937(15) & 3.405(17)  & 306897682 & 0.864 & 0.2721(11) & 3.675(14) \\
 382630105 & 0.866 & 0.5118(20) & 1.954(8)  & 358464153 & 0.873 & 0.5403(22) & 1.851(8)  & 308448166 & 0.874 & 0.2800(17) & 3.572(21)  & 410453172 & 0.875 & 0.4145(19) & 2.413(11) \\
 308540412 & 0.875 & 0.3088(25) & 3.238(26)  & 308751575 & 0.878 & 0.1480(22) & 6.8(10)  & 410452349 & 0.878 & 0.464(4) & 2.156(16)  & 372913728 & 0.886 & 0.3221(20) & 3.105(20) \\
 308402813 & 0.888 & 0.326(3) & 3.07(3)  & 272087305 & 0.891 & 0.3283(11) & 3.05(10)  & 308537434 & 0.901 & 0.2544(17) & 3.930(27)  & 382576174 & 0.902 & 0.2709(21) & 3.691(29) \\
 364397433 & 0.906 & 0.1823(19) & 5.49(6)  & 410449468 & 0.907 & 0.3944(22) & 2.536(14)  & 382577845 & 0.907 & 0.381(4) & 2.627(29)  & 287425246 & 0.909 & 0.2390(24) & 4.18(4) \\
 358467097 & 0.910 & 0.260(4) & 3.85(5)  & 308401190 & 0.911 & 0.2488(17) & 4.019(27)  & 410451789 & 0.912 & 0.1652(18) & 6.05(7)  & 308753589 & 0.912 & 0.3124(20) & 3.201(20) \\
 410451523 & 0.914 & 0.386(5) & 2.59(4)  & 308310889 & 0.915 & 0.2655(22) & 3.77(3)  & 268964926 & 0.916 & 0.564(4) & 1.774(13)  & 410452270 & 0.918 & 0.3369(26) & 2.968(23) \\
 410451134 & 0.919 & 0.2409(16) & 4.151(28)  & 364398882 & 0.926 & 0.3904(24) & 2.562(15)  & 341337628 & 0.926 & 0.3505(21) & 2.853(17)  & 264932638 & 0.931 & 0.560(4) & 1.785(12) \\
 358467394 & 0.932 & 0.294(5) & 3.40(5)  & 364398282 & 0.936 & 0.3846(21) & 2.600(14)  & 372912197 & 0.937 & 0.340(3) & 2.945(28)  & 308307207 & 0.938 & 0.254(5) & 3.94(8) \\
 306638170 & 0.946 & 0.2379(13) & 4.203(23)  & 308537764 & 0.951 & 0.2363(29) & 4.23(5)  & 341555676 & 0.951 & 0.2385(23) & 4.19(4)  & 306773471 & 0.952 & 0.2089(18) & 4.79(4) \\
 382575727 & 0.954 & 0.3088(16) & 3.239(17)  & 341412789 & 0.959 & 0.300(3) & 3.34(4)  & 308603732 & 0.960 & 0.2912(24) & 3.434(28)  & 382529205 & 0.962 & 0.4236(29) & 2.361(16) \\
 308402711 & 0.962 & 0.2161(20) & 4.63(4)  & 308448137 & 0.963 & 0.2249(21) & 4.45(4)  & 262614031 & 0.964 & 0.2461(16) & 4.063(26)  & 272551492 & 0.965 & 0.1935(18) & 5.17(5) \\
 358464383 & 0.965 & 0.485(3) & 2.064(13)  & 410450536 & 0.968 & 0.2625(16) & 3.809(24)  & 261256961 & 0.968 & 0.1972(29) & 5.07(7)  & 382577942 & 0.969 & 0.437(3) & 2.288(16) \\
 410452104 & 0.970 & 0.2251(21) & 4.44(4)  & 278519822 & 0.981 & 0.444(4) & 2.252(22)  & 341043860 & 0.983 & 0.2969(28) & 3.37(3)  & 264932478 & 0.983 & 0.299(4) & 3.34(4) \\
 358463216 & 0.984 & 0.2830(16) & 3.534(20)  & 364398101 & 0.986 & 0.2031(18) & 4.92(4)  & 341113752 & 0.987 & 0.206(4) & 4.9(10)  & 308745518 & 0.989 & 0.1882(24) & 5.31(7) \\
 308606829 & 0.997 & 0.1895(18) & 5.28(5)  & 260843884 & 0.999 & 0.292(3) & 3.42(4)  & 308541083 & 1.001 & 0.2252(27) & 4.44(5)  & 382631033 & 1.006 & 0.5080(21) & 1.968(8) \\
 287154104 & 1.013 & 0.150(8) & 6.7(4)  & 382631326 & 1.013 & 0.2674(18) & 3.740(25)  & 290156111 & 1.019 & 0.239(3) & 4.18(6)  & 308615455 & 1.021 & 0.174(4) & 5.75(14) \\
 300510057 & 1.023 & 0.1588(14) & 6.30(6)  & 382625643 & 1.025 & 0.1625(15) & 6.15(6)  & 358508452 & 1.026 & 0.2079(27) & 4.81(6)  & 308744726 & 1.038 & 0.2135(26) & 4.68(6) \\
 281637074 & 1.038 & 0.116(4) & 8.6(3)  & 372913398 & 1.039 & 0.2071(29) & 4.83(7)  & 382578003 & 1.046 & 0.2664(21) & 3.75(3)  & 341263553 & 1.046 & 0.3443(17) & 2.904(14) \\
 261257136 & 1.054 & 0.152(4) & 6.56(18)  & 341104805 & 1.060 & 0.1853(29) & 5.40(8)  & 308851949 & 1.078 & 0.1538(19) & 6.50(8)  & 290159290 & 1.081 & 0.864(3) & 1.158(4) \\
 341112775 & 1.102 & 0.397(5) & 2.52(3)  & 270474003 & 1.107 & 0.1615(29) & 6.19(11)  & 308664160 & 1.110 & 0.1601(19) & 6.24(7)  & 306824373 & 1.113 & 0.1364(23) & 7.33(12) \\
 308449453 & 1.125 & 0.1856(20) & 5.39(6)  & 308307398 & 1.190 & 0.243(5) & 4.11(8)  & 404873066 & 1.249 & 0.124(3) & 8.10(23)  \\
 \hline
    \end{tabular}
    \tablefoot{We list the TIC numbers of the stars with surface modulations, their \textit{Gaia} colour index (CI), and the measured surface rotation frequencies (Freq) as well as periods ($P$). }
\end{sidewaystable*}

\newpage

\section{All the g-mode pulsators in NGC\,2516}\label{appandix_sec:all_gdor_figures}

For each of the \gdornumberwithcleargmodepattern g-mode pulsators, we show the amplitude spectrum with identified period spacing pattern, the best-fitting TAR results and the posterior distributions of $f_\mathrm{rot}$ and $\Pi_0$. We also show all the high-resolution spectra for these stars. 

\newpage
\begin{figure*}
\centering
\includegraphics[width = 0.9\linewidth]{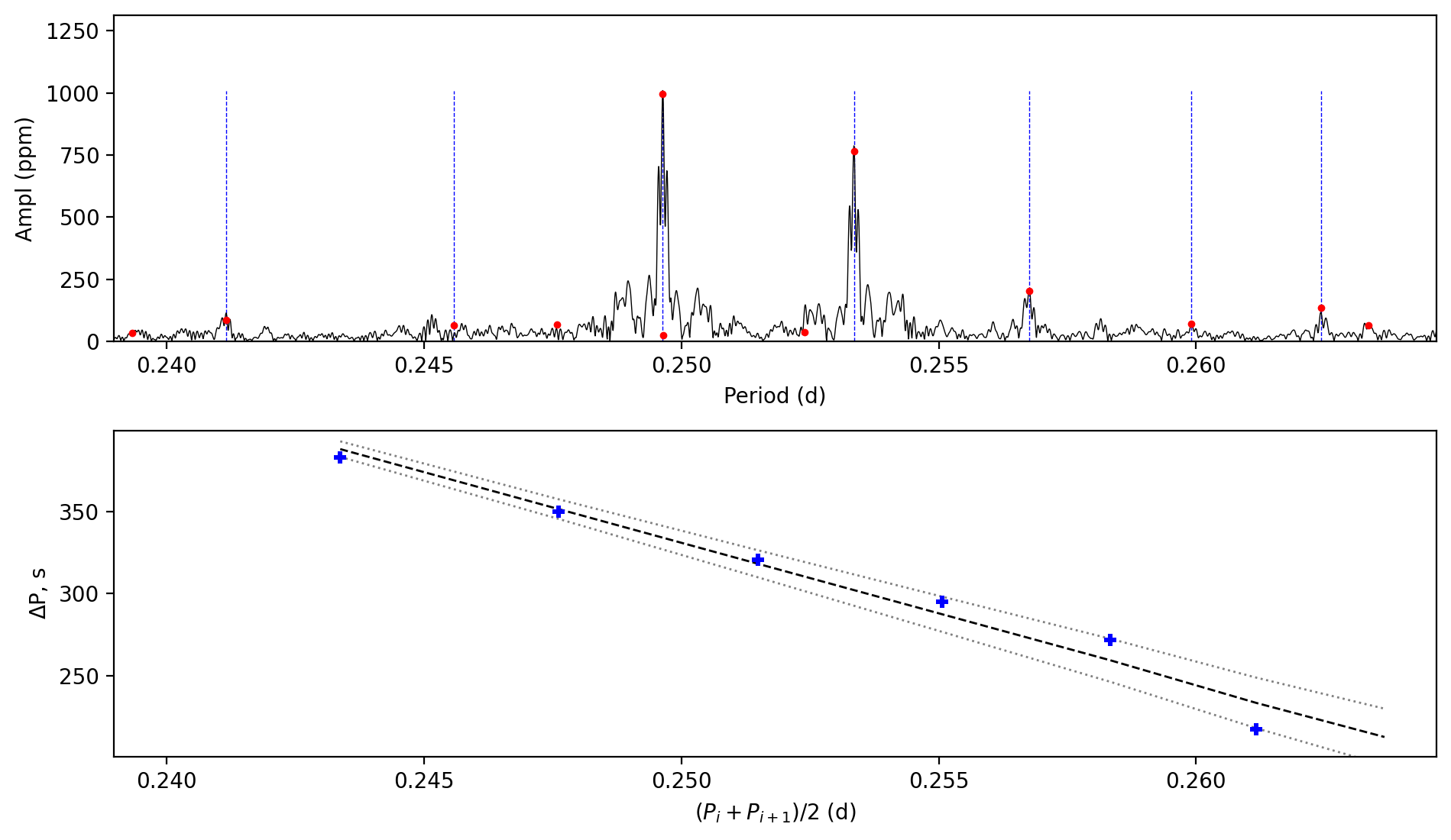}
\caption{Period spacing pattern(s) of TIC\,281582674. Top: amplitude spectrum as a function of period. Red dots are the peaks from the prewhitening process. Vertical lines mark the identified modes. Bottom: period spacing as a function of period. The black line is a linear fitting not the TAR fitting, and the grey lines exhibit the uncertainty region. The period is plotted as the mean value of the two consecutive peaks.}
\label{appendix_fig:TIC281582674_period_spacing}
\end{figure*}

\begin{figure*}
\centering
\includegraphics[width = 0.6\linewidth]{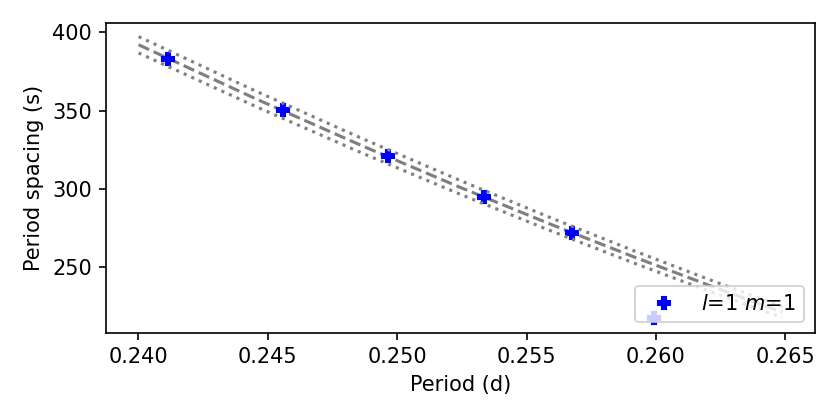}
\caption{Period spacing pattern(s) from the TAR fitting of TIC\,281582674. The dashed line marks the best-fitting model while the dotted lines exhibit the uncertainty region.}
\label{appendix_fig:TIC281582674_period_spacing_TAR}
\end{figure*}

\begin{figure*}
\centering
\includegraphics[width = 0.6\linewidth]{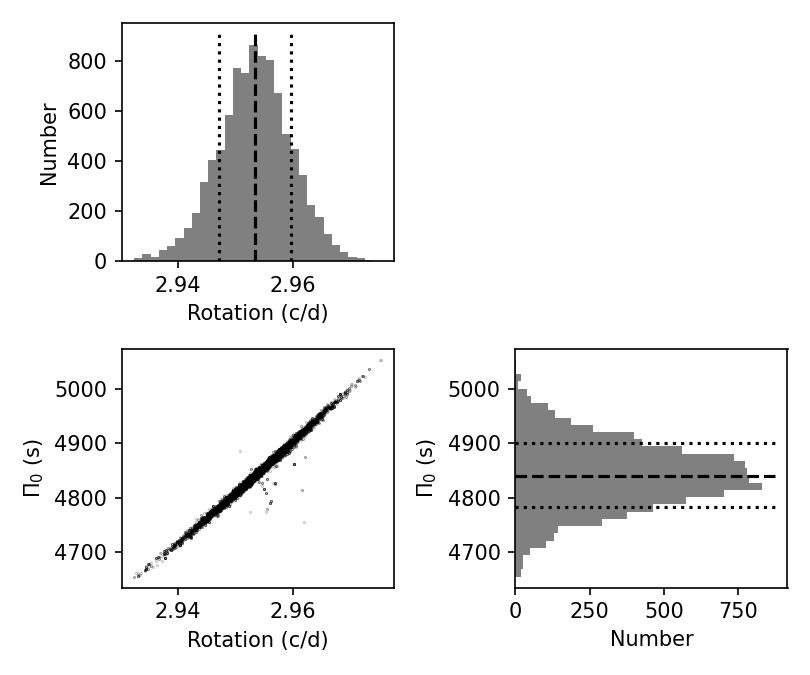}
\caption{Posterior distributions of the TAR fitting of TIC\,281582674.}
\label{appendix_fig:TIC281582674_period_spacing_TAR_corner}
\end{figure*}

\begin{figure*}
\centering
\includegraphics[width = 0.9\linewidth]{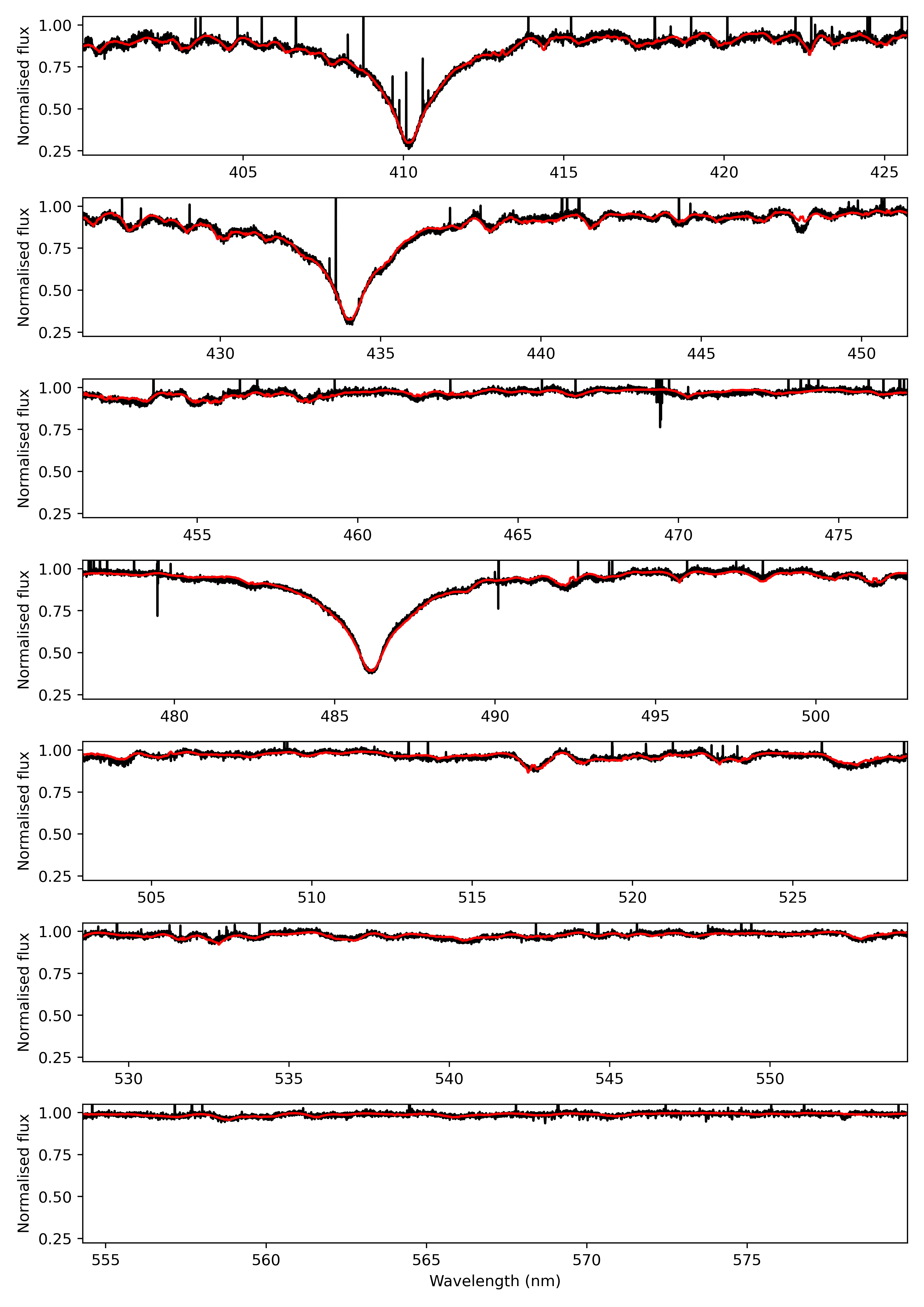}
\caption{Spectrum of TIC\,281582674 observed at 2023-03-12T23:47:34.131. The red line is the best-fitting theoretical spectrum. }
\label{appendix_fig:TIC281582674_spectrum_at_2023-03-12T23:47:34.131}
\end{figure*}

\begin{figure*}
\centering
\includegraphics[width = 0.9\linewidth]{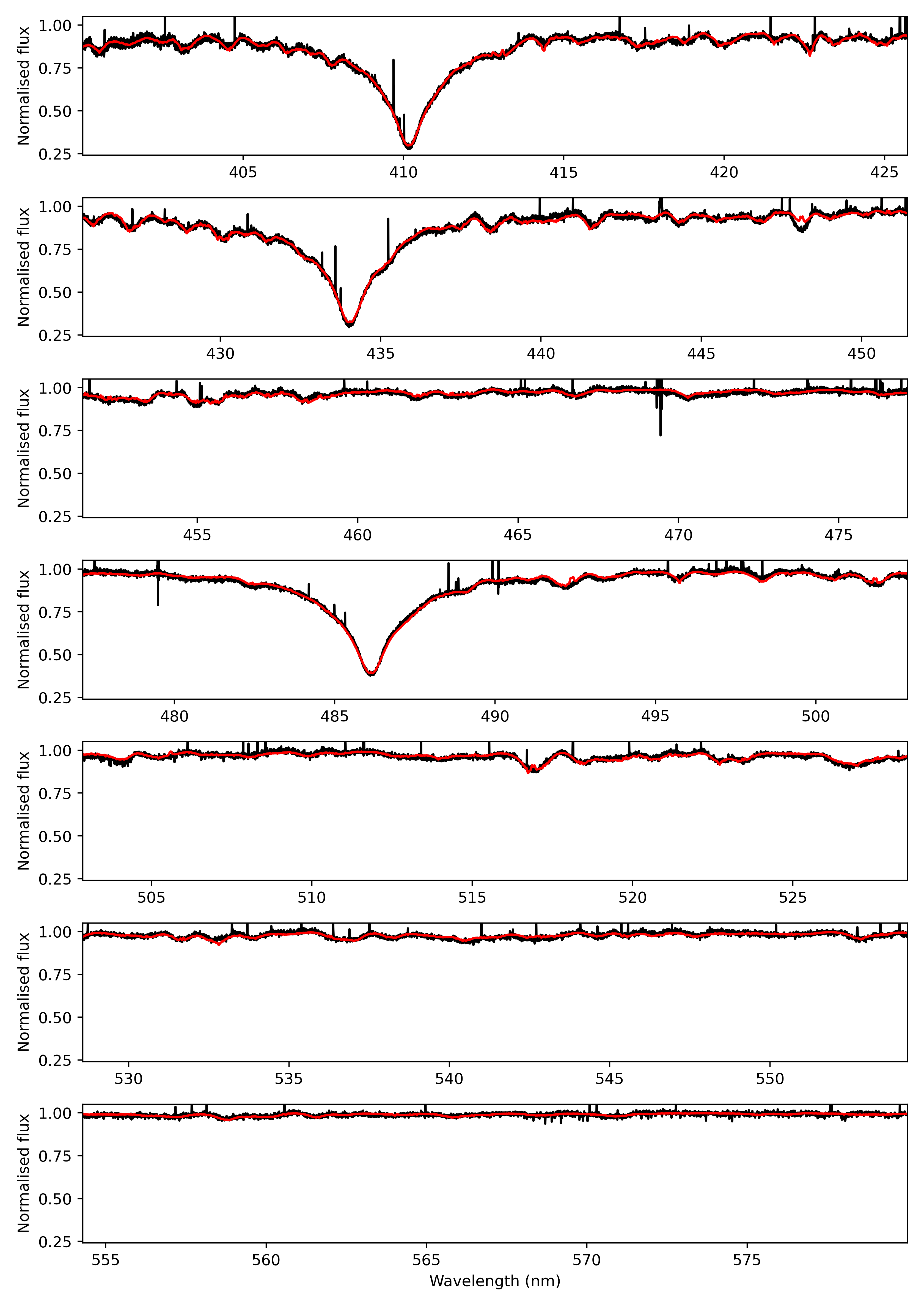}
\caption{Spectrum of TIC\,281582674 observed at 2023-03-16T02:04:10.433. The red line is the best-fitting theoretical spectrum. }
\label{appendix_fig:TIC281582674_spectrum_at_2023-03-16T02:04:10.433}
\end{figure*}

\newpage
\newpage
\begin{figure*}
\centering
\includegraphics[width = 0.9\linewidth]{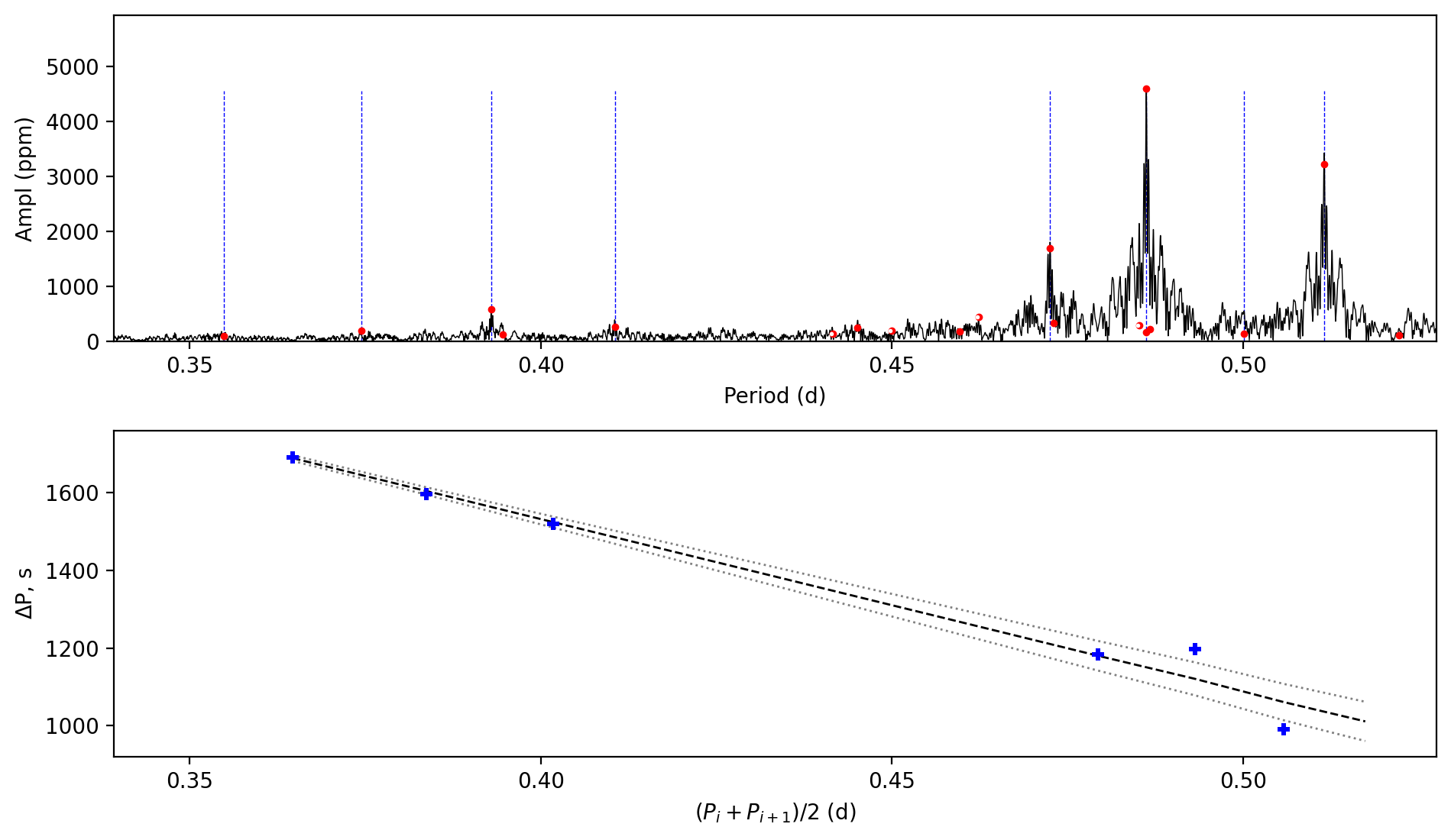}
\caption{Period spacing pattern(s) of TIC\,308307454. Top: amplitude spectrum as a function of period. Red dots are the peaks from the prewhitening process. Vertical lines mark the identified modes. Bottom: period spacing as a function of period. The black line is a linear fitting not the TAR fitting, and the grey lines exhibit the uncertainty region. The period is plotted as the mean value of the two consecutive peaks.}
\label{appendix_fig:TIC308307454_period_spacing}
\end{figure*}

\begin{figure*}
\centering
\includegraphics[width = 0.6\linewidth]{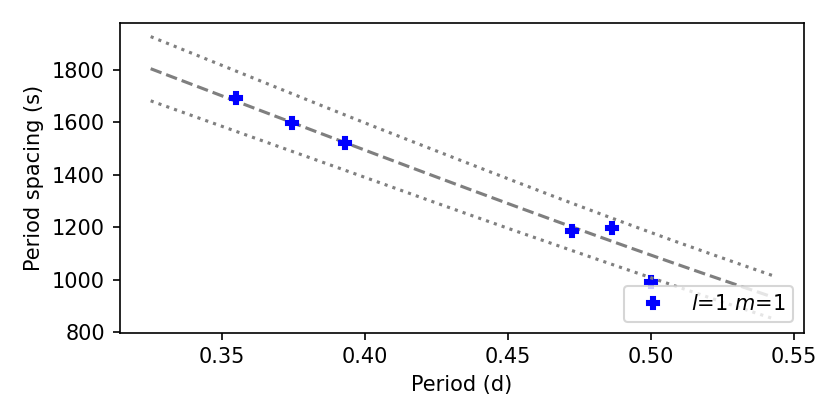}
\caption{Period spacing pattern(s) from the TAR fitting of TIC\,308307454. The dashed line marks the best-fitting model while the dotted lines exhibit the uncertainty region.}
\label{appendix_fig:TIC308307454_period_spacing_TAR}
\end{figure*}

\begin{figure*}
\centering
\includegraphics[width = 0.6\linewidth]{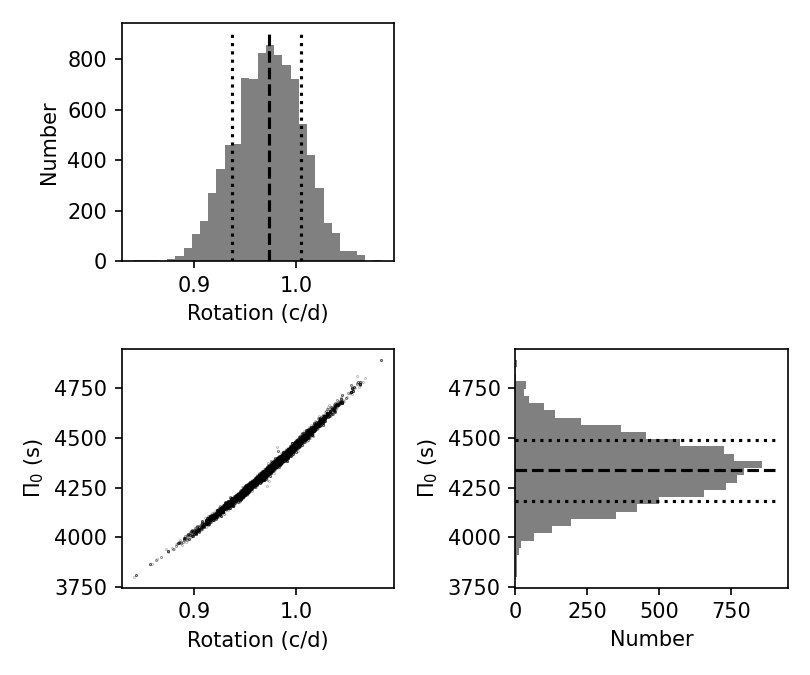}
\caption{Posterior distributions of the TAR fitting of TIC\,308307454.}
\label{appendix_fig:TIC308307454_period_spacing_TAR_corner}
\end{figure*}

\begin{figure*}
\centering
\includegraphics[width = 0.9\linewidth]{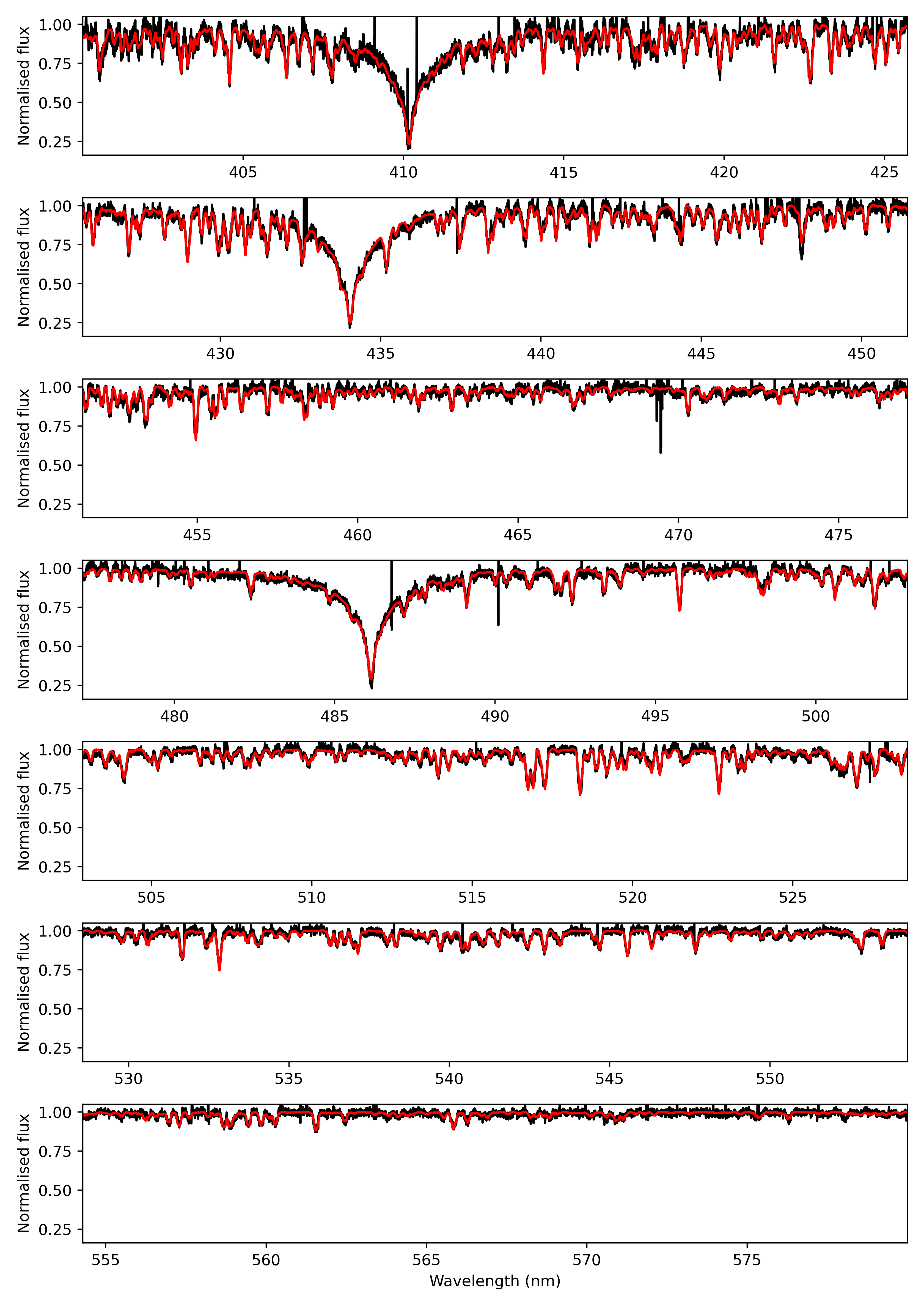}
\caption{Spectrum of TIC\,308307454 observed at 2023-03-13T00:54:35.517. The red line is the best-fitting theoretical spectrum. }
\label{appendix_fig:TIC308307454_spectrum_at_2023-03-13T00:54:35.517}
\end{figure*}

\begin{figure*}
\centering
\includegraphics[width = 0.9\linewidth]{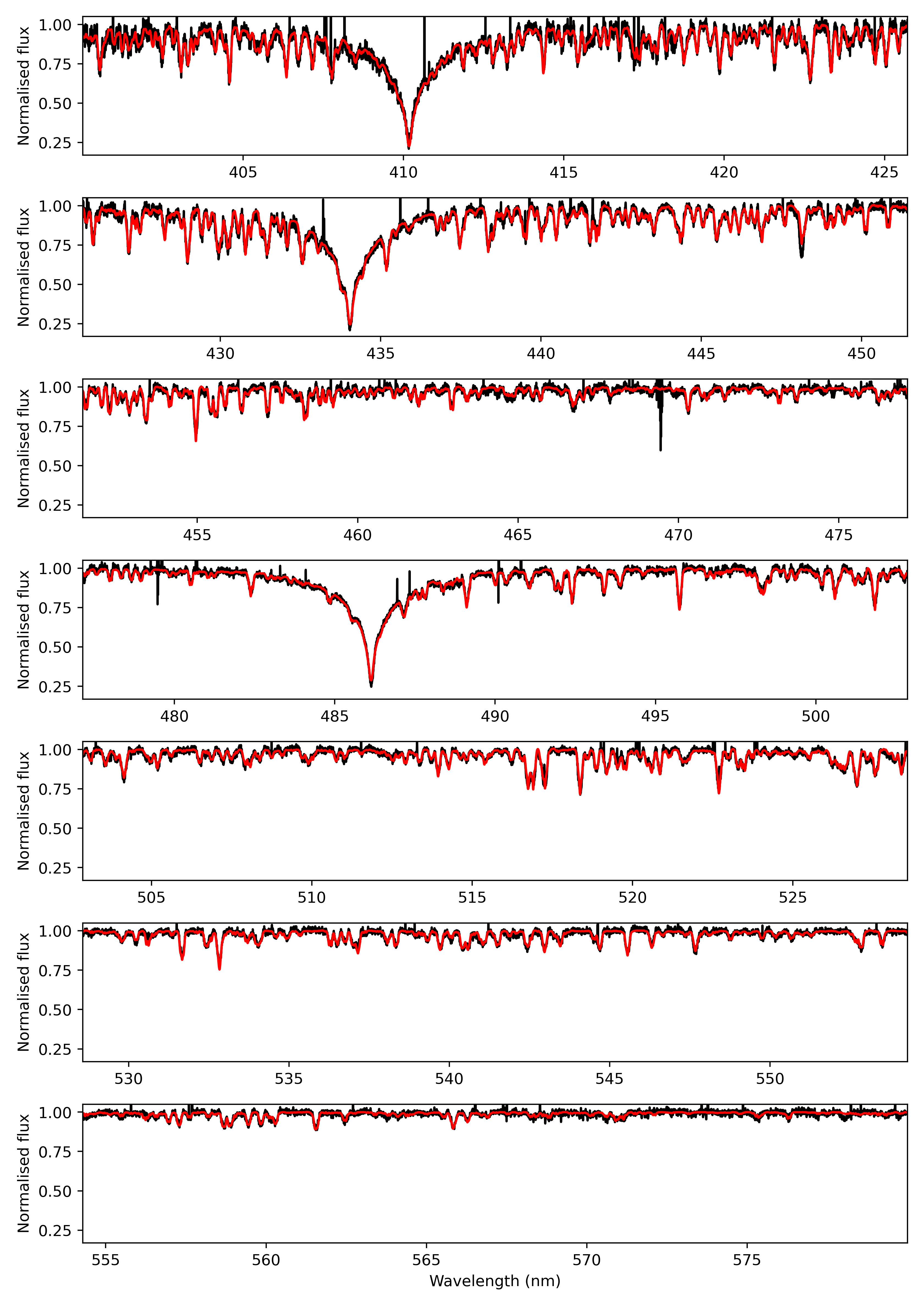}
\caption{Spectrum of TIC\,308307454 observed at 2023-03-14T04:55:16.136. The red line is the best-fitting theoretical spectrum. }
\label{appendix_fig:TIC308307454_spectrum_at_2023-03-14T04:55:16.136}
\end{figure*}

\begin{figure*}
\centering
\includegraphics[width = 0.9\linewidth]{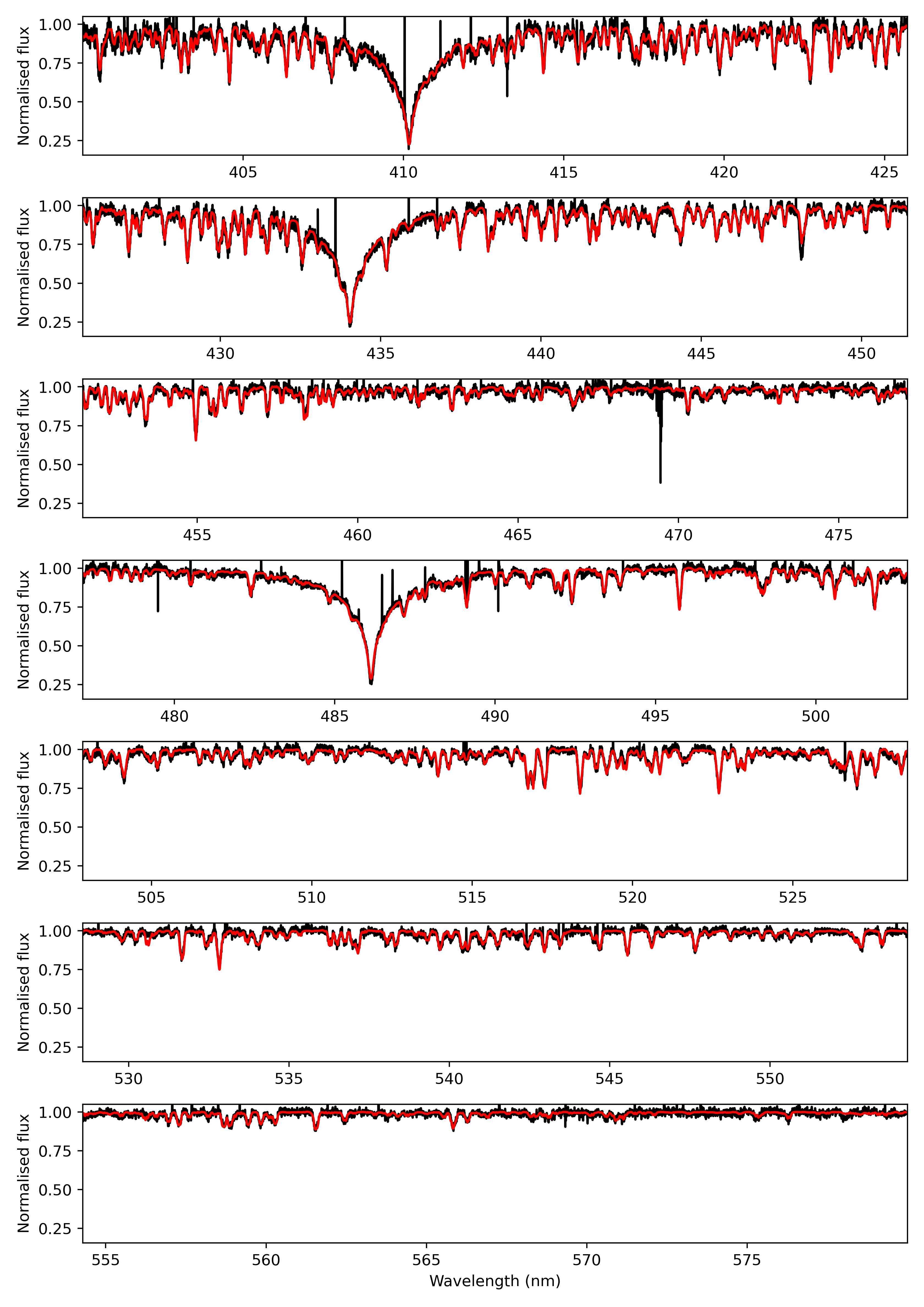}
\caption{Spectrum of TIC\,308307454 observed at 2023-03-16T03:13:33.242. The red line is the best-fitting theoretical spectrum. }
\label{appendix_fig:TIC308307454_spectrum_at_2023-03-16T03:13:33.242}
\end{figure*}

\newpage
\newpage
\begin{figure*}
\centering
\includegraphics[width = 0.9\linewidth]{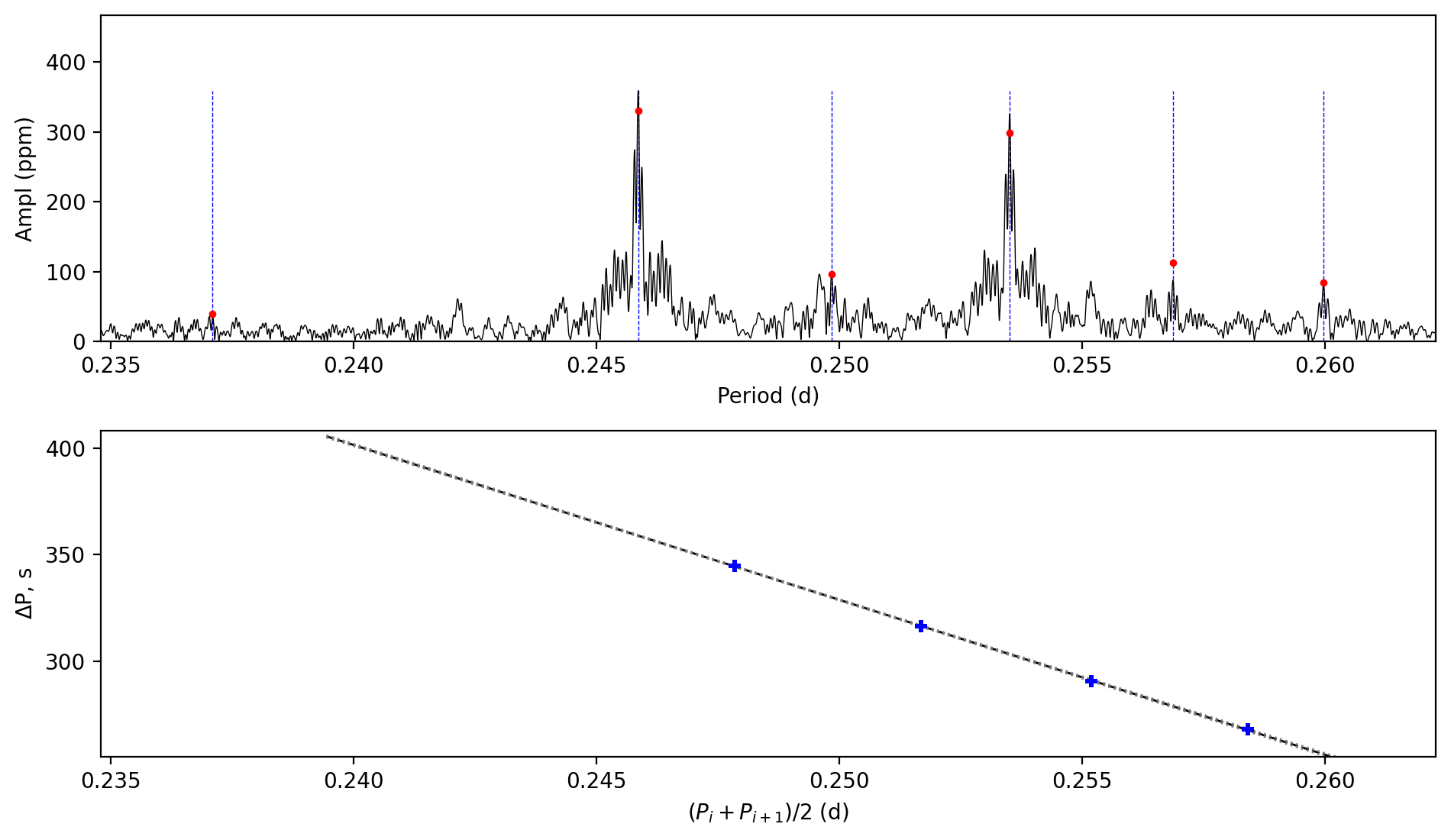}
\caption{Period spacing pattern(s) of TIC\,308992761. Top: amplitude spectrum as a function of period. Red dots are the peaks from the prewhitening process. Vertical lines mark the identified modes. Bottom: period spacing as a function of period. The black line is a linear fitting not the TAR fitting, and the grey lines exhibit the uncertainty region. The period is plotted as the mean value of the two consecutive peaks.}
\label{appendix_fig:TIC308992761_period_spacing}
\end{figure*}

\begin{figure*}
\centering
\includegraphics[width = 0.6\linewidth]{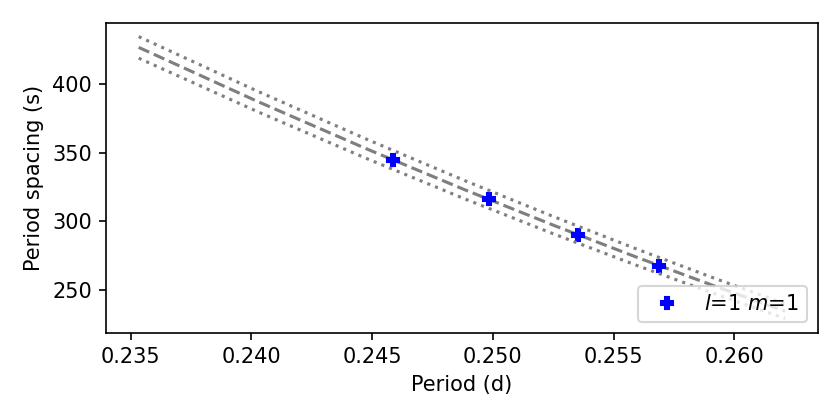}
\caption{Period spacing pattern(s) from the TAR fitting of TIC\,308992761. The dashed line marks the best-fitting model while the dotted lines exhibit the uncertainty region.}
\label{appendix_fig:TIC308992761_period_spacing_TAR}
\end{figure*}

\begin{figure*}
\centering
\includegraphics[width = 0.6\linewidth]{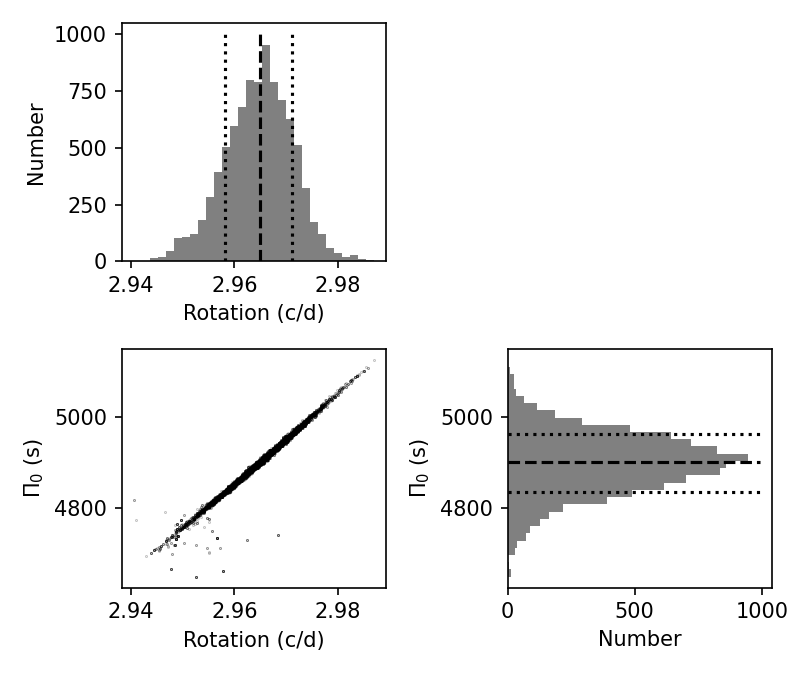}
\caption{Posterior distributions of the TAR fitting of TIC\,308992761.}
\label{appendix_fig:TIC308992761_period_spacing_TAR_corner}
\end{figure*}

\newpage
\newpage
\begin{figure*}
\centering
\includegraphics[width = 0.9\linewidth]{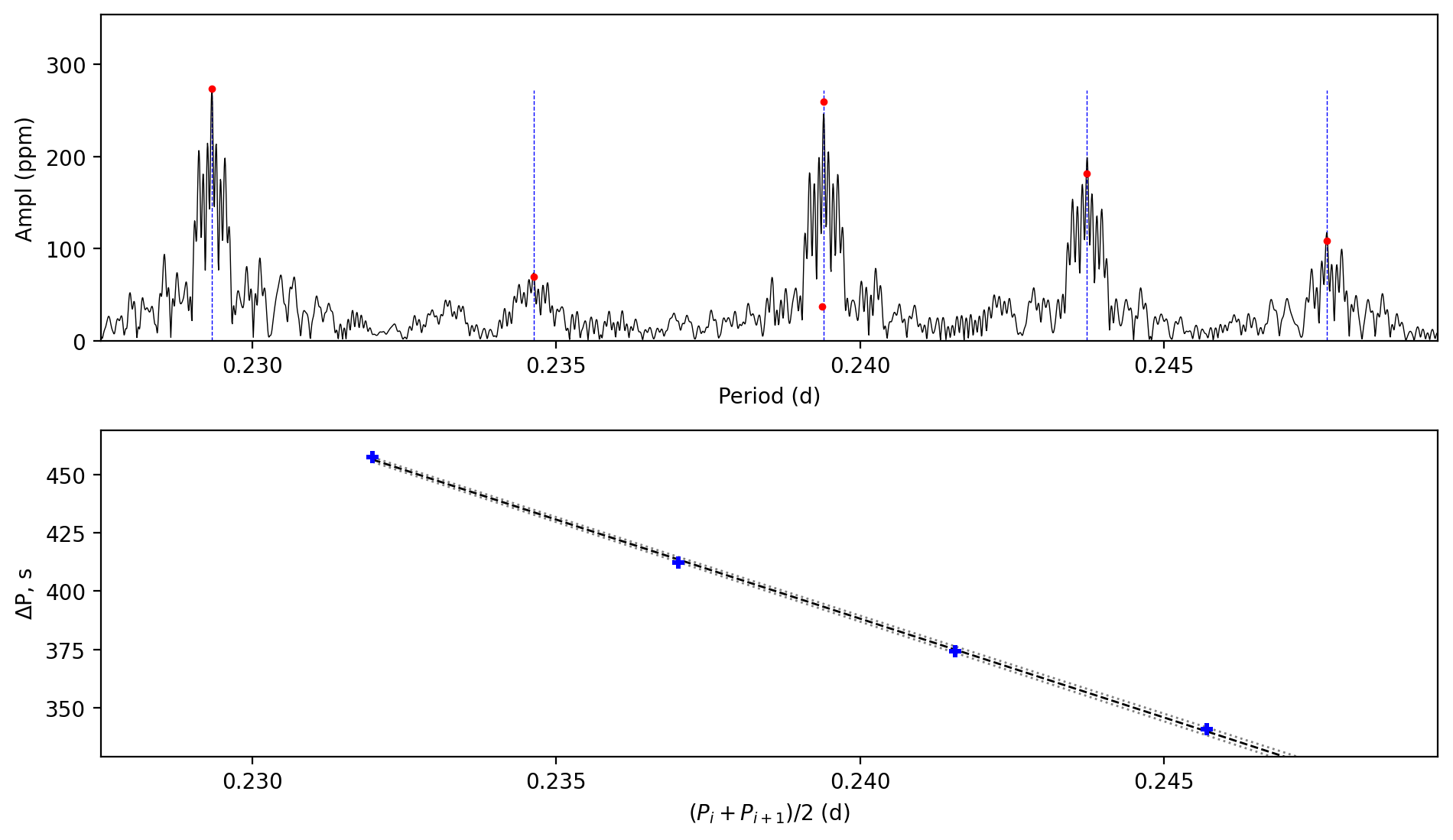}
\caption{Period spacing pattern(s) of TIC\,341043961. Top: amplitude spectrum as a function of period. Red dots are the peaks from the prewhitening process. Vertical lines mark the identified modes. Bottom: period spacing as a function of period. The black line is a linear fitting not the TAR fitting, and the grey lines exhibit the uncertainty region. The period is plotted as the mean value of the two consecutive peaks.}
\label{appendix_fig:TIC341043961_period_spacing}
\end{figure*}

\begin{figure*}
\centering
\includegraphics[width = 0.6\linewidth]{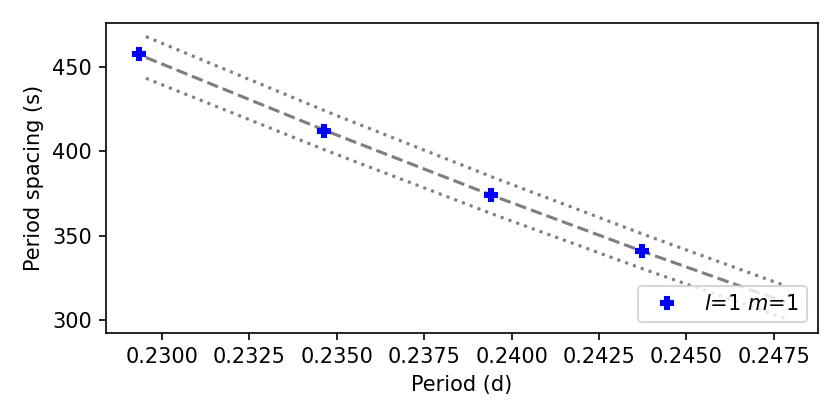}
\caption{Period spacing pattern(s) from the TAR fitting of TIC\,341043961. The dashed line marks the best-fitting model while the dotted lines exhibit the uncertainty region.}
\label{appendix_fig:TIC341043961_period_spacing_TAR}
\end{figure*}

\begin{figure*}
\centering
\includegraphics[width = 0.6\linewidth]{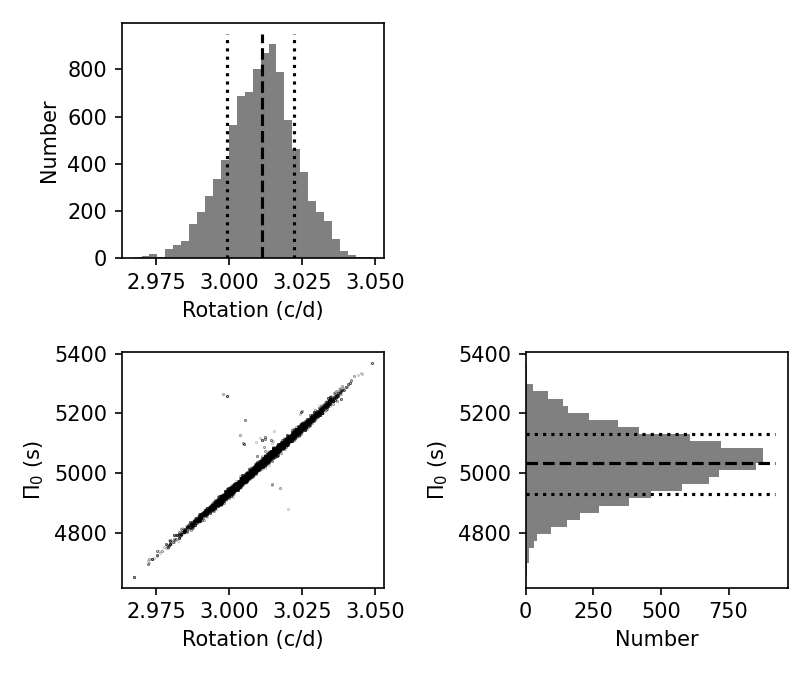}
\caption{Posterior distributions of the TAR fitting of TIC\,341043961.}
\label{appendix_fig:TIC341043961_period_spacing_TAR_corner}
\end{figure*}

\newpage
\newpage
\begin{figure*}
\centering
\includegraphics[width = 0.9\linewidth]{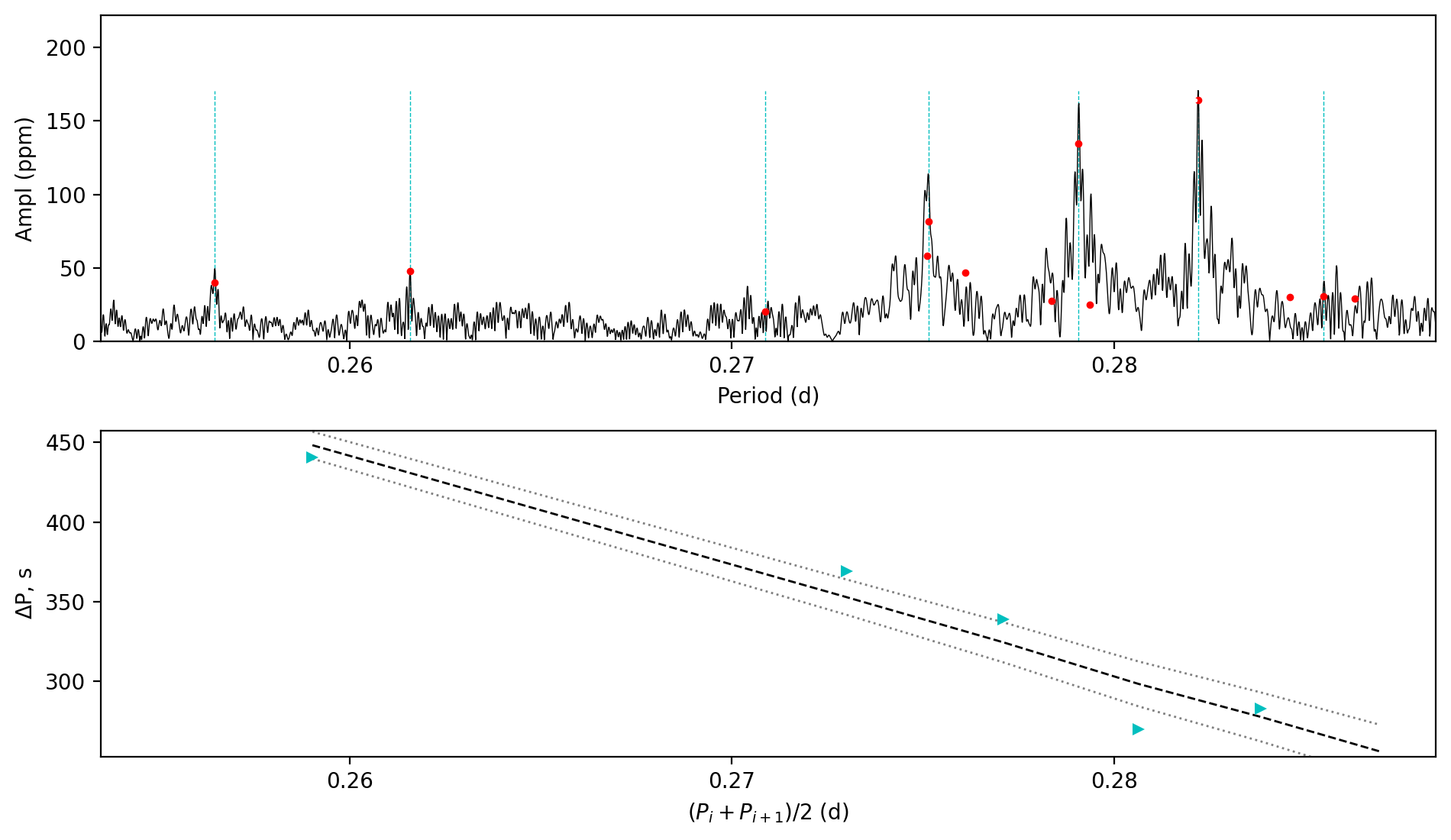}
\caption{Period spacing pattern(s) of TIC\,358466708. Top: amplitude spectrum as a function of period. Red dots are the peaks from the prewhitening process. Vertical lines mark the identified modes. Bottom: period spacing as a function of period. The black line is a linear fitting not the TAR fitting, and the grey lines exhibit the uncertainty region. The period is plotted as the mean value of the two consecutive peaks.}
\label{appendix_fig:TIC358466708_period_spacing}
\end{figure*}

\begin{figure*}
\centering
\includegraphics[width = 0.6\linewidth]{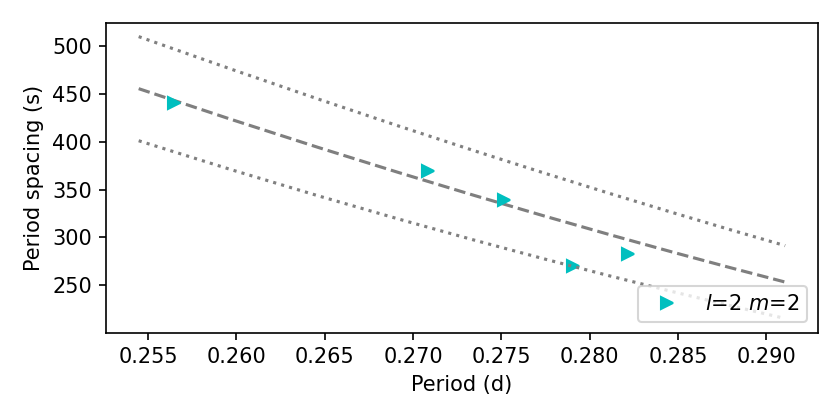}
\caption{Period spacing pattern(s) from the TAR fitting of TIC\,358466708. The dashed line marks the best-fitting model while the dotted lines exhibit the uncertainty region.}
\label{appendix_fig:TIC358466708_period_spacing_TAR}
\end{figure*}

\begin{figure*}
\centering
\includegraphics[width = 0.6\linewidth]{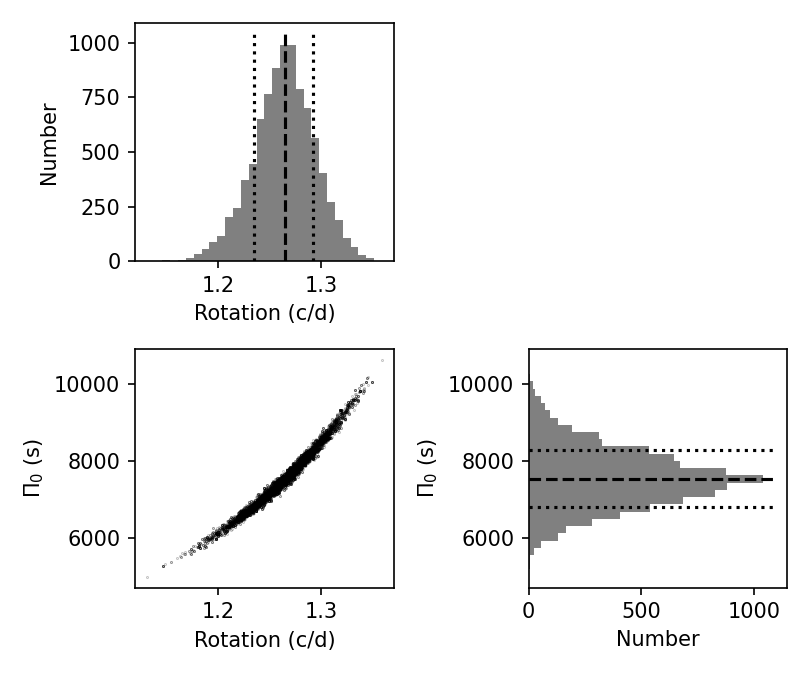}
\caption{Posterior distributions of the TAR fitting of TIC\,358466708.}
\label{appendix_fig:TIC358466708_period_spacing_TAR_corner}
\end{figure*}

\begin{figure*}
\centering
\includegraphics[width = 0.9\linewidth]{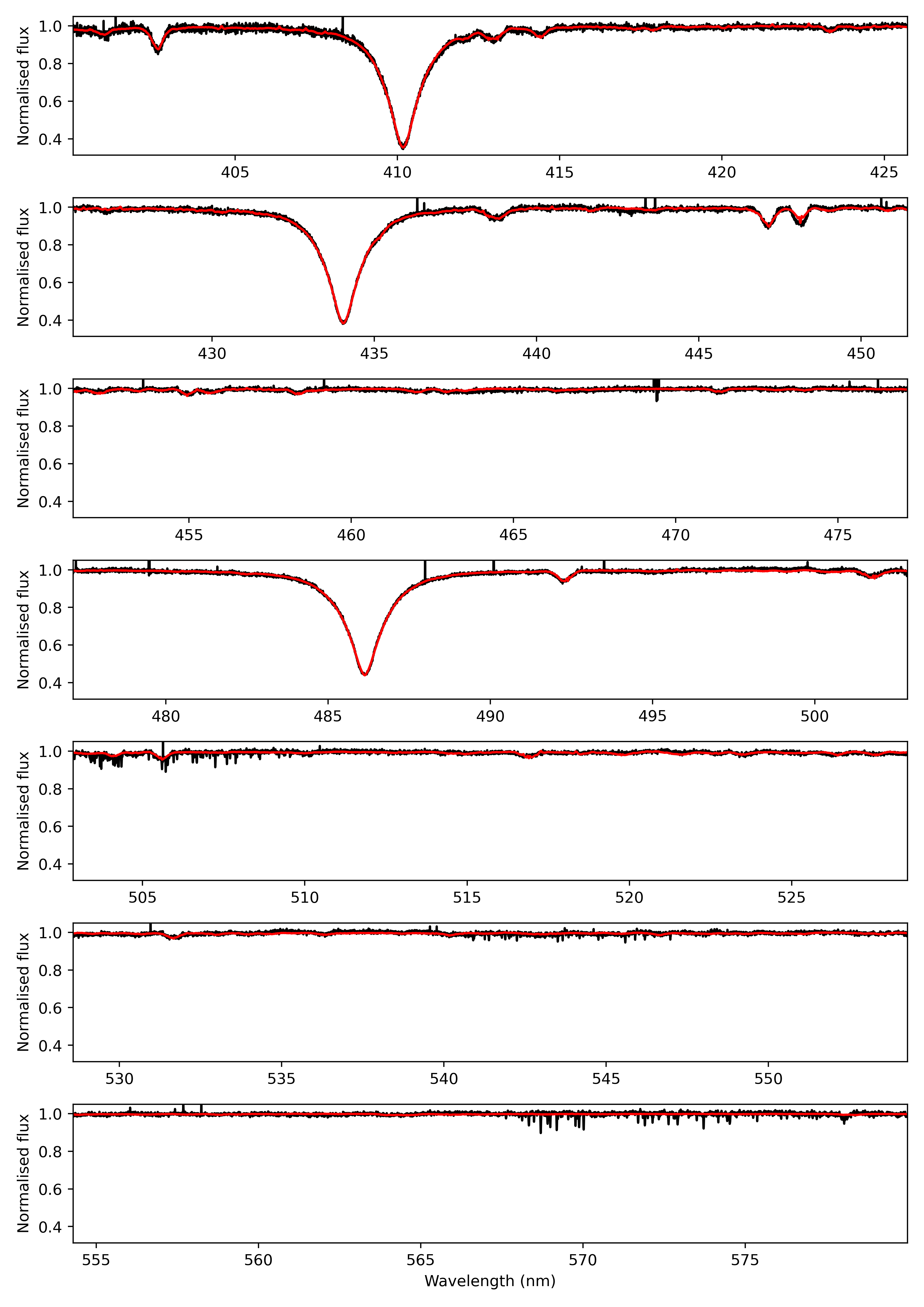}
\caption{Spectrum of TIC\,358466708 observed at 2023-03-12T06:45:37.221. The red line is the best-fitting theoretical spectrum. }
\label{appendix_fig:TIC358466708_spectrum_at_2023-03-12T06:45:37.221}
\end{figure*}

\newpage
\newpage
\begin{figure*}
\centering
\includegraphics[width = 0.9\linewidth]{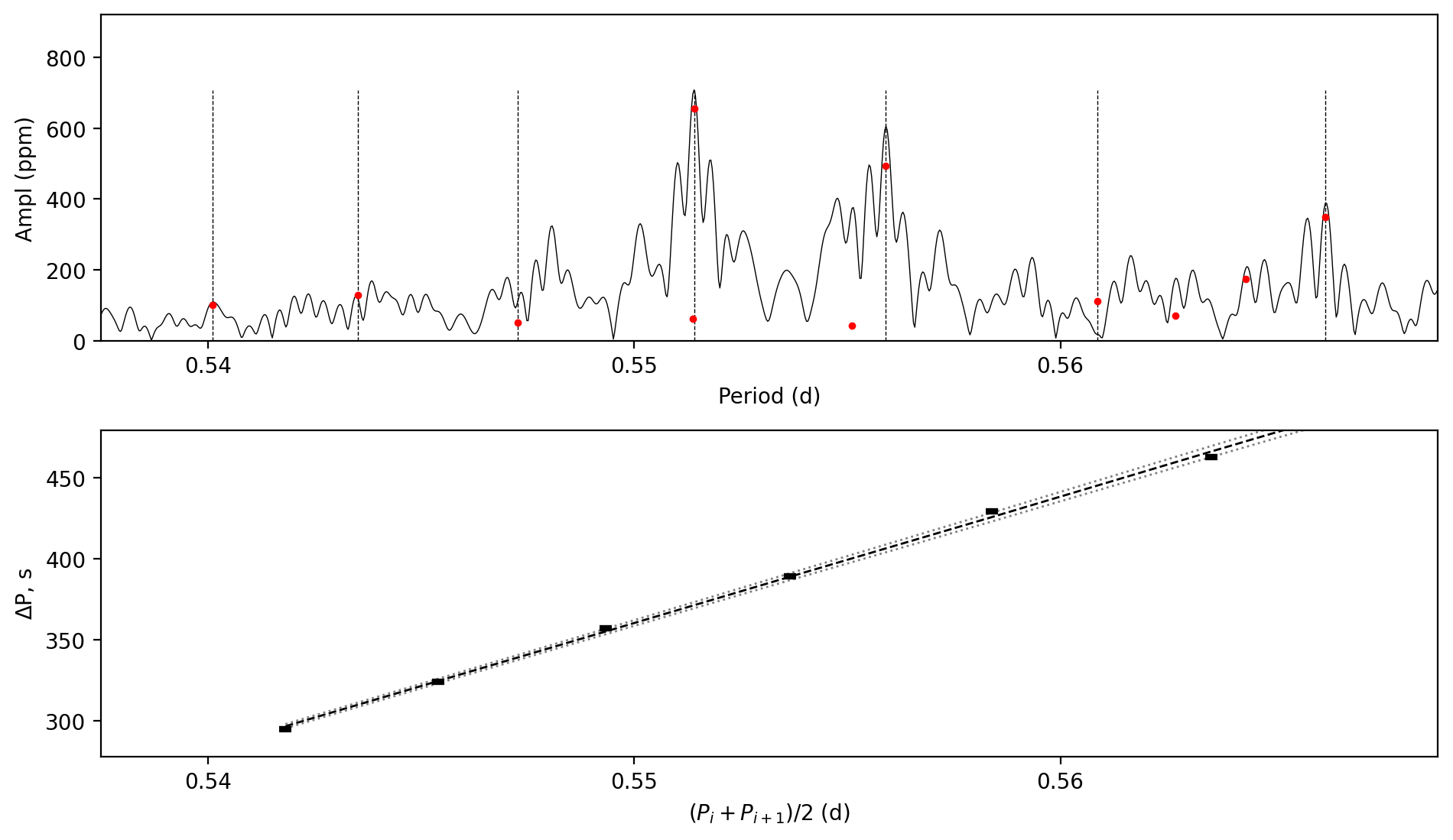}
\caption{Period spacing pattern(s) of TIC\,358466729. Top: amplitude spectrum as a function of period. Red dots are the peaks from the prewhitening process. Vertical lines mark the identified modes. Bottom: period spacing as a function of period. The black line is a linear fitting not the TAR fitting, and the grey lines exhibit the uncertainty region. The period is plotted as the mean value of the two consecutive peaks.}
\label{appendix_fig:TIC358466729_period_spacing}
\end{figure*}

\begin{figure*}
\centering
\includegraphics[width = 0.6\linewidth]{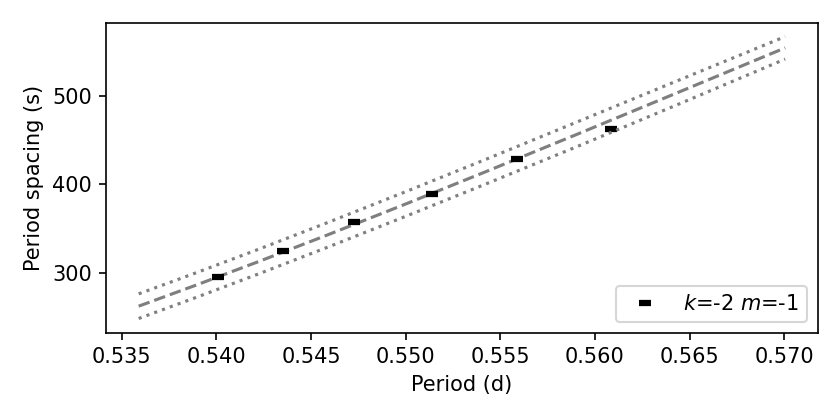}
\caption{Period spacing pattern(s) from the TAR fitting of TIC\,358466729. The dashed line marks the best-fitting model while the dotted lines exhibit the uncertainty region.}
\label{appendix_fig:TIC358466729_period_spacing_TAR}
\end{figure*}

\begin{figure*}
\centering
\includegraphics[width = 0.6\linewidth]{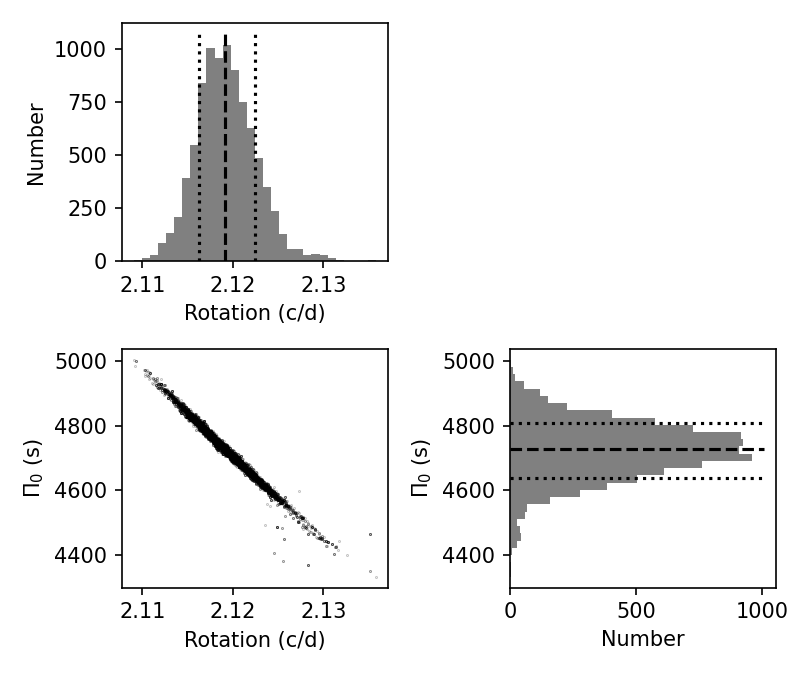}
\caption{Posterior distributions of the TAR fitting of TIC\,358466729.}
\label{appendix_fig:TIC358466729_period_spacing_TAR_corner}
\end{figure*}

\begin{figure*}
\centering
\includegraphics[width = 0.9\linewidth]{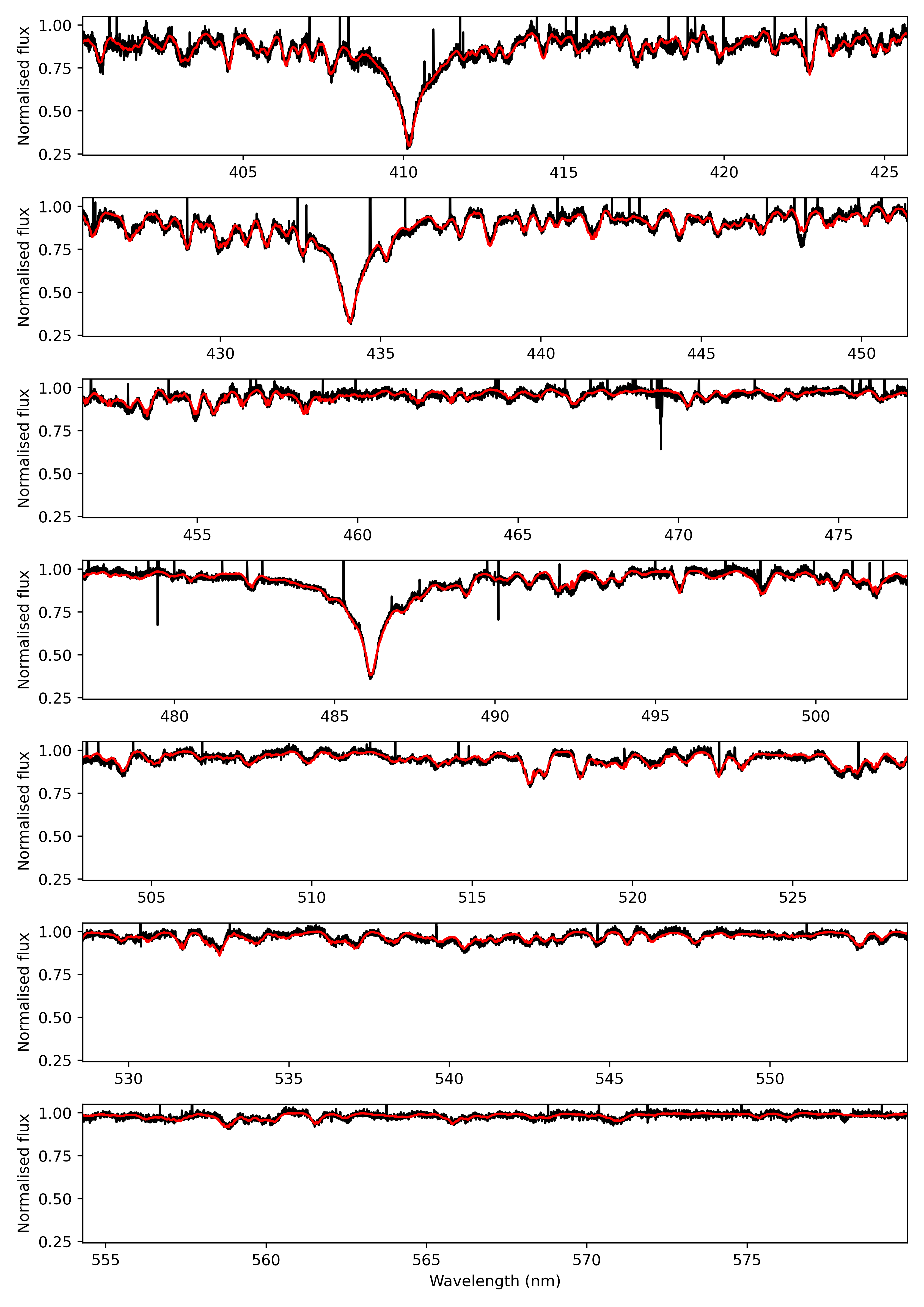}
\caption{Spectrum of TIC\,358466729 observed at 2023-03-13T02:48:18.645. The red line is the best-fitting theoretical spectrum. }
\label{appendix_fig:TIC358466729_spectrum_at_2023-03-13T02:48:18.645}
\end{figure*}

\begin{figure*}
\centering
\includegraphics[width = 0.9\linewidth]{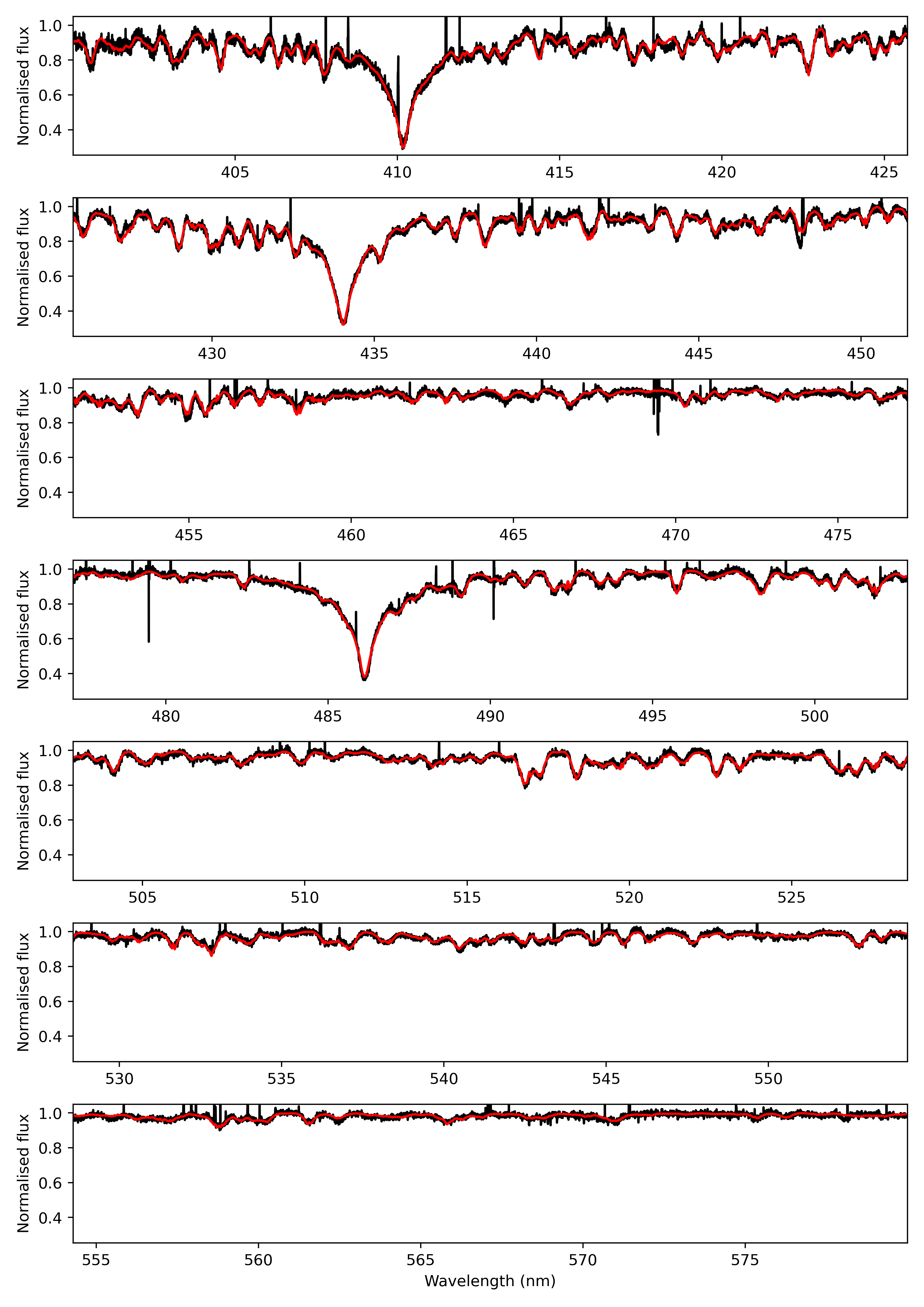}
\caption{Spectrum of TIC\,358466729 observed at 2023-03-15T02:09:35.139. The red line is the best-fitting theoretical spectrum. }
\label{appendix_fig:TIC358466729_spectrum_at_2023-03-15T02:09:35.139}
\end{figure*}

\newpage
\newpage
\begin{figure*}
\centering
\includegraphics[width = 0.9\linewidth]{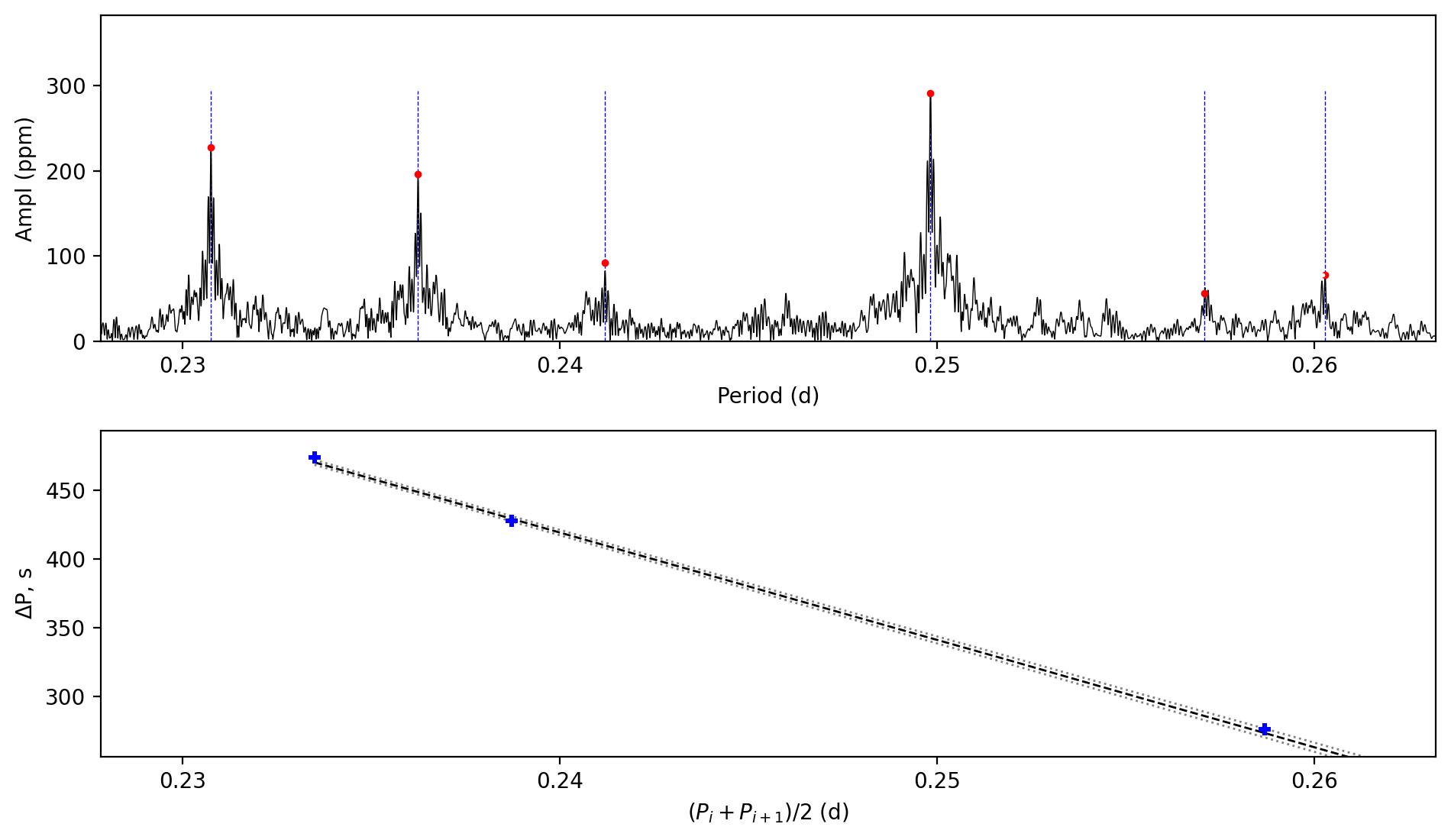}
\caption{Period spacing pattern(s) of TIC\,364398040. Top: amplitude spectrum as a function of period. Red dots are the peaks from the prewhitening process. Vertical lines mark the identified modes. Bottom: period spacing as a function of period. The black line is a linear fitting not the TAR fitting, and the grey lines exhibit the uncertainty region. The period is plotted as the mean value of the two consecutive peaks.}
\label{appendix_fig:TIC364398040_period_spacing}
\end{figure*}

\begin{figure*}
\centering
\includegraphics[width = 0.6\linewidth]{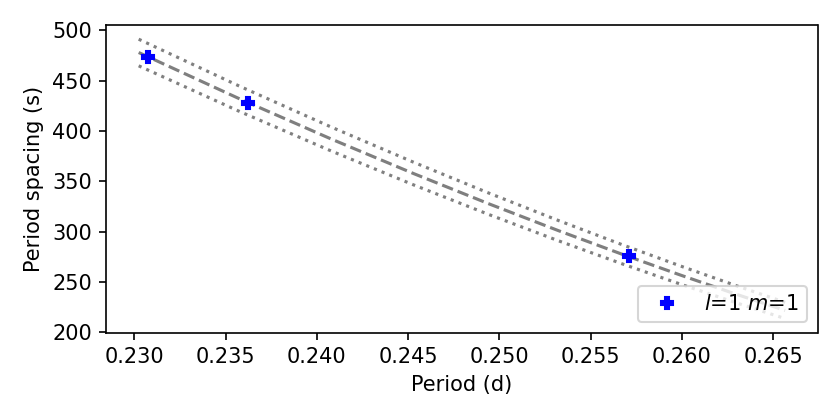}
\caption{Period spacing pattern(s) from the TAR fitting of TIC\,364398040. The dashed line marks the best-fitting model while the dotted lines exhibit the uncertainty region.}
\label{appendix_fig:TIC364398040_period_spacing_TAR}
\end{figure*}

\begin{figure*}
\centering
\includegraphics[width = 0.6\linewidth]{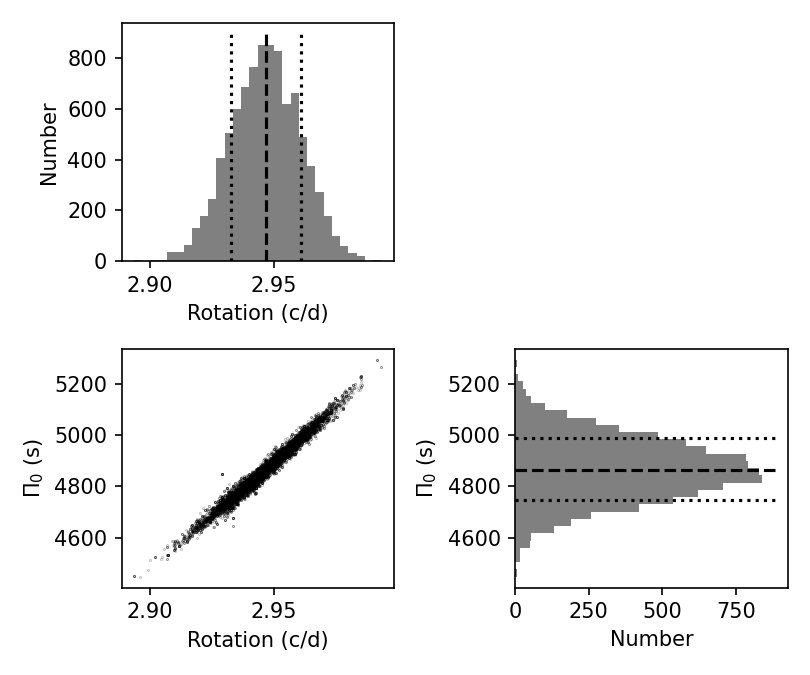}
\caption{Posterior distributions of the TAR fitting of TIC\,364398040.}
\label{appendix_fig:TIC364398040_period_spacing_TAR_corner}
\end{figure*}

\begin{figure*}
\centering
\includegraphics[width = 0.9\linewidth]{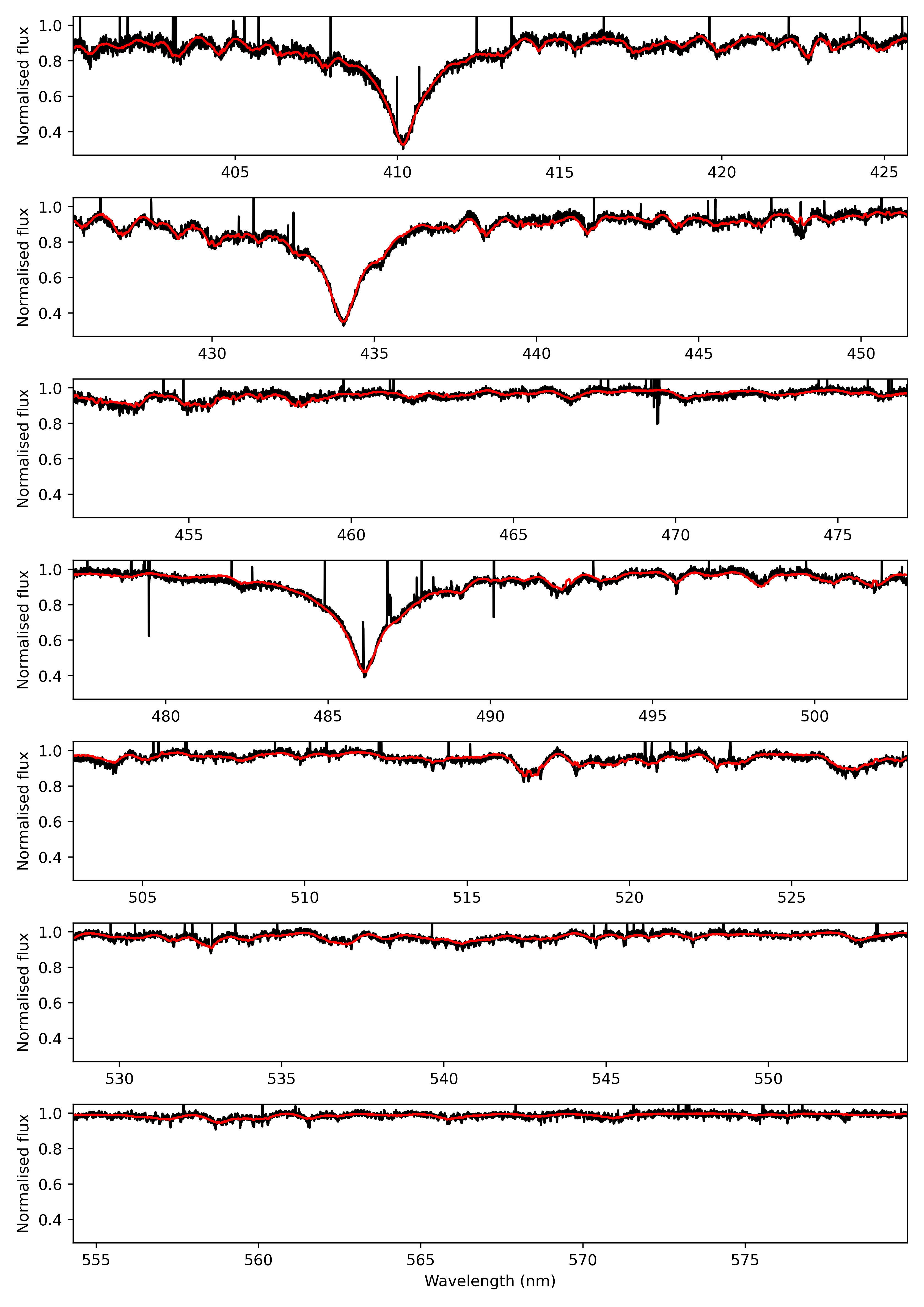}
\caption{Spectrum of TIC\,364398040 observed at 2023-03-13T00:20:15.606. The red line is the best-fitting theoretical spectrum. }
\label{appendix_fig:TIC364398040_spectrum_at_2023-03-13T00:20:15.606}
\end{figure*}

\begin{figure*}
\centering
\includegraphics[width = 0.9\linewidth]{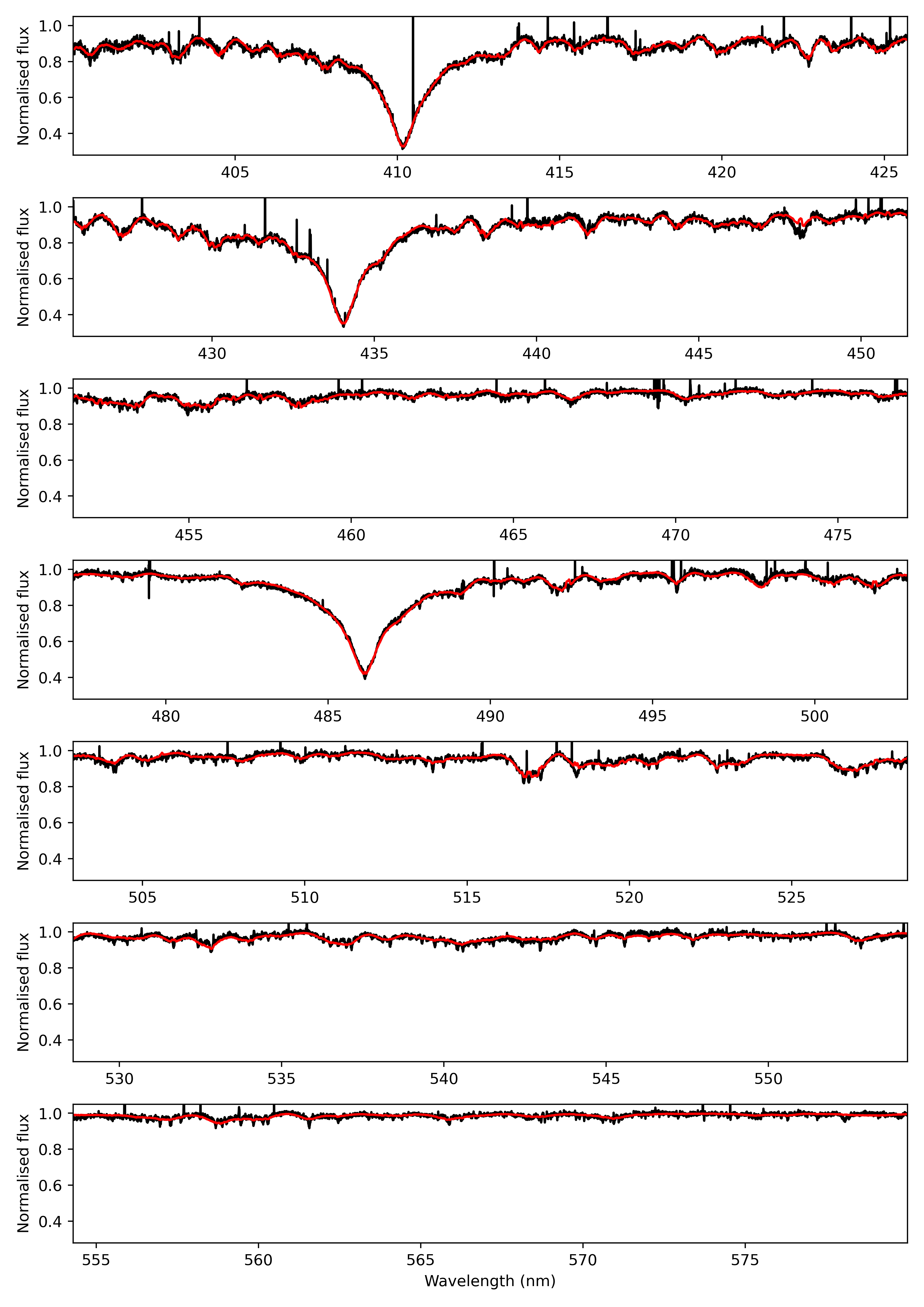}
\caption{Spectrum of TIC\,364398040 observed at 2023-03-15T01:33:49.149. The red line is the best-fitting theoretical spectrum. }
\label{appendix_fig:TIC364398040_spectrum_at_2023-03-15T01:33:49.149}
\end{figure*}

\newpage
\newpage
\begin{figure*}
\centering
\includegraphics[width = 0.9\linewidth]{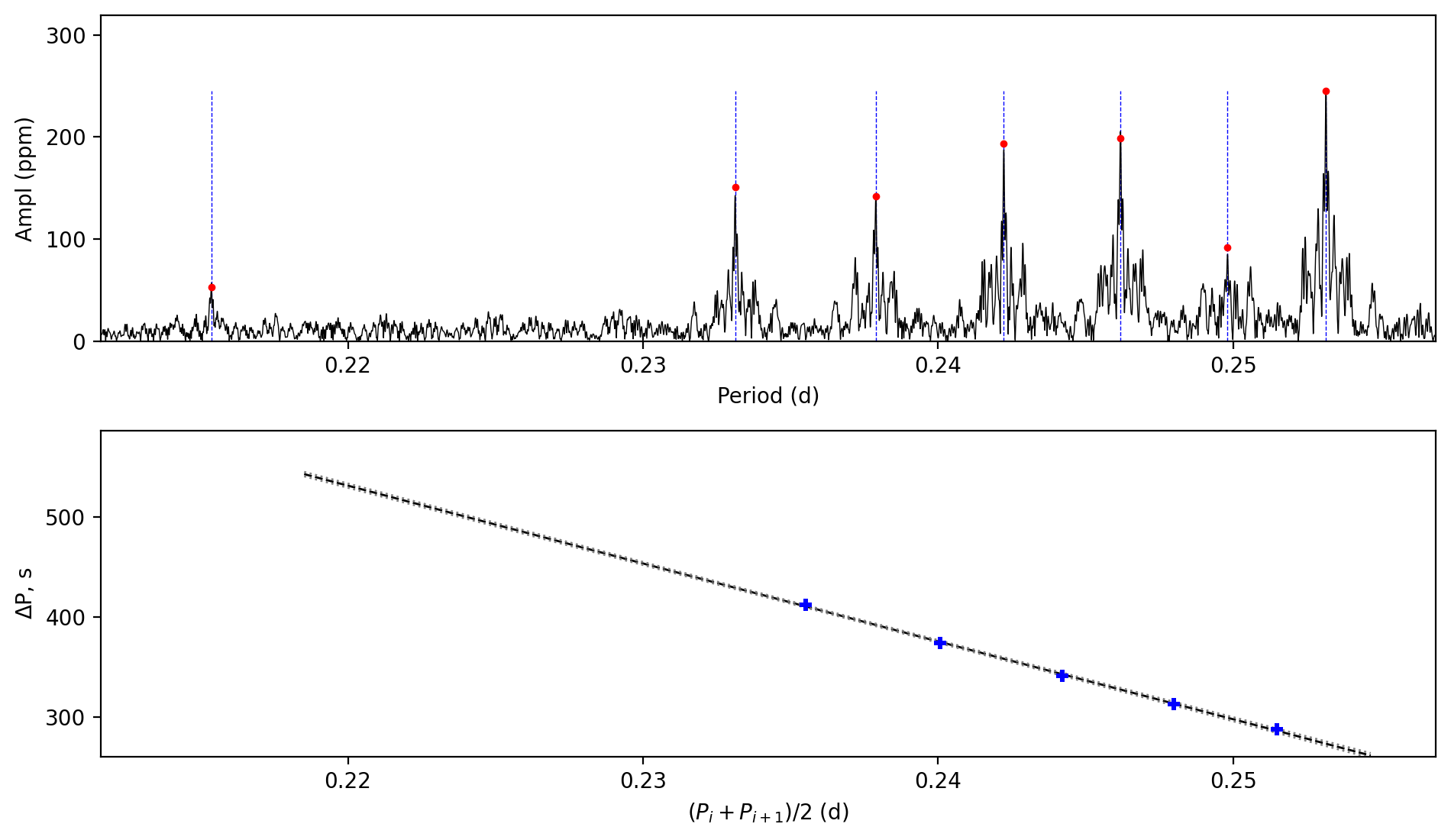}
\caption{Period spacing pattern(s) of TIC\,372912679. Top: amplitude spectrum as a function of period. Red dots are the peaks from the prewhitening process. Vertical lines mark the identified modes. Bottom: period spacing as a function of period. The black line is a linear fitting not the TAR fitting, and the grey lines exhibit the uncertainty region. The period is plotted as the mean value of the two consecutive peaks.}
\label{appendix_fig:TIC372912679_period_spacing}
\end{figure*}

\begin{figure*}
\centering
\includegraphics[width = 0.6\linewidth]{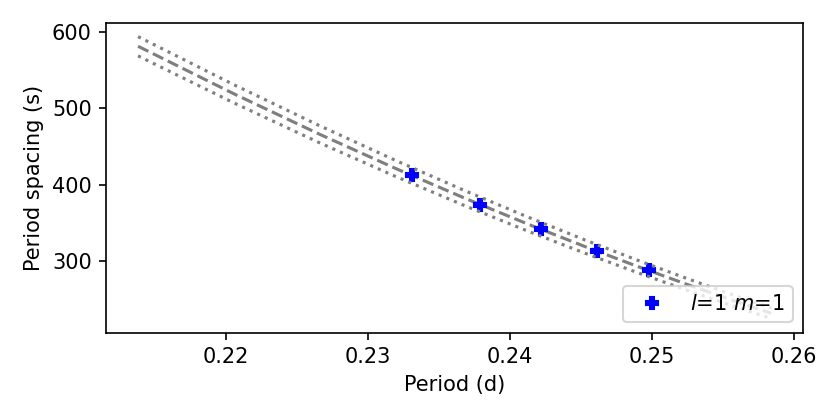}
\caption{Period spacing pattern(s) from the TAR fitting of TIC\,372912679. The dashed line marks the best-fitting model while the dotted lines exhibit the uncertainty region.}
\label{appendix_fig:TIC372912679_period_spacing_TAR}
\end{figure*}

\begin{figure*}
\centering
\includegraphics[width = 0.6\linewidth]{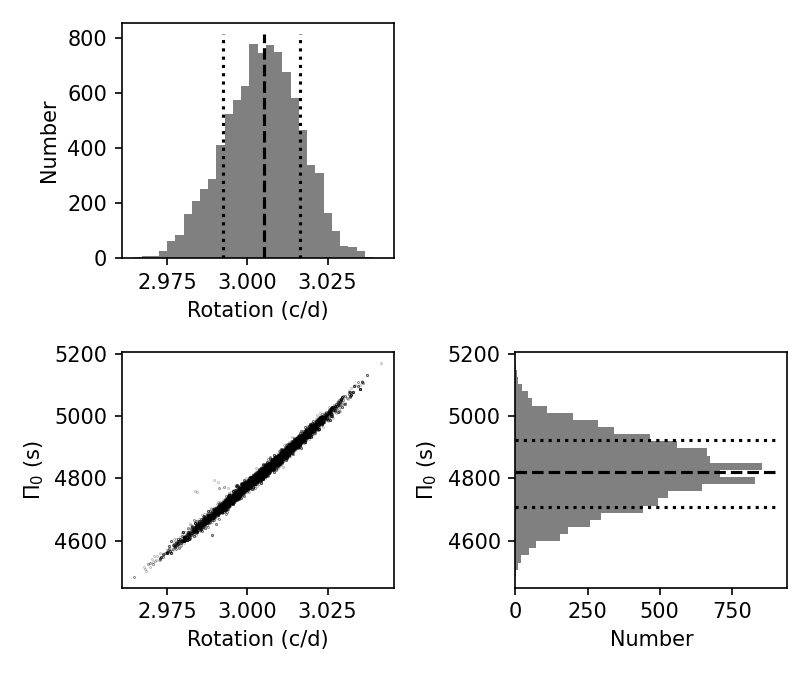}
\caption{Posterior distributions of the TAR fitting of TIC\,372912679.}
\label{appendix_fig:TIC372912679_period_spacing_TAR_corner}
\end{figure*}

\begin{figure*}
\centering
\includegraphics[width = 0.9\linewidth]{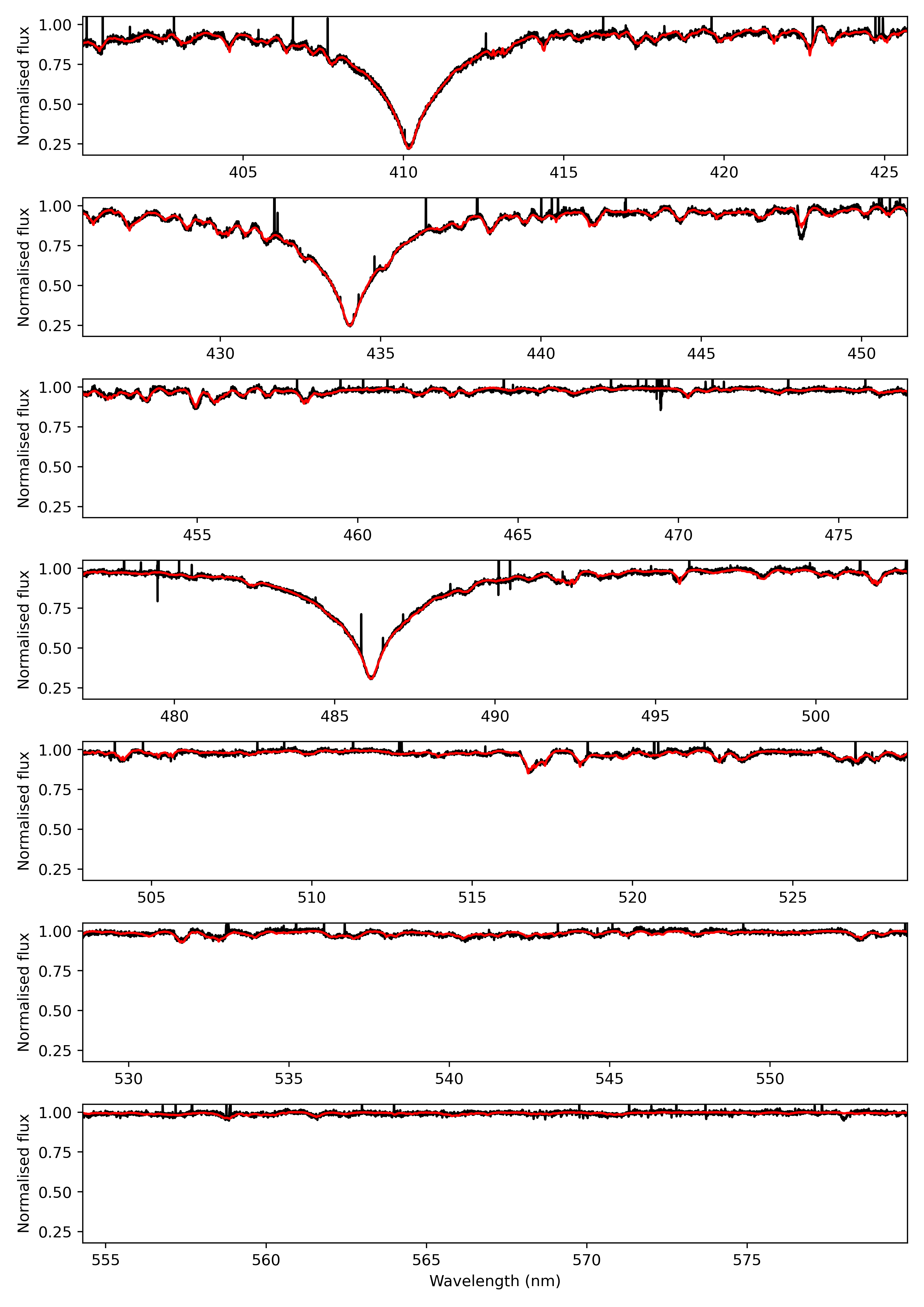}
\caption{Spectrum of TIC\,372912679 observed at 2023-03-14T01:48:11.434. The red line is the best-fitting theoretical spectrum. }
\label{appendix_fig:TIC372912679_spectrum_at_2023-03-14T01:48:11.434}
\end{figure*}

\begin{figure*}
\centering
\includegraphics[width = 0.9\linewidth]{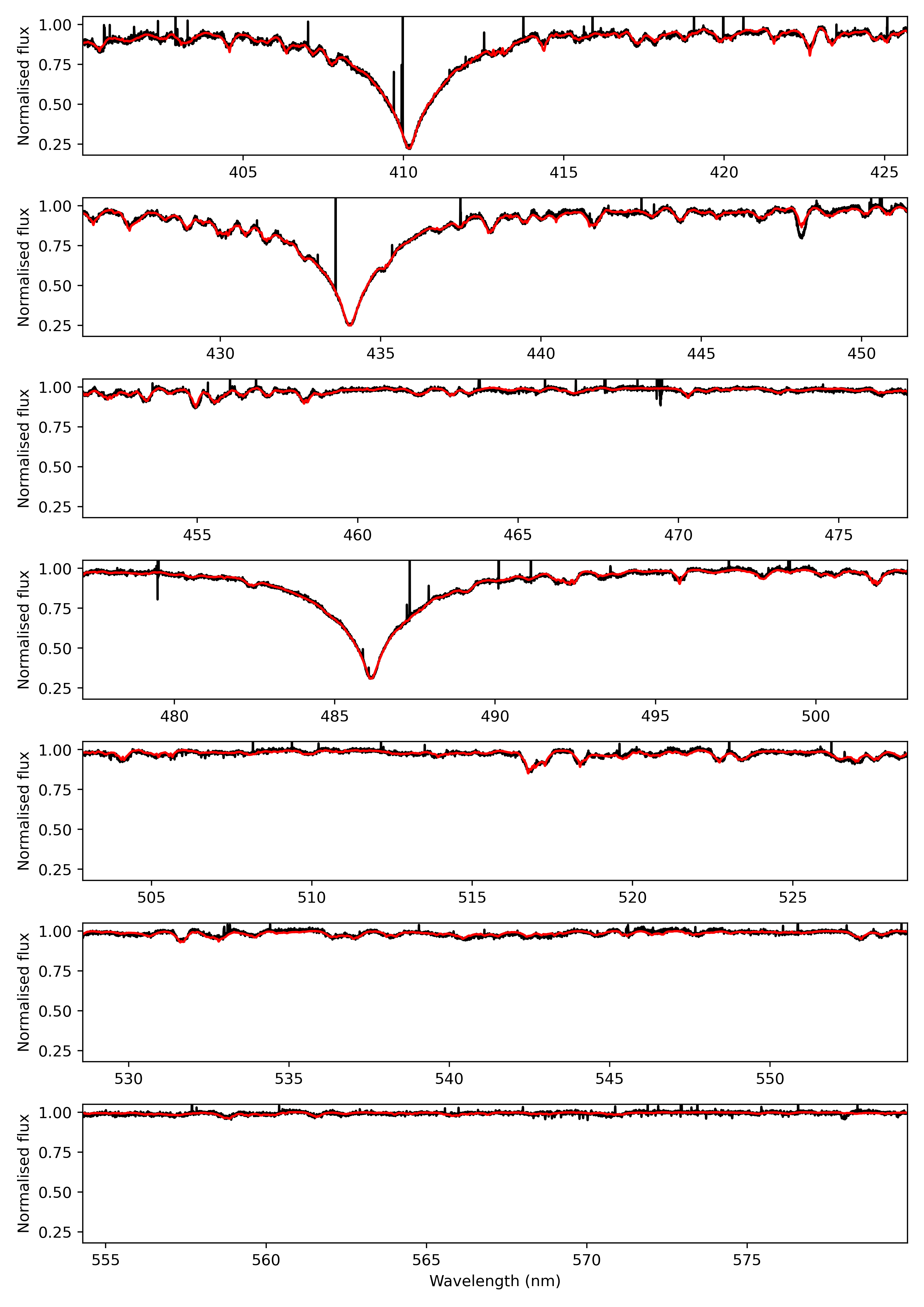}
\caption{Spectrum of TIC\,372912679 observed at 2023-03-15T01:01:24.311. The red line is the best-fitting theoretical spectrum. }
\label{appendix_fig:TIC372912679_spectrum_at_2023-03-15T01:01:24.311}
\end{figure*}

\newpage
\newpage
\begin{figure*}
\centering
\includegraphics[width = 0.9\linewidth]{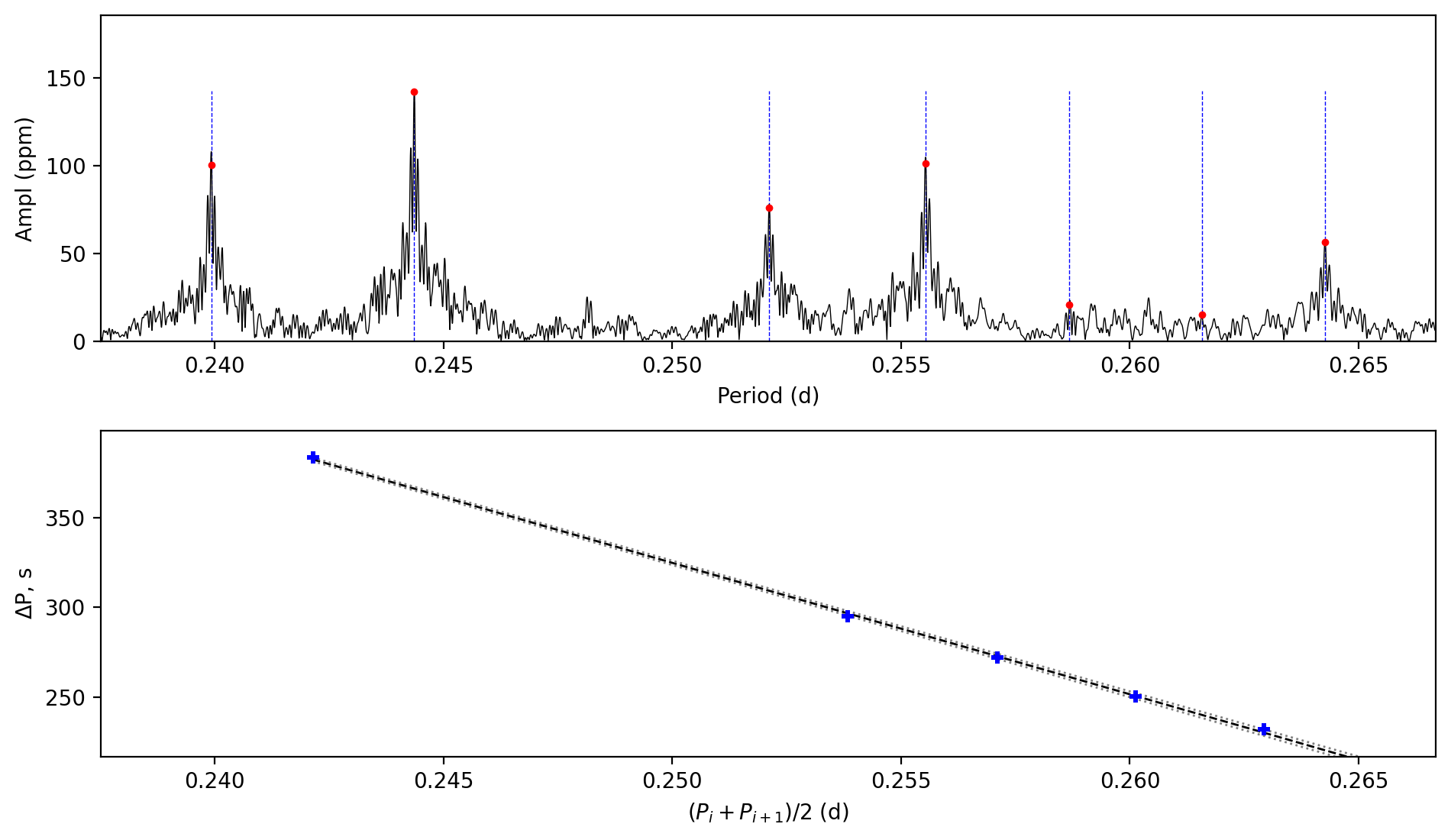}
\caption{Period spacing pattern(s) of TIC\,372913043. Top: amplitude spectrum as a function of period. Red dots are the peaks from the prewhitening process. Vertical lines mark the identified modes. Bottom: period spacing as a function of period. The black line is a linear fitting not the TAR fitting, and the grey lines exhibit the uncertainty region. The period is plotted as the mean value of the two consecutive peaks.}
\label{appendix_fig:TIC372913043_period_spacing}
\end{figure*}

\begin{figure*}
\centering
\includegraphics[width = 0.6\linewidth]{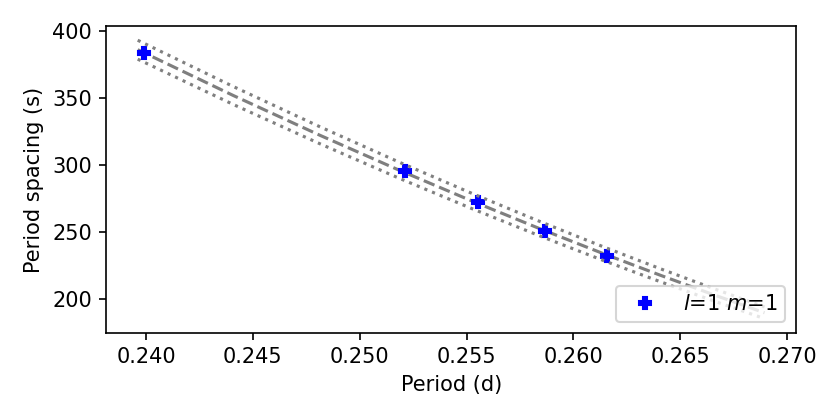}
\caption{Period spacing pattern(s) from the TAR fitting of TIC\,372913043. The dashed line marks the best-fitting model while the dotted lines exhibit the uncertainty region.}
\label{appendix_fig:TIC372913043_period_spacing_TAR}
\end{figure*}

\begin{figure*}
\centering
\includegraphics[width = 0.6\linewidth]{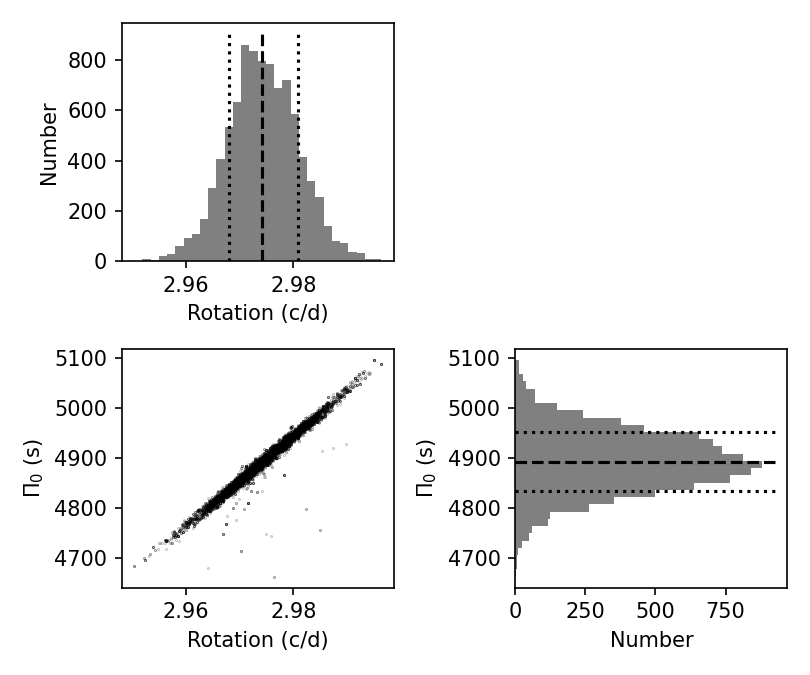}
\caption{Posterior distributions of the TAR fitting of TIC\,372913043.}
\label{appendix_fig:TIC372913043_period_spacing_TAR_corner}
\end{figure*}

\begin{figure*}
\centering
\includegraphics[width = 0.9\linewidth]{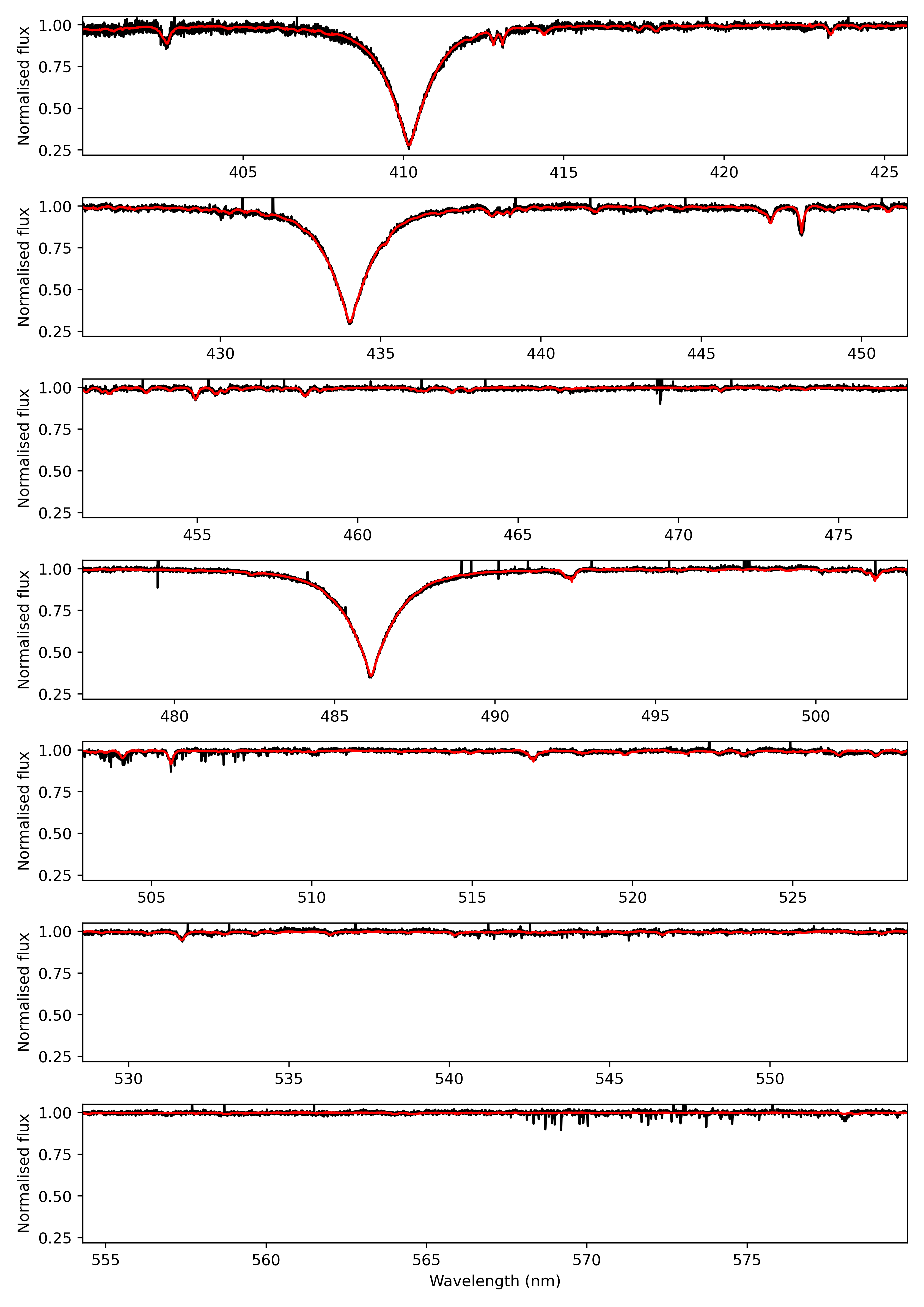}
\caption{Spectrum of TIC\,372913043 observed at 2023-03-12T06:59:31.142. The red line is the best-fitting theoretical spectrum. }
\label{appendix_fig:TIC372913043_spectrum_at_2023-03-12T06:59:31.142}
\end{figure*}

\newpage
\newpage
\begin{figure*}
\centering
\includegraphics[width = 0.9\linewidth]{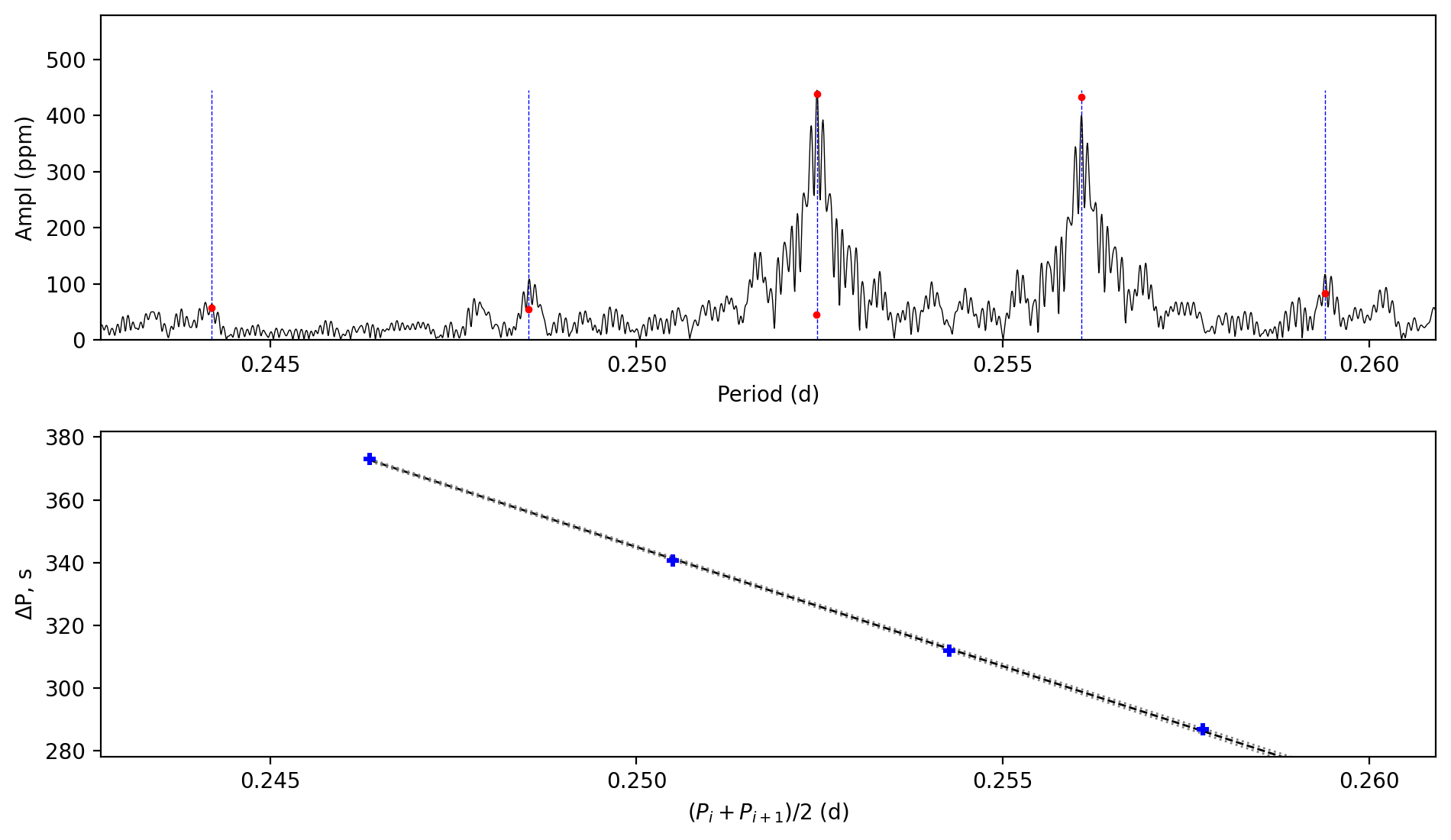}
\caption{Period spacing pattern(s) of TIC\,410451583. Top: amplitude spectrum as a function of period. Red dots are the peaks from the prewhitening process. Vertical lines mark the identified modes. Bottom: period spacing as a function of period. The black line is a linear fitting not the TAR fitting, and the grey lines exhibit the uncertainty region. The period is plotted as the mean value of the two consecutive peaks.}
\label{appendix_fig:TIC410451583_period_spacing}
\end{figure*}

\begin{figure*}
\centering
\includegraphics[width = 0.6\linewidth]{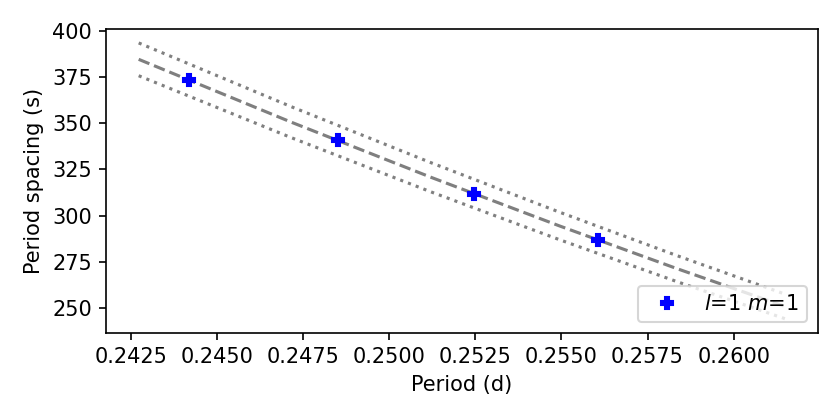}
\caption{Period spacing pattern(s) from the TAR fitting of TIC\,410451583. The dashed line marks the best-fitting model while the dotted lines exhibit the uncertainty region.}
\label{appendix_fig:TIC410451583_period_spacing_TAR}
\end{figure*}

\begin{figure*}
\centering
\includegraphics[width = 0.6\linewidth]{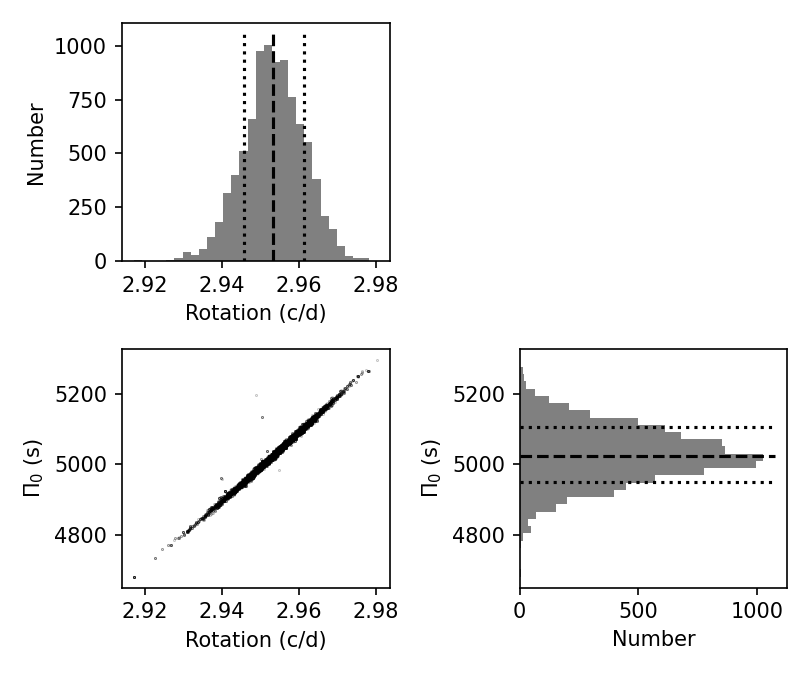}
\caption{Posterior distributions of the TAR fitting of TIC\,410451583.}
\label{appendix_fig:TIC410451583_period_spacing_TAR_corner}
\end{figure*}

\begin{figure*}
\centering
\includegraphics[width = 0.9\linewidth]{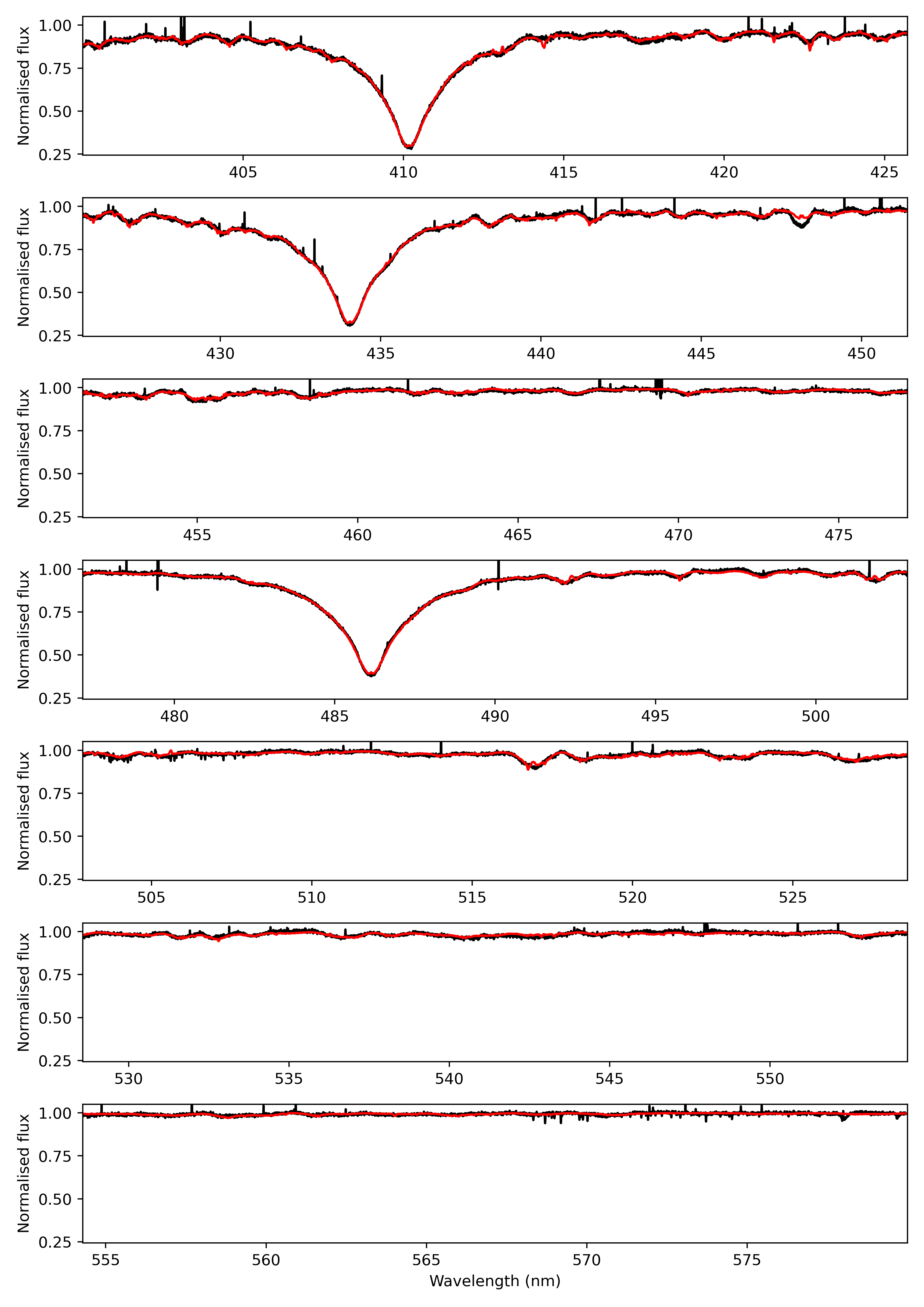}
\caption{Spectrum of TIC\,410451583 observed at 2023-03-14T04:21:29.958. The red line is the best-fitting theoretical spectrum. }
\label{appendix_fig:TIC410451583_spectrum_at_2023-03-14T04:21:29.958}
\end{figure*}

\newpage
\newpage
\begin{figure*}
\centering
\includegraphics[width = 0.9\linewidth]{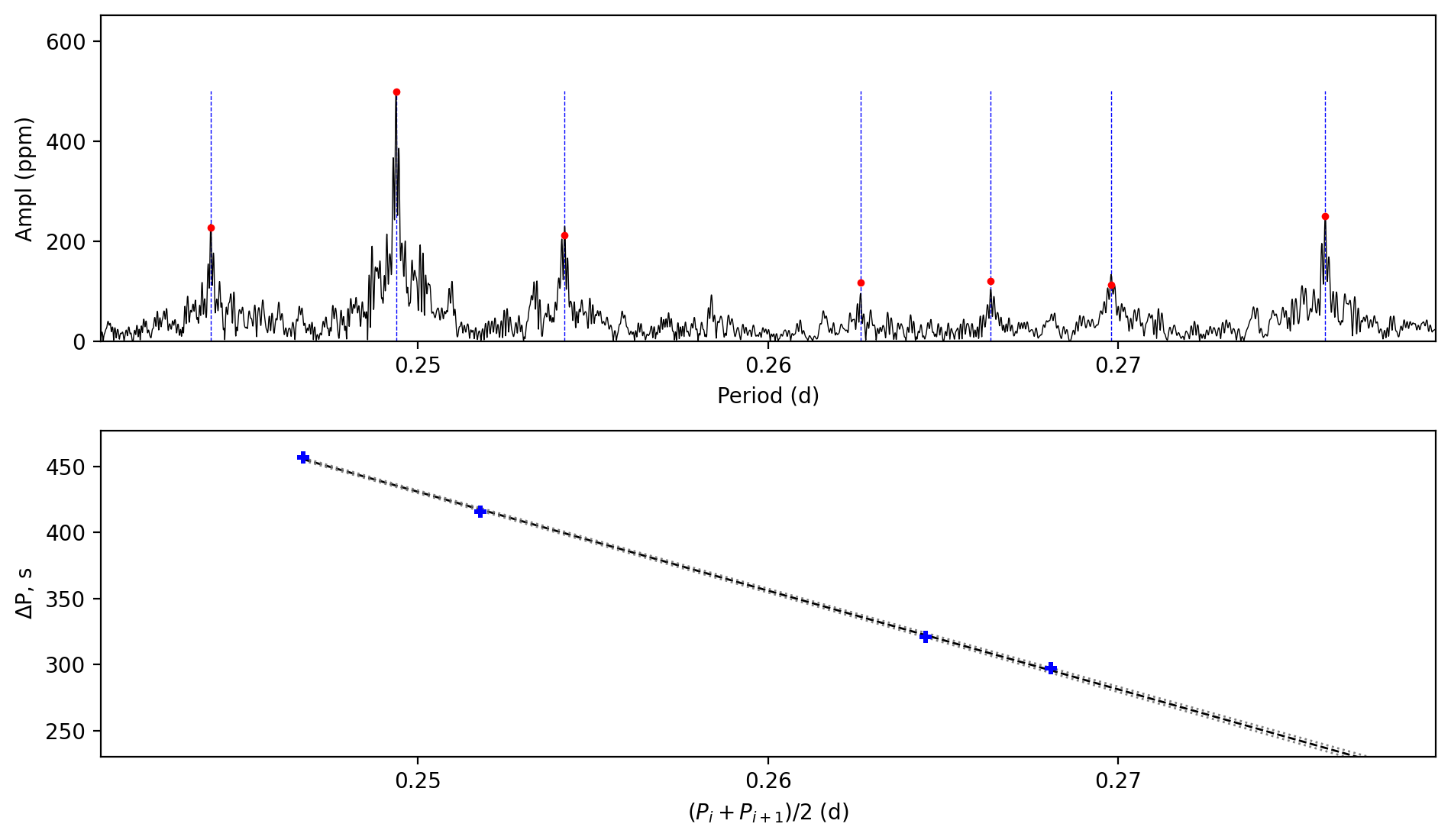}
\caption{Period spacing pattern(s) of TIC\,410452218. Top: amplitude spectrum as a function of period. Red dots are the peaks from the prewhitening process. Vertical lines mark the identified modes. Bottom: period spacing as a function of period. The black line is a linear fitting not the TAR fitting, and the grey lines exhibit the uncertainty region. The period is plotted as the mean value of the two consecutive peaks.}
\label{appendix_fig:TIC410452218_period_spacing}
\end{figure*}

\begin{figure*}
\centering
\includegraphics[width = 0.6\linewidth]{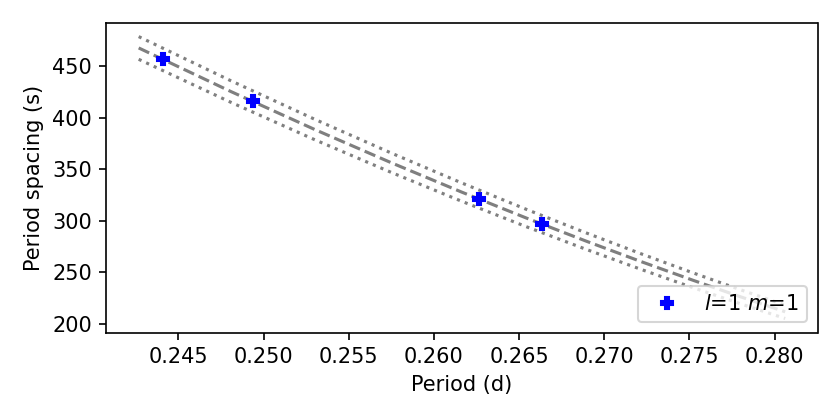}
\caption{Period spacing pattern(s) from the TAR fitting of TIC\,410452218. The dashed line marks the best-fitting model while the dotted lines exhibit the uncertainty region.}
\label{appendix_fig:TIC410452218_period_spacing_TAR}
\end{figure*}

\begin{figure*}
\centering
\includegraphics[width = 0.6\linewidth]{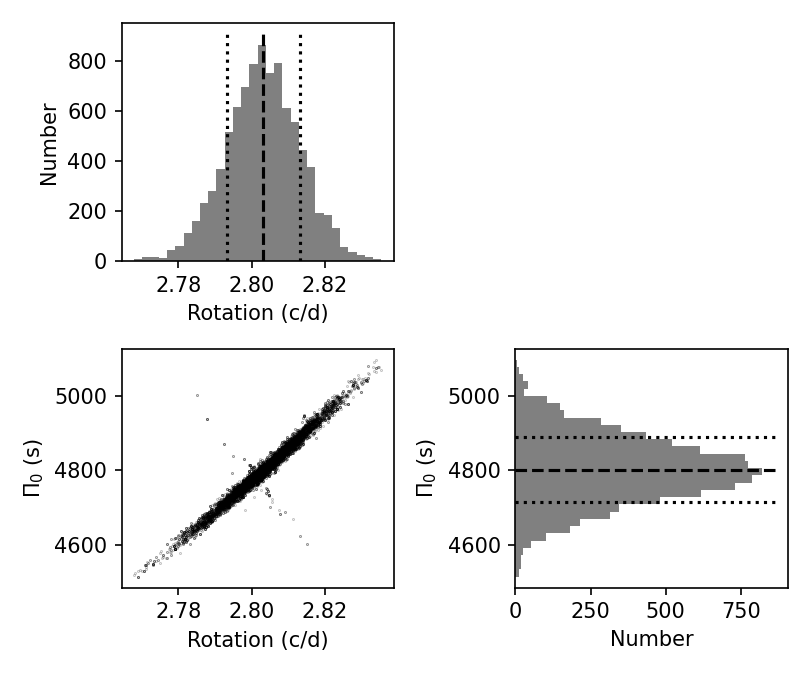}
\caption{Posterior distributions of the TAR fitting of TIC\,410452218.}
\label{appendix_fig:TIC410452218_period_spacing_TAR_corner}
\end{figure*}

\begin{figure*}
\centering
\includegraphics[width = 0.9\linewidth]{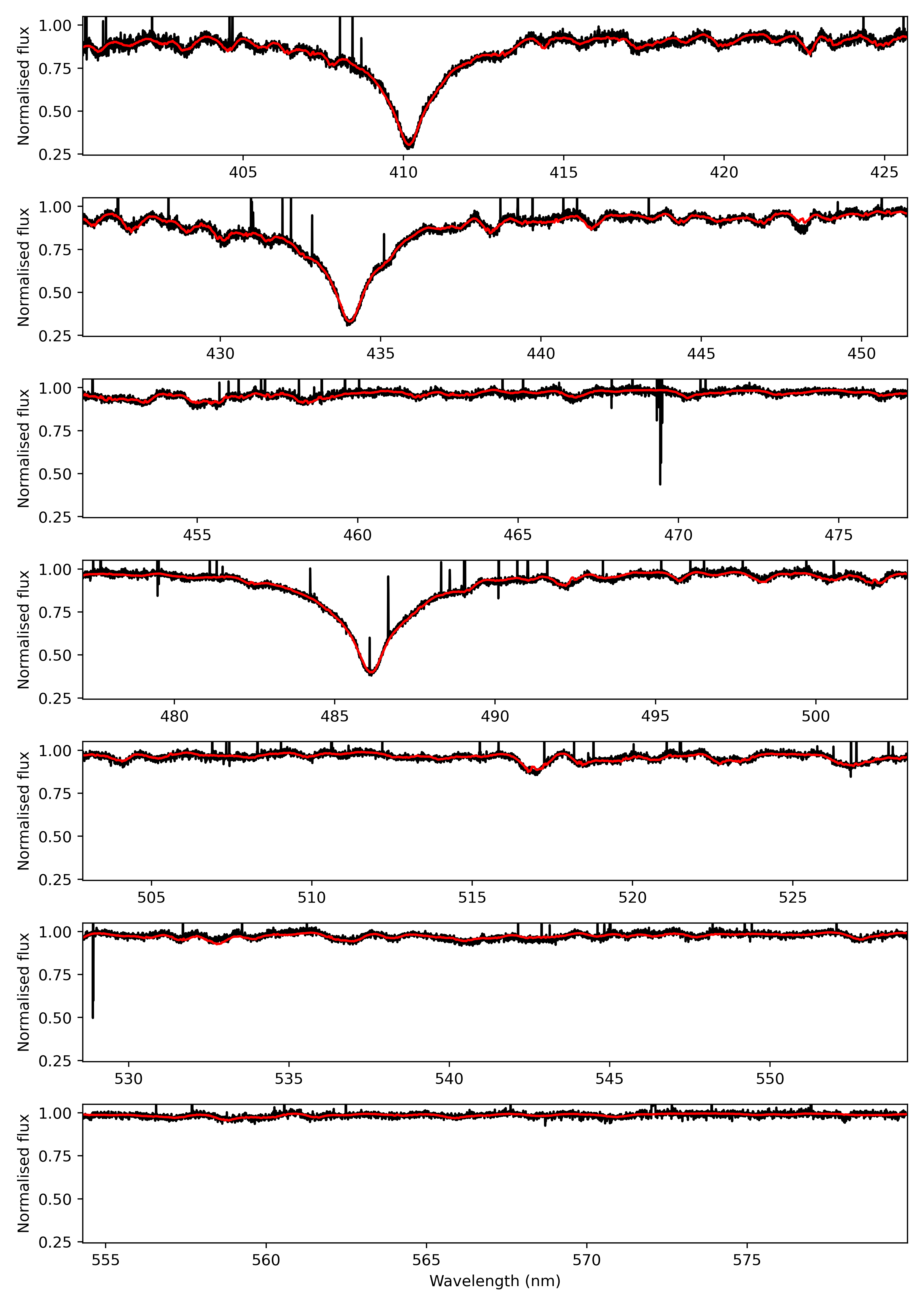}
\caption{Spectrum of TIC\,410452218 observed at 2023-03-14T02:24:11.888. The red line is the best-fitting theoretical spectrum. }
\label{appendix_fig:TIC410452218_spectrum_at_2023-03-14T02:24:11.888}
\end{figure*}

\begin{figure*}
\centering
\includegraphics[width = 0.9\linewidth]{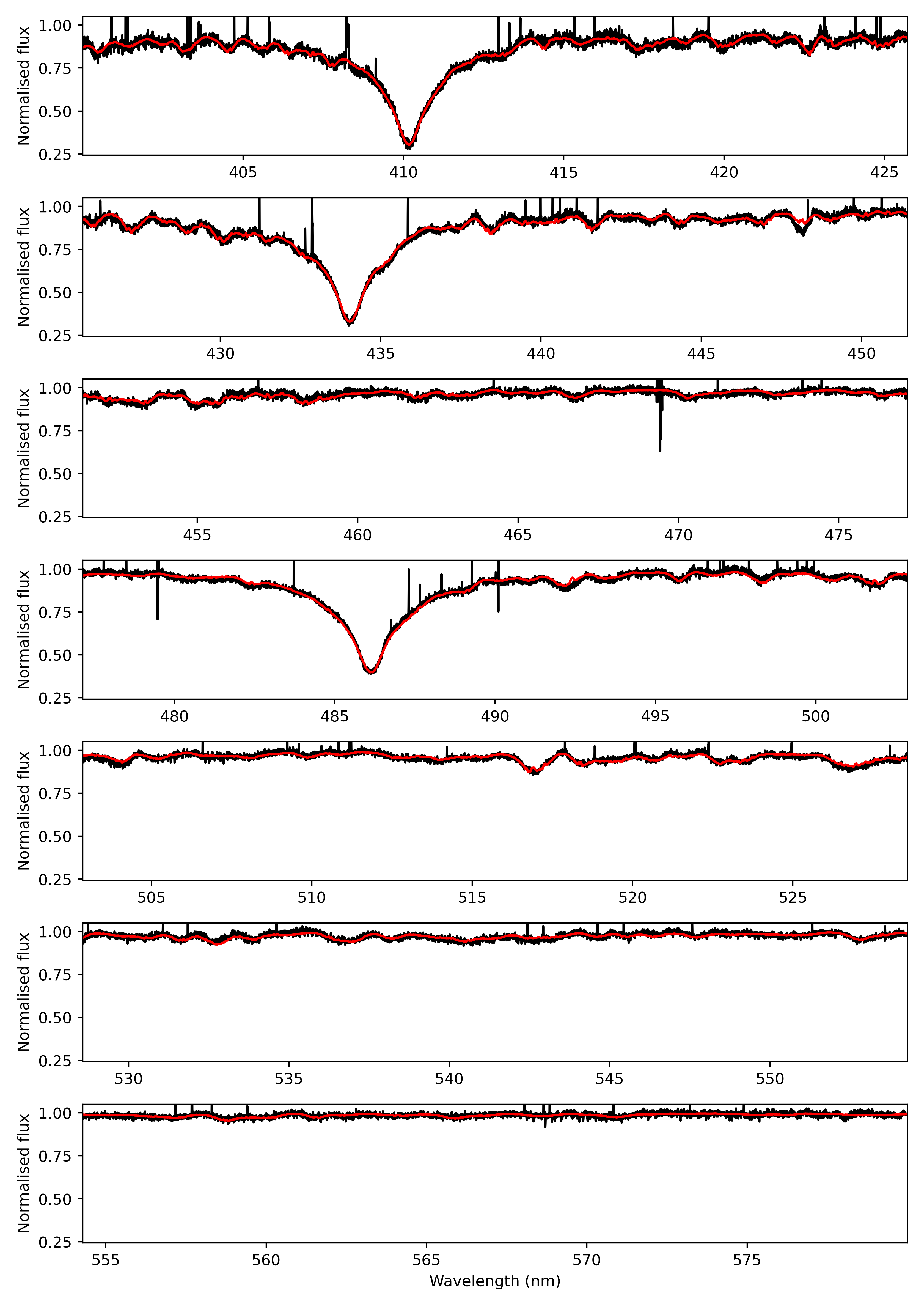}
\caption{Spectrum of TIC\,410452218 observed at 2023-03-16T02:37:05.256. The red line is the best-fitting theoretical spectrum. }
\label{appendix_fig:TIC410452218_spectrum_at_2023-03-16T02:37:05.256}
\end{figure*}

\newpage

\end{appendix}

\end{document}